\documentclass[12pt]{article}
\usepackage[latin9]{inputenc}
\usepackage{color}
\usepackage{verbatim}
\usepackage{mathtools}
\usepackage{enumitem}
\usepackage{amsmath}
\usepackage{amsthm}
\usepackage{amssymb}
\usepackage{amsfonts}
\usepackage{wasysym}
\usepackage{anyfontsize}
\usepackage{mathrsfs}
\usepackage{natbib}
\usepackage{xr-hyper}
\usepackage{url} 
\usepackage{lineno}
\newcommand{\blind}{1}

\RequirePackage[colorlinks,citecolor=green,urlcolor=blue]{hyperref}
\newcommand{\revisetext}[1]{#1}

\makeatletter

\theoremstyle{remark}
\newtheorem{rem}{Remark}
\theoremstyle{plain}
\newtheorem{thm}{Theorem}[section]

\newtheorem{coro}[thm]{Corollary}

\global\long\def\argmax{\mathop{\mathrm{argmax}}}%

\global\long\def\bM{\mathbf{M}}%

\global\long\def\bU{\mathbf{U}}%

\global\long\def\bV{\mathbf{V}}%

\newcommand\mc[1]{\multicolumn{1}{c}{#1}}

\begin{document}

	\def\spacingset#1{\renewcommand{\baselinestretch}%
		{#1}\small\normalsize} \spacingset{1}

	
	\date{}
	\if1\blind
	{
		\title{\bf Kernel meets sieve: transformed hazards models with sparse longitudinal
			covariates}
		\author{Dayu Sun\\
			Department of Biostatistics and Health Data Science,\\ Indiana University School of Medicine\\
			Zhuowei Sun  \\
			School of Mathematics, Jilin University\\
            School of Public Health, Dalian Medical University\\
			Xingqiu Zhao \thanks{Corresponding author} \\
			Department of Applied Mathematics,\\ The Hong Kong Polyetechnic University\\
			and \\
			Hongyuan Cao  \thanks{Corresponding author}\\
			Department of Statistics, Florida State University
		}
		\maketitle
	} \fi
	
	\if0\blind
	{
		\bigskip
		\bigskip
		\bigskip
		\begin{center}
			{\LARGE\bf Kernel meets sieve: transformed hazards models with sparse longitudinal
				covariates}
		\end{center}
		\medskip
	} \fi
	
	\bigskip
	\begin{abstract}
		We study the transformed hazards model with time-dependent covariates observed intermittently for the censored outcome. Existing work assumes the availability of the whole trajectory of the time-dependent covariates, 
		which is unrealistic. 
		We propose combining kernel-weighted log-likelihood and sieve maximum log-likelihood estimation to conduct statistical inference. The method is robust and easy to implement. 
		We establish the asymptotic properties of the proposed estimator and contribute to a rigorous theoretical framework for general kernel-weighted sieve M-estimators.
		Numerical studies corroborate our theoretical results and show that the proposed method performs favorably over competing methods.
		The analysis of a data set from a COVID-19 study in Wuhan identifies clinical predictors that otherwise cannot be obtained using existing methods. 
	\end{abstract}
	
	\noindent%
	{\it Keywords:}  Kernel, semiparametric model, sieve estimation, survival analysis
	\vfill
	
	\newpage
	\spacingset{1.9} 
	
	\def\arraystretch{0.6}
	\section{Introduction}
	
	The transformed hazards model provides flexible modeling of survival data by imposing an additive semiparametric structure and a known transformation on the hazard function. Specifically, given the covariate $Z(t)$, the hazard function for survival times takes the form of
	\begin{equation}
		G\{\lambda(t \mid Z(t)) \} = \alpha(t)  + \beta^T Z(t), \label{eq:transform-hazard} 
	\end{equation}
	where $\lambda(t \mid Z(t))$ is the hazard function given covariate $Z(t),$ $\alpha(t)$ is an unknown baseline hazard function, $\beta$ is the unknown regression coefficient vector and $G(\cdot)$ is a known increasing transformation function. An example of the transformation is the Box-Cox transformation \citep{Box1964}, where
	\begin{equation} 
		G(x)=(x^{s}-1)/s,\label{eq:box-cox}
	\end{equation}
	if $s>0$
	and $G(x)=\log(x)$ if $s=0.
	$
	The Box-Cox transformation includes the multiplicative proportional hazards model when $s=0$ \citep{Cox1972} and the additive hazards model when $s=1$ \citep{aalen1989linear,Lin1994}.
	
	
	The classic analysis of \eqref{eq:transform-hazard} requires the availability of the entire trajectory of $Z(t)$ \citep{Zeng2005, Lu2006,Kang2015}, which may not be realistic. In practice, $Z(t)$ is commonly measured at various sparse and intermittent time points through longitudinal studies 
	\citep{Cao2015}.
	For example, in a COVID-19 study that motivates our study, vital signs from patients were collected longitudinally at irregular time intervals, serving as covariates of hospital discharge times. For such longitudinal covariates, 
	a na\"{i}ve approach is imputing unobserved $Z(t)$ by its most recently observed value, known as the last value carried forward (LVCF). However, this approach 
	may lead to bias in estimation \citep{Kenward2009,Lachin2015,Cao2016}, as illustrated by simulation studies in Section~\ref{sec:simulation}. 
	
	To the best of our knowledge, there is no statistical methods with theoretical justification 
	for general transformed hazards model \eqref{eq:transform-hazard} with sparse longitudinal covariates.
	Some special cases were considered in the literature. 
	For example, the proportional hazards model was analyzed using kernel-smoothed partial likelihood in \cite{Cao2015, cao2021proportional}. Inference based on the additive hazards model was performed using smoothed martingale-based estimation equations \citep{Cao2021}. Both of them circumvented the estimation of the unknown nonparametric baseline hazard function. 
	The advantages of such kernel-weighting methods include easy computation and robust inference. 
	Nonetheless, these methods do not apply to 
	\eqref{eq:transform-hazard} since $\alpha(t)$ cannot be profiled out. 

	This paper proposes a sieve maximum kernel-weighted log-likelihood estimator (SMKLE) for survival analysis with sparse longitudinal covariates in a general class of transformed hazards models \eqref{eq:transform-hazard}.
	We use the kernel weighting on log-likelihood to handle intermittently observed sparse longitudinal covariates.
	The proposed kernel weighting is imposed on the individual log-likelihood contribution from each observed covariates, distinguishing itself from the conventional kernel smoothing on covariates.
	We use a sieve to estimate the nonparametric baseline hazard function $\alpha(t)$.
	To rigorously justify SMKLE and conduct statistical inference, it is crucial to study its asymptotic properties.
	However, the SMKLE belongs to a new class of estimators, called \emph{ kernel-weighted sieve M-estimators}, whose theoretical properties have not been thoroughly investigated before. 
	The intertwined effect of kernel and sieve on M-estimator poses unique theoretical challenges to the asymptotic theory derivation, which has not been considered to our best knowledge. 
	While several papers have explored the kernel-smoothing methods for various data types with sparse longitudinal covariates,
	most of them 
	belong to 
	Z-estimators based on estimating equations \citep{Cao2015,Cao2016, cao2021proportional, Cao2021, Sun2021,Sun2021recurrent, Lyu2021}.
	

	To fill research gaps, we develop general and nontrivial asymptotic theories for kernel-weighted sieve M-estimators. 
	
	Our main contributions are as follows:
	




	

	%
	%
	%
	%
	%

	(i) 
	We develop a new theorem for the almost sure convergence of empirical processes of general kernel-weighted functions uniform in kernel bandwidth $h$ and a sieve space that varies with sample size $n$.
	Our theorem shows that the uniform convergence rate is $O(\sqrt{n^{1-\nu}h})$ where $\nu$ quantifies the dimension of the sieve space as $O(n^{\nu})$.
	This result serves as a cornerstone for the asymptotic theory of the SMKLE and lays the theoretical foundation for general kernel weighting methods with sparse longitudinal covariates.
	One challenge is that traditional Donsker class arguments for sieve estimation do not apply directly because the kernel-weighted functions are not square-integrable.
	To overcome this challenge, our proof combines 
	the theory of kernel estimation, bracketing number arguments, 
	and maximal inequalities of empirical processes \citep{Vaart1996, Einmahl2005}. 

	(ii)  We prove the strong consistency of the SMKLE and derive the convergence rate of the overall SMKLE as $n^{\min\{ (1-\nu-a)/2,\kappa\nu\}}$ where $h = O(n^{-a})$ and $\kappa \ge 2$ is an integer such that the derivative of the baseline hazard function, $\alpha$, is Lipschitz continuous up to order $\kappa$. 
	This convergence rate of the SMKLE is different from that of the kernel density or regression estimation \citep{Tian2005,Tsybakov2009,Cao2016} in that it is characterized by 
	the interplay between the kernel bandwidth and the dimension of the sieve space.
	One of the major challenges in establishing a strong consistency and convergence rate is to 
	explicitly bound 
	several uniform 
	approximation errors between the 
	expected kernel-weighted functions and their corresponding limits when the bandwidth goes to zero as tight as possible.
	To address this challenge, we carefully develop various tight inequalities involving the bandwidth, 
	the smoothness of the underlying functions, and the size of the parameter space. 
	We discover that the approximation error bounds determine the optimal rate of convergence. 
	
	
	
	
	(iii) We derive the explicit form of the asymptotic normality of $\hat{\beta}$.
	Since the kernel is involved, 
	general theorems on the asymptotic normality of semiparametric M-estimators are not directly applicable 
	\citep{Wellner2007,Kosorok2008,Ding2011}.
	We establish a new general 
	asymptotic normality result with rate $\sqrt{nh}$ that applies to kernel-weighted sieve M-estimators and general kernel-based methods.
	More importantly, the asymptotic normality of $\hat{\beta}$ yields an optimal convergence rate of $\hat{\beta}$ at $o(n^{3/8})$, 
	improving the corresponding rate $o(n^{1/3})$ 
	in \citet{Cao2015}  and \citet{Cao2021}.
	This improvement confirms that likelihood-based SMKLE could be more efficient than the methods based on estimating equations in \citet{Cao2015} and \citet{Cao2021}, further reinforced by simulation studies in Section~\ref{sec:simulation}.

	The rest of the paper is organized as follows.
	Section~\ref{sec:estimation} introduces notation and proposes the SMKLE.
	In Section~\ref{sec:asym}, we rigorously investigate the asymptotic properties of the new 
	SMKLE, whose proofs are relegated in the supplementary materials.
	We introduce essential notation and regularity conditions in Section~\ref{subsec:notation}.
	In Section~\ref{subsec:uibCR}, we show a general theorem for the 
	uniform convergence results of two general kernel-weighted sieve functions in Theorem~\ref{prop:wrate}.
	In Section~\ref{subsec:asymres}, we prove the strong consistency, rate of convergence, and asymptotic normality of SMKLE in Theorems~\ref{them:consistency}-\ref{them:normality}.
	Extensive 
	simulation studies in Section \ref{sec:simulation} demonstrate that
	the proposed method works well in 
	finite samples and has favorable performances over 
	existing 
	methods.
	In Section \ref{sec:application}, we apply the proposed method to analyze the time to hospital discharge in a longitudinal observational study of COVID-19 conducted in Wuhan, China.
	Concluding remarks can be found in Section \ref{sec:conclusion}.

	

	%

	\section{Estimation procedure\label{sec:estimation}}
	
	\subsection{Notation}\label{notation}
	

	Suppose that we have $n$ independent and identically distributed random sample. For the $i$th subject, 
	let $T_i$ denote the failure time, $C_i$ be the censoring time, and $Z_i(t)$ be the time-dependent covariate process. 
	We assume that the censoring is coarsened at random so that $T_i$ and $C_i$ are
	conditionally independent given $Z_i(\cdot), i =1, \ldots, n.$
	Denote 
	$X_i=\min\left(T_i,C_i\right)$ and 
	$\Delta_i = 1\{T_i\le C_i\}.$ 
	For the $i$th subject, the longitudinal covariates $Z_i\left(t\right)$ are observed at $R_{ij}, j =1, \ldots, J_i,$ which can differ across subjects.  
	The measurement times $\{R_{ij}, j =1, \ldots, J_{i}\}$ are assumed 
	to be independent of $Z_i\left(t\right), i =1, \ldots, n.$ Using counting process notation, we have
	\begin{equation*}\label{counting}
		N_i(t) = \sum_{j=1}^{J_i}1\{R_{ij} \le t \} = N_i^*(t \wedge \tau).
	\end{equation*}
	Here, the underlying uncensored observation process $N_i^*(t) = \sum_{j=1}^{\infty}1\{R_{ij}\le t \}$ and $\tau$ is the end of study time.   
	Denote the observed data as 
	$O_{i}\coloneqq\{(X_{i},Z_{i}(R_{ij}),R_{ij},\Delta_{i}),j=1,\ldots,J_{i}\}$,
	$i=1,\ldots,n.$ 

	\subsection{Kernel-smoothed log-likelihood function}
	
	A natural inference method for model
	\eqref{eq:transform-hazard} is the maximum likelihood estimation 
	as that in \citet{Zeng2005}.
	Under model \eqref{eq:transform-hazard}, the log-likelihood function of the unknown parameter and unknown non-parametric function $\theta\coloneqq(\beta,\alpha(t))$ is
	\begin{equation}
		\ell_{n}(\theta)=n^{-1}\sum_{i=1}^{n} \left[ \Delta_{i}\log\left\{
		H\left(\alpha(X_{i})+\beta^{\top}Z_{i}(X_{i})\right)\right\}
		-\int_{0}^{X_{i}}H\left(\alpha(t)+\beta^{\top}Z_{i}(t)\right)dt\right],\label{eq:orilogll}
	\end{equation}
	where $H$ is the inverse function of $G$.
	
	The log-likelihood function \eqref{eq:orilogll} requires knowledge of the
	whole trajectory of $Z_{i}\left(t\right)$ for $t\in\left[0, X_i \right]$ for subject $i.$ 
	In practice, we only have values of 
	$Z_{i}(t)$ observed at
	$R_{ij}, j=1\ldots,J_{i}$.
	To accommodate the sparse longitudinal covariates, we propose the kernel
	smoothing/weighting for the log-likelihood function \eqref{eq:orilogll} as follows.
	\begin{multline}
		\tilde{\ell}_{n}(\theta)=\frac{1}{n}\sum_{i=1}^{n}\sum_{j=1}^{J_{i}} \bigg[K_{h}\left(X_{i}-R_{ij}\right)I\left(R_{ij}\le
		X_{i}\right)\Delta_{i}\log\left\{
		H\left(\alpha\left(X_{i}\right)+\beta^{\top}Z_{i}\left(R_{ij}\right)\right)\right\} \bigg]
		\\ -\frac{1}{n}\sum_{i=1}^{n}\sum_{j=1}^{J_{i}}\left[\int_{0}^{X_{i}}K_{h}\left(t-R_{ij}\right)I\left(R_{ij}\le t\right)H\left(\alpha\left(t\right)+\beta^{\top}Z_{i}\left(R_{ij}\right)\right)dt\right],\label{eq:smoothlogll}
	\end{multline}
	where $K_{h}\left(\cdot\right)=2K(\cdot/h)/h$, $I(\cdot)$ is the indicator function and the kernel function $K(t)$ is a symmetric probability density function
	with mean $0$ and finite variance. In equation \eqref{eq:smoothlogll}, we consider the within-subject smoothing to log-likelihood \eqref{eq:orilogll}.  This approach borrows information from nearby observations in the longitudinal covariate process from the same subject. The contribution of the observed longitudinal covariate closer in time to the failure time receives more weight to the log-likelihood function. We aggregate information by summing up the contributions from all $n$ subjects to obtain \eqref{eq:smoothlogll}.


	\subsection{Sieve estimation by
		B-splines} 
	
	Another challenge in 
	optimizing
	\eqref{eq:smoothlogll} 
	is the infinite
	dimensionality of $\alpha(t)$.
	The SMKLE overcomes this challenge by restricting
	$\alpha(t)$ to a smaller 
	space with finite dimensionality,
	which is called the sieve space for $\alpha(t)$.
	In the following, we adopt B-splines \citep{Schumaker2007} as the sieve method \citep{Shen1994,Shen1997,Chen2007} to approximate $\alpha(t)$ due to its prevalent use in the literature \citep{Ding2011,Zhao2017a}; but the proposed framework can easily be generalized to other sieve methods, such as the Bernstein polynomial \citep{Zhou2017, Zhao2020}.
	
	In the finite closed interval $\left[0,\tau\right]$, 
	let $\mathcal{T}_{n}=\left\{
	t_{i}\right\} _{1}^{m_{n}+2l}$ with
	$0=t_{1}=\cdots=t_{l}<t_{l+1}<\cdots<t_{m_{n}+l}<t_{m_{n}+l+1}=\cdots=t_{m_{n}+2l}=\tau$
	be a sequence of knots that partition $\left[0,\tau\right]$ into $m_{n}+1$
	subintervals $\left[t_{l+i},t_{l+i+1}\right]$ for $i=0,\dots,m_{n}$.
	Here, $m_{n} \to \infty$ as $n\to\infty$.
	Define $\mathcal{A}_{n,l}$, the class of B-splines of order $l$ with the knots
	sequence $\mathcal{T}_{n}$, as
	\begin{equation}
		\mathcal{A}_{n,l}=\left\{
		\alpha_{n}\left(t\right)\coloneqq\sum_{k=0}^{q_{n}}\gamma_{nk}B_{k}\left(t\right)=\boldsymbol{\gamma}_{n}^{\top}\mathbf{B}\left(t\right);\sum_{k=0}^{q_{n}}\gamma^{2}_{nk}<\infty\right\} , \label{eq:seivespace}
	\end{equation}
	where $q_{n}=m_{n}+l$, $\left\{ B_{k}\left(t\right),0\le k\le
	q_{n}\right\} $ is the basis of the B-spline class,
	$\mathbf{B}\left(t\right)\coloneqq\left(B_{0}\left(t\right),\ldots,B_{q_n}\left(t\right)\right)^{\top}$, and
	$\boldsymbol{\gamma}_{n}\coloneqq\left(\gamma_{n0},\ldots,\gamma_{nq_{n}}\right)^{\top}$.
	We 
	approximate
	$\alpha\left(t\right)$ by
	$\alpha_{n}\left(t\right)=\boldsymbol{\gamma}_{n}^{\top}\mathbf{B}\left(t\right)\in \mathcal{A}_{n,l}$.
	
	Equipped with $\mathcal{A}_{n,l}$, for $\vartheta_{n}\coloneqq\left(\beta,\gamma_{n}\right)$, \eqref{eq:smoothlogll} becomes
	\begin{multline*}
		\tilde{\ell}_{n}\left(\vartheta_{n}\right)=\frac{1}{n}\sum_{i=1}^{n}\left[\int_{0}^{X_{i}}K_{h}\left(X_{i}-r\right)\Delta_{i}\log\left\{
		H\left(\boldsymbol{\gamma}_{n}^{\top}\mathbf{B}\left(X_{i}\right)+\beta^{\top}Z_{i}\left(r\right)\right)\right\}
		dN_{i}\left(r\right) \right]\\
		-\frac{1}{n}\sum_{i=1}^{n}\left[\int_{0}^{X_{i}}\int_{0}^{t}K_{h}\left(t-r\right)H\left(\boldsymbol{\gamma}_{n}^{\top}\mathbf{B}\left(t\right)+\beta^{\top}Z_{i}\left(r\right)\right)dN_{i}\left(r\right)dt\right]. 
	\end{multline*}
	We can directly estimate
	$\vartheta_{n}$ by $\hat{\vartheta}_{n}=(\hat{\beta},\hat{\boldsymbol{\gamma}}_{n})=\argmax_{\vartheta_{n}\in\mathbb{R}^{p}\times\mathbb{R}^{q_{n}}}\tilde{\ell}_{n}\left(\vartheta_{n}\right)$, which could be solved by generic optimization algorithms.
	Then, we obtain the SMKLE for $\theta$ by
	$\hat{\theta}_{n}\coloneqq(\hat{\beta},\hat{\alpha}_{n}(t)\coloneqq\hat{\boldsymbol{\gamma}}_{n}^{T}\mathbf{B}\left(t\right))$.

	\subsection{Data-driven bandwidth selection\label{subsec:bandselect}}
	
	For SMKLE, it is vital to select a
	proper bandwidth to achieve 
	good performance \citep{Cao2015a,Cao2015}.
	We propose to select a bandwidth that minimizes the cross-validated (CV)
	negative kernel-weighted log-likelihood function.
		For $M$-fold cross-validation, we divide the whole dataset into $M$ subsets
		$\mathbf{D}^{\left(k\right)}$, i.e., $M$ folds, $k=1,\ldots,M$, with roughly
		equal sample size $n^{\left(k\right)}$.
		For each fold, we obtain
		$\hat{\theta}_{n^{\left(-k\right)}}^{-\left(k\right)}\left(h\right)$ from the
		remaining $K-1$ folds, i.e., the training set $\mathbf{D}^{-\left(k\right)}$
		with a bandwidth $h$ where $n^{\left(-k\right)}$ is the sample size of
		$\mathbf{D}^{-\left(k\right)}$.
		Then, we select $h$ minimizing the CV loss
		$-\sum_{k=1}^{M}\tilde{\ell}_{n^{\left(k\right)}}\left(\hat{\theta}_{n^{\left(-k\right)}}^{-\left(k\right)}\left(h\right)|\mathbf{D}^{\left(k\right)}\right)$
		where
		$\tilde{\ell}_{n^{\left(k\right)}}\left(\hat{\theta}_{n^{\left(-k\right)}}^{-\left(k\right)}\left(h\right)|\mathbf{D}^{\left(k\right)}\right)$
		is the smoothed log-likelihood function evaluated at
		$\hat{\theta}_{n^{\left(-k\right)}}^{-\left(k\right)}\left(h\right)$ for the
		$k$th testing dataset $\mathbf{D}^{\left(k\right)}$.
	We can also use the CV error to evaluate the goodness-of-fit of a model and
	select a model with the smallest CV error. 
	Moreover, we can choose the number of knots of B-splines by CV. \revisetext{More practical details on implementing cross-validation to select an appropriate bandwidth can be found in Section \ref{sec:simulation}.}

		\subsection{Competing risk data with administrative censoring}
		\label{sec:competing}
		Despite being motivated by right-censored survival data, the proposed method also applies to competing risk data with administrative censoring, referred to as complete censoring case, by redefining the risk set \citep{Fine1999}.
		\citet{Fine1999} argued that, in the subdistribution model for competing risk data with administrative censoring, a subject who has not failed from the cause of interest or has not been censored remains in the risk set. This can be done by setting the censoring time of a subject who has not experienced the event of interest as their administrative censoring times. The analysis of competing risk data with time-dependent covariates also motivates the proposed SMKLE. Even if covariates are fully observed before a subject's death, covariates  cannot be collected between the death from other causes and the administrative censoring. Hence, log-likelihood \eqref{eq:orilogll} is still inapplicable even with the redefined risk set and the proposed method is still needed for analyzing competing risk data. 
	
	\section{Asymptotic theory\label{sec:asym}}
	
	In this section, we investigate the asymptotic properties of the proposed SMKLE.
	After introducing notations and regularity conditions in Section~\ref{subsec:notation},
	we present a uniform-in-bandwidth convergence rate theorem regarding the general kernel-weighted sieve functions in Section~\ref{subsec:uibCR}, which is the foundation for our theoretical development. 
	We then show the strong consistency and rate of convergence of $\hat{\theta}_{n}$, and asymptotic normality of $\hat{\beta}$ in Section \ref{subsec:asymres}.
	The proofs are collected 
	in the supplementary materials.
	
	\subsection{Notation and regularity conditions}
	\label{subsec:notation}
	
	Let $S_{T|Z}\left(\cdot\right)$ and $S_{C|Z}\left(\cdot\right)$ be the
	conditional survival functions for $T$ and $C$ given $Z$, respectively. Under the conditional independence assumption between the failure time $T$ and the censoring time $C$, we have $S_{T\wedge C|Z}\left(\cdot\right)= S_{T|Z}\left(\cdot\right)S_{C|Z}\left(\cdot\right)$.
	Let $\mathscr{B}_{p}$ denote the collection of Borel sets in $\mathbb{R}^{p}$,
	and $\mathscr{B}_{1}\left[0,\tau\right]\coloneqq \left\{
	A\cap\left[0,\tau\right]:A\in\mathscr{B}_{1}\right\} $.
	Let $\mathscr{C}_{p}$ denote the collection of Borel sets in
	$\mathbb{C}_{p}\coloneqq\left\{ f:\left[0,\tau\right]\to\mathbb{R}^{p}\right\} $.
	Let $\upsilon_{0}$ be the Lebesgue measure.
	For $A_{1}\in\mathscr{B}_{1}\left[0,\tau\right]$ and $A_{2}\in\mathscr{C}_{p}$,
	we define $\upsilon_{1}\left(A_{1}\times
	A_{2}\right) \coloneqq \int_{A_{1}}\Pr\left(Z\left(u\right)\in
	A_{2}\right)d\upsilon_{0}\left(u\right).$ 
	Consequently, 
	$\upsilon_{1}\left(A_{1}\times\mathbb{C}_{p}\right)=\upsilon_{0}\left(A_{1}\right)$.
	Denote the parameter space as $\Theta \coloneqq \mathcal{B}\times\mathcal{A}$ where $\mathcal{B}$ and $\mathcal{A}$ are the parameter spaces of $\beta$ and
	$\alpha$, respectively.
	Denote the overall sieve space for $\Theta$ as $\Theta_{n} \coloneqq \mathcal{B}\times\mathcal{A}_{n,l}$.
	We write the true parameter
	$\theta_{0} \coloneqq (\boldsymbol{\beta}_{0},\alpha_{0})$.
	For a vector $x = (x_1, \dots, x_p)^\top,$ its $l_2$ norm is denoted as $\Vert x \Vert_{2} = (x_1^2+\ldots + x_p^2)^{1/2}.$ For a function $f(\cdot):[0, \tau]\to\mathbb{R},$ its $L_2$ norm is defined as $\Vert f\Vert_{2} = (\int f(x)^2dx))^{1/2}$. We abuse the notation of $\Vert \cdot \Vert_2$ when the context is clear.  
	Define the distance between $\theta_{1},\theta_{2}\in\Theta$ by
	$d(\theta_{1},\theta_{2})\coloneqq\{ \Vert\alpha_{1}-\alpha_{2}\Vert_{2}^{2}+\Vert
	\beta_{1}-\beta_{2}\Vert _{2}^{2}\} ^{1/2}$.
	
	To establish the asymptotic properties of $\hat{\theta}_{n}$, we need the
	following regularity conditions:
	\begin{enumerate}[label=(C\arabic*)]
		\item
		\label{enu:kernel}The kernel function $K$ satisfies the following
		conditions:
		\begin{enumerate}[label=(\roman*)]
			\item
			The support of $K$ is
			$\left(-1,1\right)$;
			\item $K$ is symmetric about 0 and Lipschitz continuous;
			\item
			$\bar{K} \coloneqq \left\Vert
			K\right\Vert_{\infty}=\sup_{s\in(-1,1)}\left|K\left(s\right)\right|<\infty$;
			\item
			$\int_{0}^{1}sK\left(s\right)ds<\infty$;
			\item $K$ is non-decreasing  on $[-1,0]$ and non-increasing on $[0,1]$. 
		\end{enumerate}
		\item
		\label{enu:sieveoder} The maximum spacing of the knots for $\mathcal{A}_{n,l}$, 
		$\max_{l+1\le i\le
			q_{n}+1}\left|t_{i}-t_{i-1}\right|=O\left(n^{-\nu}\right)$, for
		$\nu\in\left(0,0.5\right)$, i.e.,
		$q_{n}=m_{n}+l=O\left(n^{\nu}\right)$.
		\item
		\label{enu:alpha_para_space}
		The parameter space $\mathcal{A}$ is the collection of uniformly bounded
		functions
		$\alpha$ on $\left[0,\tau\right]$ with bounded $k$th derivatives
		$\alpha^{\left(k\right)}$, $k=1,\ldots,\kappa$.
		For all $\alpha\in\mathcal{A}$, the
		$k$th derivative $\alpha^{\left(k\right)}$ satisfies the following Lipschitz continuity
		condition,
		$\left|\alpha^{\left(k\right)}\left(s\right)-\alpha^{\left(k\right)}\left(t\right)\right|\le
		L_{0}\left|s-t\right|$ for $s,t\in\left[0,\tau\right]$ where $\kappa$ is a
		positive
		integer such that $3\le\kappa+1\le l$ and $L_{0}\le\infty$ is a Lipschitz constant.
		\item
		\label{enu:beta_para_space}
		The parameter space $\mathcal{B}$ is a compact subspace of
		$\mathbb{R}^{p}$.
		The true parameter
		$\theta_{0}=\left(\beta_{0},\alpha_{0}\right)\in\Theta^{o}$
		where $\Theta^{o}$ is the interior of $\Theta$.
		\item
		\label{enu:zbounded}
		With probability one, $Z\left(t\right)$ has a uniformly bounded total variation on $[0,\tau]$.
		In other words, for the $j$th element of $Z(t)$ denoted by $Z_{\cdot 
			j}$, $\Pr\left(\int_{0}^{\tau}\left|dZ_{\cdot j}\left(t\right)\right|+\left|Z_{\cdot 
			j}\left(0\right)\right|\le z_{0}\right)=1$ for some constant $z_{0}>0$ and all $j=1,\ldots,p$.
		Furthermore, 
		$\sqrt{E\left[\left\Vert
			Z\left(t\right)-Z\left(s\right)\right\Vert _{2}^{2}\right]}\le
		L\left|t-s\right|$ for some 
		constant $L$ and all $t,s\in[0,\tau]$.
		\item
		\label{enu:identifiability}
		If there exists some vector $\beta$ such that
		$\beta^{\top}Z\left(t\right)=\alpha\left(t\right)$
		for a deterministic function $\alpha\left(t\right)$ on $t\in\left[0,\tau\right]$ with
		probability 1, then $\beta=0$ and $\alpha\left(t\right)=0$ for $t\in\left[0,\tau\right]$. 
		\item
		\label{enu:H_cond}$H\left(\cdot\right)$ is strictly increasing, continuous, and three-times differentiable.
		In addition, with probability 1,
		$\inf_{t,r\in\left[0,\tau\right]}H\left(\alpha\left(t\right)+\beta^{\top}Z\left(r\right)\right)>0$
		for $\theta$ in a neighborhood of $\theta_{0}$ in $\Theta$.
		\item
		\label{enu:ratefun}
		$E\left[dN^{*}\left(t\right)|\mathcal{F}_{t-}\right]=\mu\left(t\right)dt,$
		where $\mathcal{F}_{t}$ is 
		$\sigma\left\{
		\Delta,X,Z\left(s\right),0\le s\le t\right\} $, a $\sigma$-algebra including all information of $\Delta$, $X$, and $Z(s)$ up to time $t$. 
		Furthermore, $\mu\left(t\right)$ is Lipschitz continuous
		and
		there exists a constant $0<c_{0}<\infty$ such that
		$1/c_{0}\le\mu\left(t\right)\le c_{0}$.
		\item
		\label{enu:censor} $C$ is independent of $T$ given
		$Z\left(t\right)$
		for $t\in\left[0,\tau\right]$.
		Moreover, with probability one,
		\[
		\inf_{Z\left(t\right),t\in\left[0,\tau\right]}\Pr\left(C\ge\tau|Z(t)\right)=\inf_{Z\left(t\right),t\in\left[0,\tau\right]}\Pr\left(C=\tau|Z(t)\right)
		>0.
		\]
		\item
		\label{enu:finiteobs}
		There exists a finite positive integer $J_{0}$ such
		that
		$\Pr\left(J_i<J_{0}\right)=1, i = 1, \ldots, n$.
		\item
		\label{enu:UZinequality}
		There exists $\eta^{*}\in\left(0,1\right)$ such that 
		$b^{\top}\mathrm{Var}\left(Z\left(U\right)|U\right)b\ge
		\eta^{*}b^{\top}E\left[Z\left(U\right)Z\left(U\right)^{\top}|U\right]b$ for all
		$b\in\mathbb{R}^{p}$ with probability $1$, where $(U,Z)$ 
		are defined on $\upsilon_{1}(\mathbb{R}^{+}\times\mathcal{Z})$
		where $\mathbb{R}^{+}$ denotes the set of positive real numbers and $\mathcal{Z}$ is the sample space of $Z(t)$.
	\end{enumerate}
	\begin{rem}
		\label{rem:cond}
		Condition \ref{enu:kernel} is a 
		mild assumption on the kernel function, satisfied by most 
		commonly used kernels
		such as the Epanechnikov kernel.
		Conditions \ref{enu:sieveoder},
		\ref{enu:alpha_para_space} and \ref{enu:beta_para_space} are also
		typical for sieve estimation and semiparametric models.
		The assumption of uniformly bounded variation in condition \ref{enu:zbounded} is common for time-dependent covariates \citep{Dupuy2006,Zeng2021a,Wong2022}, implying the uniform elemental boundedness of $Z(t)$.
		Many common stochastic processes satisfy the Lipschitz continuity specified in Condition \ref{enu:zbounded}, e.g., the Wiener process and the homogeneous Poisson process.
		Condition~\ref{enu:identifiability} is essential for the identifiability of the
		model and implies that
		$E\left[Z\left(t\right)Z^{\top}\left(t\right)\right]$ is nonsingular for any $t\in [0,\tau]$.
		Conditions \ref{enu:H_cond} and \ref{enu:censor} are common assumptions for survival data in transformed hazards models \citep{Zeng2005}.
		Common transformation functions, such as the Box-Cox transformation, satisfy \ref{enu:H_cond}.
		Condition \ref{enu:ratefun} implies that the measurement times are non-informative. 
		The Lipschitz continuity and the positive lower bound of $\mu\left(t\right)$ in \ref{enu:ratefun} are needed to ensure the identifiability of $\theta$ together with \ref{enu:identifiability} (see Remark 3.2 in \citet{Wellner2007} for a more detailed discussion).
		Condition \ref{enu:finiteobs} implies the sparseness of longitudinal measurements, different from dense longitudinal measurements. 
		Condition \ref{enu:UZinequality} is a technical assumption. 
		It is 
		a generalization from 
		Condition C14 in
		\citet{Wellner2007} to the time-dependent covariate case. 
	\end{rem}
	%
	Our theoretical results heavily depend on the modern empirical process theory \citep{Vaart1996,Vaart1998,Kosorok2008}.
	Let $\mathbb{P}_{n}$ and $\mathbf{P}$ denote the empirical and the true probability measure, respectively.
	The empirical process is defined as $\mathbb{G}_{n}\coloneqq\sqrt{n}(\mathbb{P}_{n}-\mathbf{P})$.
	Hereafter, for a class $\mathcal{G}$ of pointwise measurable real-valued
	functions and
	some function $\psi$, define the sup norm $\left\Vert
	\psi\right\Vert_{\mathcal{G}}\coloneqq\sup_{g\in\mathcal{G}}\left|\psi\left(g\right)\right|$.
	Let $\Vert f\Vert_{\infty}\coloneqq\sup\{\left|f\left(x\right)\right|:x\text{ in the support of }f\}$ be the supremum norm.
	
	
	\subsection{The uniform-in-bandwidth convergence rate of the empirical processes of general kernel-weighted sieve functions}
	
	\label{subsec:uibCR}
	
	We establish the uniform-in-bandwidth convergence rate of empirical processes of general kernel-weighted functions over a sieve space, where the bandwidth ranges in an interval that becomes shorter with increased sample size. 
	This result, summarized in Theorem~\ref{prop:wrate}, provides the foundation for the proof of asymptotic theorems, especially, the strong consistency and convergence rate.
	Theorem~\ref{prop:wrate} shares 
	similar spirits with the theorems in \citet{Einmahl2005}, but 
	is essentially different from \citet{Einmahl2005} in the following three aspects.
	First, while \citet{Einmahl2005} focused on the classic kernel density and regression estimation, we analyze a more general and complicated semiparametric model with sparse longitudinal covariates.
	Second, the result in \citet{Einmahl2005} is uniform in a subset of real coordinate space with finite dimension, whereas our theory concerns a semiparametric sieve space whose dimension goes to infinity as $n$ increases.
	The sieve space plays a vital role in uniform convergence, leading to 
	different forms of the convergence rate from those in \citet{Einmahl2005}.
	Finally, the proofs in \citet{Einmahl2005} relied on its uniform covering number condition (K.iii), which is usually challenging to verify for semiparametric models as the covering numbers must be calculated for all probability measures.
	Instead, our proof uses the bracketing number 
	evaluated for the true probability measure only and is more commonly used in the semiparametric 
	literature \citep{Cao2015,Zhao2017,Zhou2017}.

	We consider two classes of general kernel-weighted sieve functions,
	$\mathcal{W}_{nh}^{1}$ and $\mathcal{W}_{nh}^{2}$, defined as follows.
	For some measurable, strictly increasing and three-times
	differentiable functions $\varphi_{1},\varphi_{2}:\mathbb{R}\to\mathbb{R}$, define
	\[
	\mathcal{W}_{nh}^{1}\coloneqq\left\{ O\mapsto\int_{0}^{X}K_{h}\left(X-r\right)\Delta\cdot \varphi_{1}\left(\alpha_{n}\left(X\right)+\beta^{\top}Z\left(r\right)\right)dN\left(r\right);\alpha_{n}\in\mathcal{A}_{n,l},\beta\in\mathcal{B}\right\} 
	\]
	and
	\[
	\mathcal{W}_{nh}^{2}\coloneqq\left\{ O\mapsto\int_{0}^{X}\int_{0}^{t}K_{h}\left(t-r\right)\varphi_{2}\left(\alpha_{n}\left(t\right)+\beta^{\top}Z\left(r\right)\right)dN\left(r\right)dt;\alpha_{n}\in\mathcal{A}_{n,l},\beta\in\mathcal{B}\right\} ,
	\]
	which are indexed by $\left(\beta,\alpha_{n}\right)$.
	The classes $\mathcal{W}_{nh}^{k}$, $k=1,2$, are general enough 
	to be used for other
	kernel-smoothing methods in the literature.
	Without the nonparametric component in the sieve space, the functions of class
	$\mathcal{W}_{nh}^{1}$ were used to construct estimating equations \citep{Cao2015, Chen2017, Li2020}. 
	The class $\mathcal{W}_{nh}^{2}$ with $\varphi_{2}(x)=x$ has been explored in
	\cite{Cao2021}.
	When 
	$\varphi_{1}\left(\cdot\right)=\log\left(H\left(\cdot\right)\right)$
	and
	$\varphi_{2}\left(\cdot\right)=H\left(\cdot\right)$,
	$\mathcal{W}_{nh}^{1}$ and $\mathcal{W}_{nh}^{1}$ apply to the two terms in the kernel-smoothed log-likelihood \eqref{eq:smoothlogll} over $\Theta_{n}$.
	
	
	%
	The next theorem gives the uniform-in-bandwidth
	convergence rate of $\mathcal{W}_{nh}^{1}$ and $\mathcal{W}_{nh}^{2}$.
	\begin{thm}
		\label{prop:wrate}
		Suppose that Conditions \ref{enu:kernel}-\ref{enu:zbounded} and
		\ref{enu:ratefun}-\ref{enu:finiteobs} hold. 
		Then, for $\nu$ defined in Condition \ref{enu:sieveoder}, with probability one,
		\[
		\limsup_{n\to\infty}\sup_{c_{1}\log n/n^{1-\nu}\le h\le1}\sqrt{n^{1-\nu}h}\left\Vert \mathbb{P}_{n}-\mathbf{P}\right\Vert _{\mathcal{W}_{nh}^{\iota}}=d_{\iota}<\infty,
		\]
		for some constants $c_{1}>0$ and $d_{\iota}>0, \iota=1,2$.
	\end{thm}
	
	\begin{coro}
		\label{coro:wconver}
		Suppose that conditions of Theorem~\ref{prop:wrate} hold. If $h_{n}\to0$ and
		$n^{1-\nu}h_{n}/\log(n)\to\infty$ as $n\to\infty$, then with probability one, $\Vert \mathbb{P}_{n}-\mathbf{P}\Vert_{\mathcal{W}_{nh_{n}}^{\iota}} \to 0$
		as $n\to\infty$ for $\iota=1,2$.
	\end{coro}
	
	\begin{rem}
		The main difference between Theorem~\ref{prop:wrate} and those in \citet{Einmahl2005} for kernel density estimation is the additional term, $n^{\nu}$, which reflects the influence of nonparametric component $\alpha$ and sieve estimation.
		The continuity conditions in \ref{enu:zbounded} and \ref{enu:ratefun} are not
		necessary for Theorem~\ref{prop:wrate} as they do not concern the
		continuity of the functions in $\mathcal{W}_{nh}^{1}$ and $\mathcal{W}_{nh}^{2}$.
	\end{rem}
	
	Theorem~\ref{prop:wrate} is one of our main theoretical
	contributions and is general enough to investigate the asymptotic properties of other
	kernel-smoothing methods. 
	Corollary \ref{coro:wconver} 
	will be used to prove the consistency of $\hat{\theta}_{n}$.
	
	\subsection{The asymptotic properties of SMKLE}
	
	\label{subsec:asymres}
	
	We show the strong consistency of $\hat{\theta}_{n}$ in the next theorem.
	\begin{thm}[Consistency]
		\label{them:consistency}
		Suppose Conditions \ref{enu:kernel}-\ref{enu:finiteobs} hold and recall that $\nu$ is defined in Condition \ref{enu:sieveoder}. If $n^{1-\nu}h_{n}/\log(n)\to\infty$ and
		$h_{n}\to0$ as $n\to\infty$,
		$d(\hat{\theta}_{n},\theta_{0})\to0$ almost surely (a.s.).
	\end{thm}
	\begin{rem}
		\label{rem:depcov}
		Condition \ref{enu:ratefun} implies that the observation processes are external
		and independent of covariates and survival times.
		With minor modifications to the proof, this could be relaxed to the case where
		the observation processes and the survival times are independent conditionally
		on the covariates \citep{Cao2016,Cao2021,Sun2021}.
		The conditions on $h_{n}$ are weaker than those in \citet{Cao2015}, which requires $n^{1/3}h_{n}\to0.$ 
	\end{rem}
	%
	%
	\begin{thm}[Rate of convergence]
		\label{them:rate}
		Suppose Conditions \ref{enu:kernel}-\ref{enu:UZinequality} holds and $h_{n}=O\left(n^{-a}\right)$.
		If $0.4\left(1-\nu\right)\le a<1-\nu$, then
		$$n^{\min\{ (1-\nu-a)/2,\kappa\nu\}
		} d(\hat{\theta}_{n},\theta_{0})=O_{p}(1).
		$$
		The optimal rate of convergence is $n^{3\kappa/\left(10\kappa+3\right)}$
		when $\nu=3/(3+10\kappa)$ and $a=4\kappa/(10\kappa+3)$.
	\end{thm}
	\begin{rem}
		The Lipschitz continuity assumptions stated in \ref{enu:zbounded} and
		\ref{enu:ratefun} are needed for the rate of convergence. For the consistency, continuity suffices.  
		\label{rem:order}
		Theorem \ref{them:rate} indicates an interesting interplay 
		of the kernel and
		sieve on estimation as the choices of $a$ and $\nu$ are
		intertwined with each other.
		The kernel smoothing bias determines the optimal convergence rate via $a=0.4(1-\nu)$, which is the lower bound of the allowable range of $a$. To ensure this ``bias'' term vanishes asymptotically in Step 3 of the proof of Theorem~\ref{them:rate}, it is necessary to impose the constraint $a \ge 0.4(1-\nu)$, which is a vital part of determining the optimal convergence rate.
		The optimal rate is $n^{3/10}$ with $\kappa \to \infty$, which is slower than 
		$n^{1/2}$, the optimal rate of sieve MLE with time-independent covariates \citep{Lu2009,Zhao2017a}.
		Due to
		the loss of information from sparsely observed covariates, the first term in the rate of convergence is $n^{\left(1-\nu-a\right)/2}$ rather than $n^{\left(1-\nu\right)/2}$, causing the slower rate. 
		The second term $n^{\kappa\nu}$ reflects the approximation error of the sieve for the true parameter space and remains the same for other B-spline-based methods.
		Compared with the step-function approximation to the nonparametric component \citep{Cao2015}, the sieve approximation yields a smaller ``bias'' term in the proof of Theorem 3.4, allowing for a wider range of $a$ to ensure the asymptotic vanishment of the ``bias'' term. 
	\end{rem}
	Although the convergence rate of $\hat{\theta}_{n}$ is slower than  
	$O(\sqrt{nh_{n}})$, $\hat{\beta}$ 
	still converges at rate 
	$\sqrt{nh_{n}}$, 
	which coincides with that of typical kernel density estimation 
	\citep{Tsybakov2009,Einmahl2005}.
	Before proceeding further, for the simplicity of notation, we abbreviate $H(\theta;t,Z(r))\coloneqq
	H(\alpha(t)+\beta^{\top}Z(r))$,
	$H^{\prime}(\theta;t,Z(r))\coloneqq
	H^{\prime}(\alpha(t)+\beta^{\top}Z(r))$
	where $H^{\prime}$ is the first order derivative of $H$.
	Besides, let $H^{(1)}(\theta;t,Z(r))\coloneqq H^{\prime}(\theta;t,Z(r))/H(\theta;t,Z(r))$ and $H^{(2)}(\theta;t,Z(r))\coloneqq H^{\prime}(\theta;t,Z(r))^{2}/H(\theta;t,Z(r))$.
	Denote
	\[ 
	s^{\left(k\right)}\left(\theta,t\right)\coloneqq E\left[Z^{\otimes k}\left(t\right)S_{T\wedge C|Z}\left(t\right)H^{(2)}\left(\theta;t,Z(t)\right)\right],
	\]
	and
	$$
	S^{\left(k\right)}\left(\theta,t\right)\coloneqq\frac{1}{n}\sum_{i=1}^{n}\int_{0}^{t}I\left(X_{i}\ge t\right)K_{h_{n}}\left(t-u\right)Z_{i}^{\otimes k}\left(u\right)H^{(2)}\left(\theta;t,Z_{i}(t)\right)dN_{i}\left(u\right).
	$$
	Here, for a
	vector $b$, $b^{\otimes0}=1$, $b^{\otimes1}=b$ and $b^{\otimes2}=bb^{\top}$.
	The following theorem provides the asymptotic normality of $\hat{\beta}$.
	\begin{thm}[Asymptotic
		Normality] \label{them:normality}Suppose that \ref{enu:kernel}-\ref{enu:UZinequality} hold and $h_{n}=O\left(n^{-a}\right)$.
		If $0.4(1-\nu)\le a<1-\nu$ and
		$(1-a)/(4\kappa)<\nu<(1-a)/2$, 
		we have
		$$\sqrt{nh_{n}}(\hat{\beta}-\beta_{0})\to_{d}N(0,\Xi^{-1}\Omega
		\Xi^{-1}),$$
		where $\Xi=\Sigma$,
		$\Omega=4\int_{0}^{1}K^{2}\left(u\right)du\Sigma$ and 
		\[
		\Sigma \coloneqq \int_{0}^{\infty}E\left[S_{T\wedge C|Z}\left(t\right)H^{(2)}\left(\theta_{0};t,Z(t)\right)\left(\frac{s^{\left(1\right)}\left(\theta_{0},t\right)}{s^{\left(0\right)}\left(\theta_{0},t\right)}-Z\left(t\right)\right)^{\otimes2}\right]\mu\left(t\right)dt.
		\]
	\end{thm}
	We can consistently estimate the asymptotic variance of $\hat{\beta}$ by
	$\hat{\Xi}^{-1}\hat{\Omega}\hat{\Xi}^{-1}/\left(nh_{n}\right)$, where
	\begin{multline*}
		\hat{\Xi} \coloneqq \frac{1}{n}\sum_{i=1}^{n} \Bigg[\int_{0}^{X_{i}}K_{h_{n}}\left(X_{i}-r\right)\Delta_{i}H^{(1)}(\hat{\theta}_{n};X_{i},Z_{i}(r))^{2} \Bigg.\\
		\Bigg.\times\left\{ \frac{S^{(2)}\left(\hat{\theta}_{n},X_{i}\right)}{S^{(0)}\left(\hat{\theta}_{n},X_{i}\right)}-\bar{Z}\left(X_{i}\right)^{\otimes2}\right\} dN_{i}\left(r\right)  \Bigg],
	\end{multline*}
	\begin{multline*}
		\hat{\Omega} \coloneqq \frac{h_{n}}{n}\sum_{i=1}^{n}\Bigg\{ \int_{0}^{X_{i}}K_{h_{n}}\left(X_{i}-r\right)\Delta_{i}H^{(1)}\left(\hat{\theta}_{n};X_{i},Z_{i}\left(r\right)\right)\left(\bar{Z}\left(X_{i}\right)-Z_{i}\left(r\right)\right)dN_{i}\left(r\right)\Bigg.\\
		-\int_{0}^{X_{i}}\int_{0}^{t}K_{h_{n}}\left(t-r\right)\Bigg.H^{\prime}\left(\hat{\theta}_{n};t,Z_{i}\left(r\right)\right)\left(\bar{Z}\left(t\right)-Z_{i}\left(r\right)\right)dN_{i}\left(r\right)dt\Bigg\} ^{\otimes2},
	\end{multline*}
	and $\bar{Z}\left(t\right) \coloneqq S^{\left(1\right)}\left(\hat{\theta}_{n},t\right)/S^{\left(0\right)}\left(\hat{\theta}_{n},t\right)$.
	
	\begin{rem}
		\label{rem:asym}
		The constraints of $a$ and $\nu$ in Theorem \ref{them:normality} imply 
		the $o(n^{3/8})$ convergence rate of $\hat{\beta},$ 
		when $h=O(n^{-1/4})$ and $q_n=o(n^{-3/8})$.
		This 
		is faster than $o(n^{1/3})$, the corresponding rate 
		without sieve in \citep{Cao2015,Cao2021}.
		Technically, the convergence rate is $n^{(1-a)/2}$. The smaller the lower bound of $a$, the faster the rate. Using the sieve estimation, we reduce the lower bound of $a$ from 0.4 to $0.4(1-\nu),$ improving efficiency. 
		
		
		
			However, $\hat{\beta}$ does not achieve semiparametric efficiency as its rate of convergence is slower than $\sqrt{n}$ \citep[][Lemma 25.23]{Vaart1998}.
			Moreover, both \citet{Wellner2007} and \citet{Sun2024} recognized that if the limit of the loss function differs from that of the log-likelihood function, the resulting M-estimator may not be asymptotically efficient even if the score of the Euclidean parameter is orthogonal to the nuisance parameter space.
			In fact, the limit of the kernel-smoothed log-likelihood function \eqref{eq:smoothlogll} is different from that of the log-likelihood function with the fully observed covariates.
			%
			Intuitively, the loss of efficiency is due to the sparsity in covariate observation.
	\end{rem} 
	\begin{rem}
		From Theorems~\ref{them:consistency}-\ref{them:normality}, the order of the sieve space dimension, $O(n^{\nu})$, 
		governs the effect of sieve estimation on the asymptotic behavior of the SMKLE. 
		The asymptotic theorems can 
		be generalized to other sieve methods by characterizing the sieve space dimension. 
		For example, \citet{Zhou2017} 
		used $O(n^{\nu})$ as the dimension of Bernstein polynomials. 
		Therefore, all our theoretical results also apply to the SMKLE with Bernstein polynomials.	
	\end{rem}
	\begin{rem}
		\label{rem:fullkernel}
			When the event is not terminal, that is, the covariates can be observed after the occurrence of the event of interest, we can improve the performance of the SKMLE by adopting the \textit{full-kernel} smoothing scheme, which removes $I(R_{ij}\le X)$ and $I(R_{ij}\le t)$ in equation \eqref{eq:smoothlogll}. For the full-kernel scheme, the form of the rate of convergence remains the same as that in Theorem~\ref{them:rate}. However, the allowable range of $a$ is expanded from $0.4(1-\nu) \le a \le 1 - \nu$ to $0.25(1-\nu) \le a \le 1 - \nu$ due to a smaller order of the ``bias'' term. Hence, the optimal convergence rate of $\hat{\theta}_{n}$ increases to
			$n^{3\kappa/\left(8\kappa+3\right)}$, which converges to $n^{3/8}$ when $\kappa \to\infty$, larger than its counterparts $n^{3/10}$ in Theorem~\ref{them:rate}. 
			The asymptotic variance of the full-kernel scheme becomes half of its counterpart in Theorem~\ref{them:normality}, indicating information gain and higher efficiency.
			In the supplementary materials, we provide a more detailed discussion of the full-kernel scheme.
	\end{rem}
	
	%
	%

	\section{Simulation studies\label{sec:simulation}}
	
	In this section, we carried out extensive simulation studies to investigate the
	finite sample performances of the proposed method. 
	We considered time-dependent covariate $Z_{1}\left(t\right)$ and
	time-invariant covariate $Z_{2}$ with maximum follow up time $\tau=1$.
	We generated $Z_{1}\left(t\right)$ by a piecewise step function
	$Z^{*}\left(t\right).$ 
	Specifically, 
	$Z_{1}\left(t\right)=2\left\{\Phi\left(Z^{*}\left(t\right)\right)-0.5\right\}$
	where $Z^{*}\left(t\right)=\sum_{k=1}^{20}\mathbf{I}\left\{
	\left(k-1\right)/20\le t<k/20\right\} z_{k}$ and $\left\{ z_{k}\right\}
	_{k=1}^{20}$ follow the multivariate normal distribution with mean zero and
	the covariance matrix $\left\{ e^{-\left|j-k\right|/20}\right\}
	_{j,k=1,\ldots,20}$.
	The time-invariant covariate $Z_{2}$ was set to be 1 if $\sum_{k=1}^{20}2(\Phi(z_k)-0.5)/20 + U^{*}>0$ and zero otherwise, where $U^{*}$ follows a standard normal distribution.
	The time-invariant $Z_{2}$ mimics a baseline binary 
	variable, for example, gender.
	The failure time $T$ was generated by the Box-Cox transformation
	\eqref{eq:box-cox}. Specifically, 
	$ 
	\left\{ \left(\lambda\left(t\mid Z(t)\right)\right)^{s}-1\right\} /s=\alpha\left(t\right)+\beta_{1}Z_{1}\left(t\right)+\beta_{2}Z_{2},
	$
	where $\alpha\left(t\right)=0.75\{(s+1)/2+t(1-\sin(2\pi(t-0.25)))\}$, $\beta_{1}=1$ and
	$\beta_{2}=-0.5$.
	Here, we set $\alpha\left(t\right)$ to be dependent on $s$ so that the
	resultant hazard functions are relatively comparable for different $s$.
	We varied $s$ to be 0, 0.25, 0.5, 0.75 
	%
	and 1.
	Specifically, $T$ was generated from the proportional hazards model when $s=0$,
	and the additive hazards model when $s=1$.
	For each $s$, we generated 1000 replicates of datasets with a sample size 200
	or 400.
	The observation times for covariates were generated from a nonhomogeneous
	Poisson process with the rate function 
	$\mu\left(t\right)=8(1+0.5\sin(4\pi
	t))$.
	The censoring time was generated as $\min\left(1,C^{*}\right)$ where $C^{*}\sim
	Unif\left(\underline{C},1.05\right)$ to achieve $20\%$ and $30\%$ censoring
	rates by adjusting $\underline{C}$, respectively.
	
	The Epanechnikov kernel $K\left(t\right)=0.75\left(1-t^{2}\right)_{+}$ was used. Results based on other kernel functions are similar and thus omitted.
	The order of B-splines was chosen to be 3, and the inner knots of B-splines were
	chosen 
	to be $\left(1/3,2/3\right)$ with boundary knots $\left(0,1\right)$.
	Nevertheless, when the censoring rate was high, the estimation of $\alpha$ near
	the upper boundary knot suffered from high volatility by the B-spline approximation.
	To overcome this, we employed natural splines \citep{Schumaker2007}, which 
	belongs to B-splines but confines
	the spline polynomials at the boundary 
	points to be linear and hence reduces
	the oscillation at the boundary knots.
	The optimization of \eqref{eq:smoothlogll} was done by the algorithm \texttt{SLSQP} \citep{Kraft1994} in R package \texttt{nloptr} \citep{Johnson2022}. We also compared our method to LVCF and the kernel-weighted methods in
	\citep{Cao2015} for the proportional hazards model and \citep{Cao2021} for the additive hazards model. 
	
	\revisetext{The asymptotic theorems in Section 3 indicate that the bandwidth $h$ should decay polynomially as the sample size $n$ increases.
	Preliminary simulation studies 
    have shown that the SMKLE balances estimation bias and standard errors best for bandwidth in the range of $(n^{-0.5}, n^{-0.4}).$ 
    We then constrain $h$ between $h_L = \max\{n^{-0.5}, \min_{i=1,\ldots,n,j=1,\ldots,J_i}(X_{i} - R_{ij})\}$ and $h_U = \min\{n^{-0.4}, \max_{i=1,\ldots,n,j=1,\ldots,J_i}(X_{i} - R_{ij})\}$.
	The lower limit $h_L$ 
    ensures that the kernel-smoothed log-likelihood includes at least one $(X_i, R_{ij})$ pair to prevent the likelihood from being zero while the upper limit $h_U$ helps prevent excessive bias.
    We then generate eleven candidate bandwidths for cross-validation (CV) based on a geometric sequence spanning the interval $[h_L, h_U]$ and select the bandwidth that minimizes the CV loss. }
	The bandwidths for \cite{Cao2015} and \cite{Cao2021} were either 
	$n^{-0.4}$, $n^{-0.5}$ or by their proposed data-driven sample splitting (SS)
	methods.
	
	Tables \ref{tab:sothers25}-\ref{tab:sothers75} present the relative bias (RB) that is the bias divided by $|\beta_{0}|$, the mean of the
	estimated standard errors (ESE), the empirical standard error (SE), and the
	coverage probability (CP) of the confidence intervals with 95\% nominal level
	of $\hat{\beta}$ by the SMKLE and LVCF. We consider 
	$s=0.25,0.5,$ and $0.75$ in Tables \ref{tab:sothers25}-\ref{tab:sothers75}, respectively.
	We observe that the proposed SMKLE has a
	small bias, its estimated standard errors are close to the empirical standard
	errors, and its coverage probabilities are close to the nominal levels. 
	As the sample size increases from 200 to 400, the RB, ESE and SE
	from the proposed method generally decrease, and the coverage probabilities 
	become closer to the nominal level.
	In contrast, the LVCF method produces a larger bias, which does not attenuate with 
	increased $n$ 
	especially for $\beta_{1}$,
	the effect of the time-dependent 
	covariate.

	\begin{table}[htb!] 
		\centering{}\caption{Simulation results for model (\ref{eq:box-cox}) 
			with $s=0.25$.\label{tab:sothers25}}
		\begin{tabular}{cccrrrrcrrrr@{}}
			\hline
			&             &            & \multicolumn{4}{c}{20\% censoring rate} &  &         \multicolumn{4}{c}{30\% censoring rate}          \\ $n$ &    Coef     &    $h$     & \mc{RB} & \mc{ESE} & \mc{SE} &  \mc{CP} &  & \mc{RB} & \mc{ESE} & \mc{SE} & \multicolumn{1}{c@{}}{CP} \\ \hline
			&             &            &                                  \multicolumn{9}{c}{Proposed SMKLE}                                  \\
			200 & $\beta_{1}$ & $n^{-0.4}$ &  -0.025 &    0.276 &   0.289 &     93.3 &  &  -0.018 &    0.292 &   0.297 &                      94.2 \\
			&             & $n^{-0.5}$ &  -0.004 &    0.327 &   0.343 &     94.3 &  &   0.004 &    0.345 &   0.362 &                      93.0 \\
			&             &     CV     &   0.002 &    0.287 &   0.311 &     92.7 &  &   0.010 &    0.302 &   0.326 &                      93.0 \\
			& $\beta_{2}$ & $n^{-0.4}$ &   0.006 &    0.308 &   0.336 &     92.8 &  &   0.014 &    0.325 &   0.337 &                      93.8 \\
			&             & $n^{-0.5}$ &  -0.040 &    0.366 &   0.397 &     92.7 &  &  -0.016 &    0.386 &   0.406 &                      94.3 \\
			&             &     CV     &  -0.026 &    0.320 &   0.360 &     92.0 &  &  -0.019 &    0.336 &   0.363 &                      93.0 \\
			400 & $\beta_{1}$ & $n^{-0.4}$ &  -0.021 &    0.211 &   0.219 &     94.4 &  &  -0.021 &    0.223 &   0.229 &                      94.6 \\
			&             & $n^{-0.5}$ &  -0.004 &    0.260 &   0.270 &     95.1 &  &  -0.003 &    0.274 &   0.287 &                      93.1 \\
			&             &     CV     &   0.008 &    0.223 &   0.245 &     93.8 &  &   0.006 &    0.233 &   0.251 &                      93.3 \\
			& $\beta_{2}$ & $n^{-0.4}$ &   0.002 &    0.236 &   0.247 &     94.1 &  &   0.029 &    0.248 &   0.265 &                      92.6 \\
			&             & $n^{-0.5}$ &  -0.036 &    0.292 &   0.298 &     94.6 &  &   0.011 &    0.308 &   0.325 &                      93.3 \\
			&             &     CV     &  -0.036 &    0.249 &   0.267 &     94.1 &  &   0.003 &    0.260 &   0.284 &                      92.5 \\
			&             &            &                                       \multicolumn{9}{c}{LVCF}                                        \\
			200 & $\beta_{1}$ &            &  -0.121 &    0.152 &   0.169 &     83.5 &  &  -0.118 &    0.160 &   0.177 &                      85.2 \\
			& $\beta_{2}$ &            &   0.096 &    0.157 &   0.199 &     86.8 &  &   0.089 &    0.165 &   0.200 &                      88.6 \\
			400 & $\beta_{1}$ &            &  -0.121 &    0.107 &   0.119 &     75.8 &  &  -0.124 &    0.112 &   0.128 &                      76.3 \\
			& $\beta_{2}$ &            &   0.094 &    0.111 &   0.140 &     84.7 &  &   0.094 &    0.116 &   0.143 &                      85.6 \\
			\hline
		\end{tabular}
	\end{table}
	\begin{table}[htb!]
		\centering{}\caption{Simulation results for model (\ref{eq:box-cox}) 
			with $s=0.5$.\label{tab:sothers50}}
		\begin{tabular}{cccrrrrcrrrr@{}}
			\hline
			&             &            & \multicolumn{4}{c}{20\% censoring rate} &  &         \multicolumn{4}{c}{30\% censoring rate}          \\ $n$ &    Coef     &    $h$     & \mc{RB} & \mc{ESE} & \mc{SE} &  \mc{CP} &  & \mc{RB} & \mc{ESE} & \mc{SE} & \multicolumn{1}{c@{}}{CP} \\ \hline
			&             &            &                                  \multicolumn{9}{c}{Proposed SMKLE}                                  \\
			200 & $\beta_{1}$ & $n^{-0.4}$ &  -0.031 &    0.298 &   0.309 &     94.0 &  &  -0.034 &    0.313 &   0.331 &                      92.0 \\
			&             & $n^{-0.5}$ &   0.003 &    0.351 &   0.371 &     92.4 &  &  -0.005 &    0.367 &   0.397 &                      92.0 \\
			&             &     CV     &  -0.003 &    0.308 &   0.332 &     93.0 &  &  -0.007 &    0.323 &   0.358 &                      91.2 \\
			& $\beta_{2}$ & $n^{-0.4}$ &   0.024 &    0.334 &   0.344 &     93.7 &  &  -0.020 &    0.352 &   0.373 &                      93.3 \\
			&             & $n^{-0.5}$ &  -0.011 &    0.396 &   0.411 &     93.7 &  &  -0.043 &    0.415 &   0.443 &                      93.2 \\
			&             &     CV     &  -0.008 &    0.345 &   0.365 &     93.7 &  &  -0.046 &    0.363 &   0.395 &                      92.9 \\
			400 & $\beta_{1}$ & $n^{-0.4}$ &  -0.024 &    0.228 &   0.235 &     95.1 &  &  -0.031 &    0.241 &   0.256 &                      92.2 \\
			&             & $n^{-0.5}$ &   0.004 &    0.279 &   0.293 &     93.1 &  &  -0.009 &    0.294 &   0.319 &                      92.0 \\
			&             &     CV     &   0.006 &    0.239 &   0.261 &     93.7 &  &  -0.004 &    0.251 &   0.280 &                      91.4 \\
			& $\beta_{2}$ & $n^{-0.4}$ &   0.003 &    0.257 &   0.264 &     94.3 &  &  -0.017 &    0.271 &   0.286 &                      93.7 \\
			&             & $n^{-0.5}$ &  -0.036 &    0.317 &   0.327 &     93.3 &  &  -0.041 &    0.333 &   0.356 &                      92.7 \\
			&             &     CV     &  -0.035 &    0.270 &   0.288 &     93.3 &  &  -0.047 &    0.282 &   0.313 &                      92.6 \\
			&             &            &                                       \multicolumn{9}{c}{LVCF}                                        \\
			200 & $\beta_{1}$ &            &  -0.122 &    0.165 &   0.195 &     83.2 &  &  -0.115 &    0.173 &   0.198 &                      85.3 \\
			& $\beta_{2}$ &            &   0.110 &    0.172 &   0.214 &     88.1 &  &   0.061 &    0.181 &   0.226 &                      88.6 \\
			400 & $\beta_{1}$ &            &  -0.126 &    0.116 &   0.133 &     76.1 &  &  -0.120 &    0.122 &   0.141 &                      79.3 \\
			& $\beta_{2}$ &            &   0.100 &    0.122 &   0.148 &     86.8 &  &   0.059 &    0.127 &   0.153 &                      89.2 \\
			\hline
		\end{tabular}
	\end{table}
	\begin{table}[htb!]
		\centering{}\caption{Simulation results for model (\ref{eq:box-cox}) 
			with $s=0.75$.\label{tab:sothers75}}
		\begin{tabular}{cccrrrrcrrrr@{}}
			\hline
			&             &            & \multicolumn{4}{c}{20\% censoring rate} &  &         \multicolumn{4}{c}{30\% censoring rate}          \\ $n$ &    Coef     &    $h$     & \mc{RB} & \mc{ESE} & \mc{SE} &  \mc{CP} &  & \mc{RB} & \mc{ESE} & \mc{SE} & \multicolumn{1}{c@{}}{CP} \\ \hline
			&             &            &                                  \multicolumn{9}{c}{Proposed SMKLE}                                  \\
			200 & $\beta_{1}$ & $n^{-0.4}$ &  -0.030 &    0.320 &   0.341 &     93.0 &  &  -0.036 &    0.341 &   0.363 &                      93.3 \\
			&             & $n^{-0.5}$ &   0.003 &    0.381 &   0.412 &     93.0 &  &  -0.015 &    0.407 &   0.417 &                      94.4 \\
			&             &     CV     &  -0.004 &    0.332 &   0.362 &     91.9 &  &  -0.015 &    0.354 &   0.378 &                      93.6 \\
			& $\beta_{2}$ & $n^{-0.4}$ &   0.001 &    0.357 &   0.381 &     93.6 &  &   0.055 &    0.380 &   0.429 &                      91.5 \\
			&             & $n^{-0.5}$ &  -0.026 &    0.425 &   0.468 &     93.0 &  &   0.023 &    0.452 &   0.492 &                      92.9 \\
			&             &     CV     &  -0.027 &    0.370 &   0.407 &     93.8 &  &   0.032 &    0.394 &   0.454 &                      91.6 \\
			400 & $\beta_{1}$ & $n^{-0.4}$ &  -0.025 &    0.246 &   0.254 &     93.7 &  &  -0.036 &    0.259 &   0.274 &                      91.3 \\
			&             & $n^{-0.5}$ &   0.008 &    0.304 &   0.323 &     93.3 &  &  -0.015 &    0.318 &   0.333 &                      93.7 \\
			&             &     CV     &   0.005 &    0.257 &   0.285 &     92.9 &  &  -0.014 &    0.270 &   0.294 &                      91.4 \\
			& $\beta_{2}$ & $n^{-0.4}$ &   0.010 &    0.275 &   0.293 &     94.3 &  &   0.044 &    0.289 &   0.312 &                      93.3 \\
			&             & $n^{-0.5}$ &  -0.035 &    0.341 &   0.372 &     93.9 &  &   0.021 &    0.358 &   0.386 &                      93.1 \\
			&             &     CV     &  -0.022 &    0.288 &   0.316 &     93.8 &  &   0.019 &    0.302 &   0.337 &                      92.7 \\
			&             &            &                                       \multicolumn{9}{c}{LVCF}                                        \\
			200 & $\beta_{1}$ &            &  -0.120 &    0.178 &   0.213 &     85.0 &  &  -0.107 &    0.187 &   0.230 &                      83.9 \\
			& $\beta_{2}$ &            &   0.094 &    0.186 &   0.237 &     85.3 &  &   0.085 &    0.195 &   0.269 &                      83.1 \\
			400 & $\beta_{1}$ &            &  -0.125 &    0.126 &   0.149 &     78.6 &  &  -0.121 &    0.132 &   0.152 &                      79.9 \\
			& $\beta_{2}$ &            &   0.091 &    0.131 &   0.162 &     87.8 &  &   0.091 &    0.138 &   0.177 &                      86.0 \\ \hline
		\end{tabular}
	\end{table}
	
	Tables \ref{tab:cox} and \ref{tab:additive} compare the proposed method and
	LVCF with the method in \cite{Cao2015} under the proportional hazards model
	and the method in \cite{Cao2021} under the additive hazards model,
	respectively.
	Overall, the SMKLE 
	has small biases and satisfactory coverage
	probabilities under the proportional and additive hazards model, comparable to results using methods proposed in \cite{Cao2015} and \cite{Cao2021}.
	For the additive hazards model, our method generally has smaller empirical
	standard errors of $\hat{\beta}$ than those 
	using the method proposed in \cite{Cao2021}.
	The gain in efficiency may result from using the full likelihood and sieve estimation in SMKLE.
	We tried 
	different numbers of knots and different $\mu(t)$. The results were similar and were thus omitted.
	
	\begin{table}[htbp!]
		\caption{Simulation results for Cox's
			proportional hazards model ($s=0$).
			\label{tab:cox}}
		\centering{}%
		\begin{tabular}{@{}cccrrrrcrrrr@{}}
			\hline
			&             &            & \multicolumn{4}{c}{20\% censoring rate} &  &         \multicolumn{4}{c}{30\% censoring rate}          \\ \cline{4-7}\cline{9-12}
			$n$ &    Coef     &    $h$     & \mc{RB} & \mc{ESE} & \mc{SE} &  \mc{CP} &  & \mc{RB} & \mc{ESE} & \mc{SE} & \multicolumn{1}{c@{}}{CP} \\ \hline
			&             &            &                                  \multicolumn{9}{c}{Proposed SMKLE}                                  \\
			200 & $\beta_{1}$ & $n^{-0.4}$ &  -0.042 &    0.252 &   0.266 &     92.1 &  &  -0.033 &    0.268 &   0.281 &                      93.2 \\
			&             & $n^{-0.5}$ &  -0.020 &    0.299 &   0.317 &     93.9 &  &  -0.005 &    0.318 &   0.332 &                      94.8 \\
			&             &     CV     &  -0.018 &    0.262 &   0.287 &     91.8 &  &  -0.007 &    0.278 &   0.305 &                      92.7 \\
			& $\beta_{2}$ & $n^{-0.4}$ &   0.031 &    0.273 &   0.282 &     95.2 &  &   0.017 &    0.290 &   0.311 &                      93.4 \\
			&             & $n^{-0.5}$ &   0.026 &    0.326 &   0.345 &     94.5 &  &   0.000 &    0.346 &   0.367 &                      93.4 \\
			&             &     CV     &   0.011 &    0.284 &   0.303 &     94.8 &  &  -0.005 &    0.300 &   0.333 &                      92.8 \\
			400 & $\beta_{1}$ & $n^{-0.4}$ &  -0.038 &    0.193 &   0.193 &     94.0 &  &  -0.031 &    0.203 &   0.201 &                      95.2 \\
			&             & $n^{-0.5}$ &  -0.020 &    0.239 &   0.239 &     94.9 &  &  -0.004 &    0.252 &   0.252 &                      95.6 \\
			&             &     CV     &  -0.013 &    0.204 &   0.215 &     93.6 &  &  -0.004 &    0.214 &   0.225 &                      94.5 \\
			& $\beta_{2}$ & $n^{-0.4}$ &   0.024 &    0.209 &   0.208 &     95.7 &  &   0.030 &    0.222 &   0.231 &                      94.1 \\
			&             & $n^{-0.5}$ &   0.020 &    0.261 &   0.262 &     94.9 &  &   0.014 &    0.276 &   0.287 &                      94.2 \\
			&             &     CV     &   0.003 &    0.221 &   0.227 &     95.0 &  &   0.009 &    0.233 &   0.250 &                      93.4 \\[6pt]
			&             &            &                                  \multicolumn{9}{c}{\citet{Cao2015}}                                  \\
			200 & $\beta_{1}$ & $n^{-0.4}$ &  -0.053 &    0.251 &   0.262 &     92.2 &  &  -0.044 &    0.266 &   0.277 &                      93.5 \\
			&             & $n^{-0.5}$ &  -0.032 &    0.297 &   0.312 &     92.6 &  &  -0.015 &    0.316 &   0.331 &                      94.0 \\
			&             &     SS     &  -0.034 &    0.281 &   0.306 &     91.8 &  &  -0.020 &    0.298 &   0.323 &                      93.1 \\
			& $\beta_{2}$ & $n^{-0.4}$ &   0.047 &    0.280 &   0.279 &     95.5 &  &   0.027 &    0.297 &   0.306 &                      94.2 \\
			&             & $n^{-0.5}$ &   0.046 &    0.331 &   0.341 &     94.8 &  &   0.008 &    0.351 &   0.362 &                      94.3 \\
			&             &     SS     &   0.040 &    0.313 &   0.325 &     94.8 &  &   0.008 &    0.331 &   0.343 &                      93.9 \\
			400 & $\beta_{1}$ & $n^{-0.4}$ &  -0.044 &    0.192 &   0.191 &     93.8 &  &  -0.036 &    0.202 &   0.200 &                      95.1 \\
			&             & $n^{-0.5}$ &  -0.025 &    0.239 &   0.237 &     94.5 &  &  -0.007 &    0.251 &   0.254 &                      95.2 \\
			&             &     SS     &  -0.029 &    0.221 &   0.231 &     93.4 &  &  -0.010 &    0.233 &   0.246 &                      94.6 \\
			& $\beta_{2}$ & $n^{-0.4}$ &   0.034 &    0.214 &   0.206 &     96.2 &  &   0.035 &    0.226 &   0.231 &                      94.5 \\
			&             & $n^{-0.5}$ &   0.031 &    0.264 &   0.261 &     95.8 &  &   0.017 &    0.280 &   0.289 &                      94.0 \\
			&             &     SS     &   0.035 &    0.245 &   0.240 &     96.2 &  &   0.014 &    0.259 &   0.270 &                      93.7 \\[6pt]
			&             &            &                                       \multicolumn{9}{c}{LVCF}                                        \\
			200 & $\beta_{1}$ &            &  -0.131 &    0.139 &   0.158 &     80.1 &  &  -0.125 &    0.147 &   0.161 &                      83.1 \\
			& $\beta_{2}$ &            &   0.087 &    0.141 &   0.177 &     87.0 &  &   0.072 &    0.147 &   0.181 &                      88.6 \\
			400 & $\beta_{1}$ &            &  -0.135 &    0.098 &   0.109 &     68.6 &  &  -0.129 &    0.103 &   0.110 &                      74.0 \\
			& $\beta_{2}$ &            &   0.083 &    0.099 &   0.118 &     87.7 &  &   0.087 &    0.104 &   0.124 &                      87.9 \\ \hline
		\end{tabular}
	\end{table}
	\begin{table}[htbp!]
		\caption{Simulation results for the
			additive hazards model ($s=1$).
			\label{tab:additive}}        
		\centering{}%
		\begin{tabular}{@{}cccrrrrcrrrr@{}}
			\hline
			&             &            & \multicolumn{4}{c}{20\% censoring rate} &  &         \multicolumn{4}{c}{30\% censoring rate}          \\ \cline{4-7}\cline{9-12}
			$n$ &    Coef     &    $h$     & \mc{RB} & \mc{ESE} & \mc{SE} &  \mc{CP} &  & \mc{RB} & \mc{ESE} & \mc{SE} & \multicolumn{1}{c@{}}{CP} \\ \hline
			&             &            &                                  \multicolumn{9}{c}{Proposed SMKLE}                                  \\
			200 & $\beta_{1}$ & $n^{-0.4}$ &  -0.033 &    0.382 &   0.389 &     93.1 &  &  -0.042 &    0.409 &   0.407 &                      92.8 \\
			&             & $n^{-0.5}$ &   0.007 &    0.473 &   0.467 &     93.5 &  &  -0.018 &    0.511 &   0.489 &                      92.8 \\
			&             &     CV     &  -0.010 &    0.396 &   0.417 &     91.8 &  &  -0.026 &    0.432 &   0.431 &                      92.8 \\
			& $\beta_{2}$ & $n^{-0.4}$ &   0.038 &    0.408 &   0.429 &     95.3 &  &   0.036 &    0.442 &   0.454 &                      94.6 \\
			&             & $n^{-0.5}$ &   0.040 &    0.500 &   0.507 &     95.8 &  &   0.023 &    0.545 &   0.540 &                      95.1 \\
			&             &     CV     &   0.018 &    0.420 &   0.451 &     93.8 &  &   0.025 &    0.463 &   0.474 &                      94.4 \\
			400 & $\beta_{1}$ & $n^{-0.4}$ &  -0.029 &    0.281 &   0.292 &     93.0 &  &  -0.032 &    0.299 &   0.289 &                      95.5 \\
			&             & $n^{-0.5}$ &  -0.003 &    0.355 &   0.365 &     94.3 &  &  -0.021 &    0.382 &   0.358 &                      95.3 \\
			&             &     CV     &  -0.001 &    0.299 &   0.321 &     93.0 &  &  -0.015 &    0.316 &   0.312 &                      94.8 \\
			& $\beta_{2}$ & $n^{-0.4}$ &   0.023 &    0.307 &   0.320 &     94.0 &  &   0.005 &    0.327 &   0.341 &                      94.0 \\
			&             & $n^{-0.5}$ &   0.015 &    0.385 &   0.406 &     92.9 &  &  -0.010 &    0.414 &   0.419 &                      94.6 \\
			&             &     CV     &  -0.005 &    0.324 &   0.350 &     93.8 &  &  -0.012 &    0.343 &   0.363 &                      93.3 \\[6pt]
			&             &            &                                  \multicolumn{9}{c}{\citet{Cao2021}}                                  \\
			200 & $\beta_{1}$ & $n^{-0.4}$ &  -0.045 &    0.483 &   0.437 &     96.0 &  &  -0.026 &    0.535 &   0.479 &                      96.4 \\
			&             & $n^{-0.5}$ &  -0.011 &    0.556 &   0.506 &     96.1 &  &  -0.004 &    0.615 &   0.576 &                      95.8 \\
			&             &     SS     &  -0.018 &    0.531 &   0.498 &     95.7 &  &  -0.009 &    0.586 &   0.561 &                      95.7 \\
			& $\beta_{2}$ & $n^{-0.4}$ &   0.035 &    0.566 &   0.479 &     97.8 &  &   0.004 &    0.627 &   0.566 &                      96.5 \\
			&             & $n^{-0.5}$ &   0.031 &    0.644 &   0.559 &     97.4 &  &  -0.010 &    0.712 &   0.664 &                      95.5 \\
			&             &     SS     &   0.035 &    0.617 &   0.531 &     97.3 &  &   0.000 &    0.681 &   0.638 &                      95.9 \\
			400 & $\beta_{1}$ & $n^{-0.4}$ &  -0.030 &    0.366 &   0.325 &     96.4 &  &  -0.025 &    0.403 &   0.347 &                      97.2 \\
			&             & $n^{-0.5}$ &  -0.010 &    0.440 &   0.392 &     96.7 &  &  -0.009 &    0.484 &   0.438 &                      96.6 \\
			&             &     SS     &  -0.013 &    0.414 &   0.382 &     96.0 &  &  -0.008 &    0.454 &   0.424 &                      96.3 \\
			& $\beta_{2}$ & $n^{-0.4}$ &   0.002 &    0.428 &   0.369 &     97.5 &  &  -0.014 &    0.470 &   0.431 &                      96.8 \\
			&             & $n^{-0.5}$ &  -0.009 &    0.510 &   0.452 &     97.6 &  &  -0.048 &    0.560 &   0.529 &                      96.6 \\
			&             &     SS     &  -0.005 &    0.481 &   0.421 &     97.9 &  &  -0.043 &    0.527 &   0.498 &                      96.4 \\[6pt]
			&             &            &                                       \multicolumn{9}{c}{LVCF}                                        \\
			200 & $\beta_{1}$ &            &  -0.119 &    0.200 &   0.236 &     85.6 &  &  -0.094 &    0.212 &   0.247 &                      88.6 \\
			& $\beta_{2}$ &            &   0.071 &    0.206 &   0.268 &     85.9 &  &   0.038 &    0.218 &   0.277 &                      89.7 \\
			400 & $\beta_{1}$ &            &  -0.121 &    0.139 &   0.160 &     81.5 &  &  -0.100 &    0.146 &   0.170 &                      85.4 \\
			& $\beta_{2}$ &            &   0.077 &    0.144 &   0.183 &     88.0 &  &   0.051 &    0.151 &   0.192 &                      87.3 \\ \hline
		\end{tabular}
	\end{table}

	\section{Application to a COVID-19 study}
	\label{sec:application}
	We apply the proposed SMKLE to a longitudinal COVID-19
	observational dataset obtained from 
	Tongji Hospital Sino-French New City campus in
	Wuhan, China, in early 2020.
	The study followed hospitalized COVID-19 patients until they were discharged or died. The main goal is to investigate the effect of risk factors on the duration of hospitalization. 
	Vital signs, including blood pressure, body temperature, and respiratory frequency, were recorded daily for each patient to monitor disease progression. 
	These covariates varied over time, leading to sparse longitudinal data. 
	Time-invariant covariates, such as gender, age at admission, and baseline neutrophil counts, were also collected. 
	Furthermore, due to the overwhelming load of the local healthcare system 
	at the early pandemic stage, the time-dependent covariates were not entirely measured as planned or properly recorded daily. This further leads 
	to intermittency and irregularity 
	in the measurement times for each patient. 
	As a result, classical survival analysis methods are not applicable 
	to this dataset. 
	
	The dataset consisted of 180 patients 
	among which 7 died and others were discharged after treatment. 
	We consider five covariates: two demographic covariates, i.e., age and sex,
	two sparse longitudinal covariates, i.e., body temperature (Temp) and
	respiratory frequencies (Resp), and a baseline covariate, neutrophil counts
	(NC), measured on the first day of the hospitalization.
		Moreover, death and discharge were competing events during the follow-up period of 44 days. 
		The maximum follow-up time is the administrative censoring time, and the data fall under the category of complete censoring data defined in \cite{Fine1999}. 
		As described in Section \ref{sec:competing}, we set subjects
		who died as censored at the maximum follow-up time, 44 days, rather than at their
		respective death times.
		By this manipulation, the proposed method infers the subdistribution hazards \citep{Fine1999} of the discharge time.
	
	We consider 
	model
	\eqref{eq:box-cox} with $s=0,0.25,0.5,0.75,1$.
	Specifically, when $s=0$ (proportional hazards
	model) and 1 (additive hazards model), we compare the SMKLE to \citet{Cao2015} and \citet{Cao2021}.
	All the observation times and event times are rescaled to $\left[0,1\right]$
	by dividing the largest follow-up time.
	Following the simulation studies, the proposed CV method selects the bandwidths, while the bandwidths for \citet{Cao2015} and
	\citet{Cao2021} are selected by their respective sample splitting methods.
	We also reported CV loss and BIC based on the kernel-weighted log-likelihood function for model selection.
	
	Tables \ref{tab:realdata01} and \ref{tab:realdataother} summarize the analysis
	results. 
	The CV loss indicates that the model with $s=0.5$ fits the data best, while the
	BIC suggests that the model with $s=0.75$ fits the data best.
	There is strong evidence that neutrophil counts were significantly negatively associated with the hazards of the subdistribution of discharge, meaning that patients with higher neutrophil counts had longer hospital stays. Evidence also supports a positive association between respiratory rates and hospitalization duration. 
	Since different goodness-of-fit criteria recommend different models, 
	we
	consider ``profiling'' or ``combining'' the results of different models to have
	a definitive hypothesis testing result.
	We use the Cauchy combination test proposed by \citet{Liu2020} and \citet{Pillai2016} to combine the five models and calculate a $p$-value for each covariate from a
	combined test statistic whose distribution can be approximated by the Cauchy
	distribution.
	By the Cauchy combination test, at the significance level of 0.05, we conclude
	that the effects of neutrophil counts (with 0.006 $p$-value) and respiratory rates
	(with 0.035 $p$-value) are significant, which are consistent with the medical literature \citep{Ikram2022,MassoSilva2022}.
	The analysis illustrates that the proposed method uncovers significant clinical risk factors that existing methods cannot make otherwise. 
	

	\begin{table}[htbp!]
		\centering{}\caption{The estimation results under the
			Cox's proportional and additive model by the SMKLE and
			martingale-based methods, respectively.
			\label{tab:realdata01}}
		\begin{tabular}{@{}cccrccc@{\extracolsep{6pt}}crc@{}}
			\hline 
			&  &  & \multicolumn{3}{c}{Proposed SMKLE} &  & \multicolumn{3}{c}{\citet{Cao2015}}\\
			\cline{4-7} \cline{8-10}
			$s$  & $h$  &  & \multicolumn{1}{c}{$\hat{\beta}$(SE)}  & p-val  &  &  & $h$  & \multicolumn{1}{c}{$\hat{\beta}$(SE)}   & pval \\
			\cline{1-7} \cline{8-10}
			0  & 0.139  & Age  & -0.018(0.014)  & 0.188  & CV loss  & BIC & 0.052  & -0.019(0.021)  & 0.354 \\
			&  & Sex  & 0.120(0.458)  & 0.794  & 2.614  & 2.311 &  & 0.932(0.692)  & 0.178 \\
			&  & Temp  & 0.098(0.352)  & 0.781  &  &  &  & -0.141(0.577)  & 0.806 \\
			&  & NC  & -0.154(0.089)  & 0.085  &   &  &  & -0.184(0.111)  & 0.096 \\
			&  & Resp  & -0.010(0.016)  & 0.532  &   &  &  & 0.000(0.019)  & 0.986 \\[6pt]
			&  &  &  &  &  &  & \multicolumn{3}{c}{\citet{Cao2021}}\\
			\cline{8-10}
			1  & 0.139  & Age  & -0.006(0.014)  & 0.673  & CV loss  & BIC & 0.052  & -0.003(0.004)  & 0.377 \\
			&  & Sex  & 0.465(0.549)  & 0.397  & 2.771  & 2.059 &  & 0.111(0.088)  & 0.209 \\
			&  & Temp  & 0.002(0.297)  & 0.994  &  &  &  & 0.002(0.050)  & 0.971 \\
			&  & NC  & -0.172(0.089)  & 0.052  &   &  &  & -0.014(0.011)  & 0.204 \\
			&  & Resp  & -0.015(0.009)  & 0.107  &   &  &  & 0.000(0.004)  & 0.998 \\
			\hline
		\end{tabular}
	\end{table}
	\begin{table}[htbp!]
		\centering{}\caption{The estimation results under Box-Cox
			transformed hazards model with $s=0.25,0.5,0.75$ by the SMKLE.
			\label{tab:realdataother}}
		\begin{tabular}{@{}lccrccc@{}}
			\hline 
			\multicolumn{1}{@{}c}{$s$}  &  $h$ &  & \multicolumn{1}{c}{$\hat{\beta}$(SE)} & p-val  & CV loss  & BIC \\
			\hline 
			0.25  & 0.119  & Age  & -0.012(0.014)  & 0.378  & 2.075  & 1.824 \\
			&  & Sex  & 0.382(0.423)  & 0.367  &  & \\
			&  & Temp  & 0.195(0.325)  & 0.548  &  & \\
			&  & NC  & -0.175(0.076)  & 0.021  &  & \\
			&  & Resp  & -0.027(0.015)  & 0.061  &  & \\[6pt]
			0.5  & 0.119  & Age  & -0.008(0.013)  & 0.531  & 2.045  & 1.766 \\
			&  & Sex  & 0.466(0.376)  & 0.215  &  & \\
			&  & Temp  & 0.158(0.277)  & 0.568  &  & \\
			&  & NC  & -0.166(0.059)  & 0.005  &  & \\
			&  & Resp  & -0.027(0.010)  & 0.009  &  & \\[6pt]
			0.75  & 0.119  & Age  & -0.004(0.011)  & 0.682  & 2.119  & 1.731 \\
			&  & Sex  & 0.499(0.351)  & 0.155  &  & \\
			&  & Temp  & 0.137(0.247)  & 0.579  &  & \\
			&  & NC  & -0.154(0.049)  & 0.002  &  & \\
			&  & Resp  & -0.025(0.015)  & 0.105  &  & \\
			\hline
		\end{tabular}
	\end{table}
	
	\section{Concluding remarks\label{sec:conclusion}}
	
	
	Time-dependent covariates collected longitudinally are ubiquitous in clinical trials and epidemiological studies. We propose a novel approach, SMKLE, that combines kernel weighting with sieve estimation for transformed hazards models. 
	This is a highly flexible and thoroughly substantiated method for analyzing survival data. Large sample theories provide valuable insights into the interplay between kernel and sieve 
	on the behavior of SMKLE.
	Extensive simulation studies validate our theoretical predictions and show favorable performance 
	over existing methods. Analyzing a COVID-19 dataset 
	identifies significant risk factors for hospital duration that were missed by using existing methods. 
	
	For survival outcome with longitudinal covariates, joint modeling is well established in the literature \citep{Wulfsohn1997}. The proposed method avoids the strong distributional assumptions and heavy computational burden of the joint modeling approach.
	To the best of our knowledge, there does not exist a joint modeling method for the transformed hazards model that we investigate in this paper. Nonetheless, joint modeling conceptually enables the incorporation the true values,  within-subject autocorrelation and the summary trends, such as slope, of the longitudinal covariates into the survival model \citep{Tsiatis2001,Tsiatis2004}. This capability enhances the interpretability of the model, allowing the investigators to assess the association between the overall trajectory and the event time \citep{Tsiatis2004}, which is not supported by the kernel-based approach.

	We have developed a rigorous theoretical framework to investigate the asymptotic properties of the kernel-weighted sieve M-estimator. This provides a solid theoretical foundation for future research. 
	For example, it is of interest to 
	study survival data with other censoring schemes, such as interval censoring
	or current status data, or other types of time-to-event data, such as recurrent
	event data or panel count data \citep{Sun2024}.
		Following \citet{Guo2023} and \citet{Sun2024a}, the proposed method can be extended to the time-varying coefficient model by employing the sieve approximation to the time-varying coefficients. 
        \revisetext{It is an open question whether the bandwidth selected by CV can achieve the optimal convergence rate in Theorem~\ref{them:rate}.}
		More research is needed and we leave this for future work.
	
	

	\bigskip
	\begin{center}
		{\large\bf SUPPLEMENTARY MATERIAL}
	\end{center}
	
	\begin{description}
		
		\item[Supplement materials:] The supplement provides the detailed discussion of the full-kernel scheme and the proofs of Theorem \ref{prop:wrate}, Corollary \ref{coro:wconver}, Theorems \ref{them:consistency}-\ref{them:normality} and lemmas. 

	\end{description}

    \subsection*{Acknowledgments}
    We would like to express our gratitude to the Editor, the Associate Editor, and two reviewers for their constructive comments that significantly improved this paper. We thank Dr. Longbo Li for providing us the COVID-19 dataset. Dayu Sun's research is supported by IUSM Startup Funds. Cao's research is partially supported by NSF DMS-2311249. Zhao's research is supported in part by the National Natural Science Foundation of China (12271459) and the Research Grant Council of Hong Kong (15306521).
\subsection*{Disclosure Statement}
The authors report there are no competing interests to declare.

	\bibliographystyle{apalike}
	
	\bibliography{asyntransbib}
\end{document}


\def\spacingset#1{\renewcommand{\baselinestretch}%
		{#1}\small\normalsize} \spacingset{1}

	\newcommand*{\addFileDependency}[1]{
		\typeout{(#1)}
		\@addtofilelist{#1}
		\IfFileExists{#1}{}{\typeout{No file #1.}}
	}
	\makeatother
	
	\newcommand*{\myexternaldocument}[1]{%
		\externaldocument{#1}[#1.pdf]%
		\addFileDependency{#1.tex}%
		\addFileDependency{#1.aux}%
	}
	
	\myexternaldocument{main}
	
	\renewcommand{\theequation}{S.\arabic{equation}}
	\renewcommand*{\thesection}{S.\arabic{section}}
	\renewcommand*{\thetable}{S.\arabic{table}}
	
	\date{}
	\if1\blind
	{
		\title{\bf Supplement to ``Kernel meets sieve: transformed hazards models with sparse longitudinal
			covariates''}
		\author{Dayu Sun\\
		Department of Biostatistics and Health Data Science,\\ Indiana University School of Medicine\\
		Zhuowei Sun  \\
		School of Mathematics, Jilin University\\
		School of Public Health, Dalian Medical University\\
		Xingqiu Zhao \thanks{Corresponding author} \\
		Department of Applied Mathematics,\\ The Hong Kong Polyetechnic University\\
		and \\
		Hongyuan Cao  \thanks{Corresponding author}\\
		Department of Statistics, Florida State University
		}
		\maketitle
	} \fi
	
	\if0\blind
	{
		\bigskip
		\bigskip
		\bigskip
		\begin{center}
			{\LARGE\bf  Supplement to ``Kernel meets sieve: transformed hazards models with sparse longitudinal
				covariates''}
		\end{center}
		\medskip
	} \fi

	\spacingset{1.3} 
	%
	\def\arraystretch{0.77}
	
	\linenumbers
	
	The references to quantities, lemmas, propositions and theorems in this supplement all begin with an ``S.''.
	Other references without ``S.'' are in the main paper.
	Notation and quantities in the main paper will not be redefined but will be recalled for readers.
	This supplementary material first discusses the details of the full-kernel scheme in Section~\ref{supp:fullkernel}.
	The proofs of Theorem~\ref{prop:wrate} and Corollary~\ref{coro:wconver} are shown in Section~\ref{supp:wrateproof}.
	The lemmas needed to prove  Theorem~\ref{prop:wrate} are collected in Section~\ref{supp:wratelemmaproof}.
	Section~\ref{sec:proof} proves Theorem~\ref{them:consistency}-\ref{them:normality}, the main asymptotic theoretical results of SMKLE.
	Section~\ref{supp:thmproof} provides the proofs of the lemmas and proposition and the other technical details that are essential for the proof of Theorem~\ref{them:consistency}-\ref{them:normality} in Section~\ref{sec:proof}.
	Finally, Section~\ref{supp:teclem} gives some preliminary technical lemmas.
	
	In this supplement, we need the following notation for the proofs hereafter.
	The symbol $\lesssim$ ($\gtrsim$) means that the left (right)-hand
	side is bounded above by the right (left)-hand side multiplied by a constant.
	The $\epsilon$-bracketing number associated with some norm $\lVert
	\:\cdot\:\rVert$, denoted by $N_{[\,]}\left(\epsilon,\mathcal{F},\left\Vert
	\:\cdot\:\right\Vert \right)$,
	is the minimum number of
	$\epsilon$-brackets, which are the sets of all functions $f\in\mathcal{F}$ with $b\le f \le d$ and $\left\Vert b-d\right\Vert
	\le\epsilon$, needed to cover $\mathcal{F}$.
	Recall that  $\mathbb{G}_{n}\coloneqq\sqrt{n}(\mathbb{P}_{n}-\mathbf{P})$ denotes the empirical process where $\mathbb{P}_{n}$ and $\mathbf{P}$ denote the empirical and the true probability measure, respectively.

	\newpage
	{ 
		\section{The full-kernel smoothing scheme and its asymptotic behaviors}
		
		\label{supp:fullkernel}
		In many studies, 
		the event of interest is often terminal, and 
		the covariates cannot be observed after the occurrence of the event of interest. 
		The termination of follow-up motivates us to consider the \textit{half-kernel} smoothing scheme that only considers observations before the event time and the time within the integrand as shown in equation (4).
		Additionally, when the event is not terminal, In the case of the non-terminal event, the covariates can still be observed after the occurrence of the event of interest. We can improve the performance of the SKMLE by the \textit{full-kernel} smoothing scheme that includes more covariates. The \textit{full-kernel} smoothing scheme utilizes both forward and backward observations and has different asymptotic behaviors from the half-kernel smoothing. 
		
		To include all the observations before and after the terminal event, we can remove \textit{both} $I(R_{ij}\le X)$ and $I(R_{ij}\le t)$ in equation (4), resulting the full-kernel smoothing scheme for SMKLE.
		Specifically, we can derive the following kernel-smoothed log-likelihood with full-kernel smoothing
		\begin{multline*}
			\tilde{\ell}_{n}(\theta)=\frac{1}{n}\sum_{i=1}^{n}\sum_{j=1}^{J_{i}} \bigg[K_{h}\left(X_{i}-R_{ij}\right) \Delta_{i}\log\left\{
			H\left(\alpha\left(X_{i}\right)+\beta^{\top}Z_{i}\left(R_{ij}\right)\right)\right\} \bigg]
			\\  -\frac{1}{n}\sum_{i=1}^{n}\sum_{j=1}^{J_{i}}\left[\int_{0}^{X_{i}}K_{h}\left(t-R_{ij}\right)H\left(\alpha\left(t\right)+\beta^{\top}Z_{i}\left(R_{ij}\right)\right)dt\right], \label{eq:full-kernel}
		\end{multline*}
		where $K_h(\cdot)=K(\cdot/h)/h$. 
		We can then obtain an SMKLE by optimizing the full kernel-smoothed log-likelihood with a sieve approximation to $\alpha$.

		The asymptotic theory results of the full-kernel scheme can demonstrate its advantages over the half-kernel scheme if covariates can be observed after the event occurrence.
		Next, we briefly discuss the asymptotic theory results of the full-kernel scheme and compare the results with those of the half-kernel scheme.
		The main difference in the asymptotic results between the half kernel and full kernel estimation is the constraints of $a$, that is, the order of the bandwidth $h=O(n^{-a})$, to ensure the ``bias'' terms vanish asymptotically, while the result formulation usually remains the same.
		For the full kernel estimation, we need to modify regularity condition (C5) to a H\"older condition that
		$\sqrt{E\left[\left\Vert
			Z\left(t\right)-Z\left(s\right)\right\Vert _{2}^{2}\right]}\le
		L(\left|t-s\right|)^{b}$
		with large enough $b>1$, indicating a higher degree of smoothness of the covariate processes.
		Then, the order of the ``bias'' terms in the proof becomes $O(h_{n}^2)$, which is smaller than $O(h_{n})$ under the half-kernel scheme.
		The smaller order of the  ``bias'' terms incurs a smaller bias of the resulting estimator and the smaller lower bound of $\alpha$ required for the convergence rate and asymptotic normality.
		
		The form of the rate of convergence under the full-kernel scheme remains the same as that in Theorem~\ref{them:rate}. However, the allowable range of $a$ is expanded from $0.4(1-\nu) \le a \le 1 - \nu$ to $0.25(1-\nu) \le a \le 1 - \nu$ due to a smaller order of the ``bias'' term.
		This optimal rate $n^{3/8}$ is larger than its counterparts $n^{3/10}$ of the half-kernel method. 
		Additionally, the optimal rate of $\hat{\beta}$ from Theorem~\ref{them:normality} becomes
		$o(n^{3/7})$, which is larger than 
		$o(n^{2/5})$ in \citet{Cao2015} and \citet{Cao2021}, where full kernel is used for estimation. 
		The optimal rate of $\hat{\beta}$ from Theorem~\ref{them:normality} becomes
		$o(n^{3/7})$. The rate from Theorem~\ref{them:normality} is larger than its corresponding rate, $o(n^{2/5})$, in \citet{Cao2015} and \citet{Cao2021}, again, reinforcing the advantages of using sieve estimation.
		Under the full-kernel scheme, $\Xi$ remains $\Sigma$ but $\Omega$ becomes 
		$$\int_{-1}^{1}K^{2}(u)du\Sigma = 2\int_{0}^{1}K^{2}(u)du\Sigma,$$
		compared with their counterparts in Theorem~\ref{them:normality}.
		Consequently, the asymptotic variance in Theorem~\ref{them:normality} of the full-kernel scheme becomes half of the half-kernel one, indicating information gain and higher efficiency. 
		Correspondingly, under the full-kernel scheme, the upper bound of $r$ in $S^{\left(k\right)}$, $\hat{\Xi}$ and $\hat{\Omega}$ should be replaced by $\tau$, including more covariates information.
		
		We further validate the asymptotic results of the full-kernel scheme by a brief simulation study.
		In Table \ref{tab:full_kernel}, we show the estimates from the full-kernel scheme under the same setting as the simulation in the main manuscript. The bandwidth is fixed at $n^{-0.5}$. Comparing Table \ref{tab:full_kernel} with Table \ref{tab:half_kernel}, we can find that the SEs of the full kernel estimators are generally smaller than those from the half-kernel scheme, which suggests improved efficiency by utilizing more covariate information.

		\begin{table}[!htb] 
			\centering{}\caption{Simulation results with full-kernel method}
			\label{tab:full_kernel}
			\begin{tabular}{ccrrrrrrrrrr@{}}
				\hline
				&            & \multicolumn{4}{c}{20\% censoring rate} &  &         \multicolumn{4}{c}{30\% censoring rate}          \\ 
				Coef     &    $n$     & \mc{RB} & \mc{ESE} & \mc{SE} &  \mc{CP} &  & \mc{RB} & \mc{ESE} & \mc{SE} & \mc{CP} \\ 
				\hline
				&            &                                  \multicolumn{9}{c}{$s=0$}                                  \\
				$\beta_{1}$ & 200 &  -0.020 &    0.235 &   0.244 &     92.7 &  &   -0.010 &    0.250 &   0.258 &                      94.3 \\
				& 400 &  -0.014 &    0.184 &   0.186 &     94.5 &  &   -0.008 &    0.195 &   0.195 &                      95.2 \\
				$\beta_{2}$ & 200 &  0.004 &    0.252 &   0.257 &     93.1 &  &  -0.006 &    0.267 &   0.290 &                      93.4 \\
				& 400 &  0.018 &    0.198 &   0.201 &     94.6 &  &  -0.014 &    0.209 &   0.222 &                      93.2 \\
				&            &                                       \multicolumn{9}{c}{$s=0.25$}                                 \\
				$\beta_{1}$ & 200 &  -0.015 &    0.257 &   0.264 &     93.9 &  &  -0.002 &    0.271 &   0.274 &                      94.0 \\
				& 400 &  -0.011 &    0.201 &   0.205 &     94.2 &  &  -0.008 &    0.211 &   0.217 &                      94.1 \\
				$\beta_{2}$ & 200 &  0.014 &    0.284 &   0.310 &     92.7&   &  0.009 &    0.300 &   0.316 &                      93.1 \\
				& 400 &  0.011 &    0.223 &   0.235 &     93.8&   &  -0.003 &    0.235 &   0.249 &                      94.4 \\
				&            &                                       \multicolumn{9}{c}{$s=0.5$}                                 \\
				$\beta_{1}$ & 200  &  -0.025 &    0.276 &   0.288 &     92.9 &  &  -0.007 &    0.289 &   0.308 &                      92.2 \\
				& 400 &  -0.012 &    0.217 &   0.222 &     94.0 &  &  -0.010 &    0.228 &   0.239 &                      93.2 \\
				$\beta_{2}$ & 200 &  -0.012 &    0.309 &   0.325 &     93.8&   &  0.041 &    0.323 &   0.356 &                      92.8 \\
				& 400 &  0.004 &    0.244 &   0.248 &     94.4&   &  0.044 &    0.255 &   0.260 &                      94.8 \\
				&            &   \multicolumn{9}{c}{$s=0.75$}                                 \\
				$\beta_{1}$  & 200 &  -0.020 &    0.295 &   0.311 &     92.3 &  &  -0.008 &    0.311 &   0.334 &                      92.4\\
				& 400 &  -0.006 &    0.232 &   0.238 &     94.6 &  &  -0.010 &    0.243 &   0.253 &                      93.5\\
				$\beta_{2}$  & 200 &  0.009 &    0.328 &   0.355 &     93.0&   &  -0.008 &    0.347 &   0.381 &                      93.0 \\
				& 400 &  0.026 &    0.259 &   0.267 &     94.0&   &  -0.001 &    0.273 &   0.294 &                      92.7 \\
				&            &   \multicolumn{9}{c}{$s=1$}                                 \\
				$\beta_{1}$ & 200 &  -0.011 &    0.348 &   0.351 &     94.3 &  &  -0.005 &    0.365 &   0.371 &                      94.4 \\
				& 400 &  -0.009 &    0.266 &   0.272 &     94.0 &  &  -0.009 &    0.272 &   0.279 &                      95.3 \\
				$\beta_{2}$ & 200 &  -0.036 &    0.372 &   0.385 &     94.3&   &  0.024 &    0.397 &   0.407 &                      95.5 \\
				& 400 &  -0.012 &    0.289 &   0.300 &     95.2&   &  0.024 &    0.305 &   0.327 &                      93.5 \\
				\hline
			\end{tabular}
		\end{table}

		\begin{table}[!htb] 
			\centering{}\caption{Simulation results with the half-kernel method}
			\label{tab:half_kernel}
			\begin{tabular}{ccrrrrrrrrrr@{}}
				\hline
				&            & \multicolumn{4}{c}{20\% censoring rate} &  &         \multicolumn{4}{c}{30\% censoring rate}          \\ 
				Coef     &    $n$     & \mc{RB} & \mc{ESE} & \mc{SE} &  \mc{CP} &  & \mc{RB} & \mc{ESE} & \mc{SE} & \mc{CP} \\ 
				\hline
				&            &                                  \multicolumn{9}{c}{$s=0$}                                  \\
				$\beta_{1}$ & 200  &-0.020 &    0.299 &   0.317 &     93.9 &  &  -0.005 &    0.318 &   0.332 &                      94.8 \\
				& 400 &  -0.020 &    0.239 &   0.239 &     94.9 &  &  -0.004 &    0.252 &   0.252 &                      95.6 \\
				$\beta_{2}$ & 200 &0.026 &    0.326 &   0.345 &     94.5 &  &   0.000 &    0.346 &   0.367 &                      93.4 \\
				& 400 &   0.020 &    0.261 &   0.262 &     94.9 &  &   0.014 &    0.276 &   0.287 &                      94.2 \\
				&            &                                       \multicolumn{9}{c}{$s=0.25$}                                 \\
				$\beta_{1}$ & 200 &  -0.004 &    0.327 &   0.343 &     94.3 &  &   0.004 &    0.345 &   0.362 &    93.0 \\
				& 400  &  -0.004 &    0.260 &   0.270 &     95.1 &  &  -0.003 &    0.274 &   0.287 &                      93.1 \\
				$\beta_{2}$ & 200 &  -0.040 &    0.366 &   0.397 &     92.7 &  &  -0.016 &    0.386 &   0.406 &    94.3 \\
				& 400  &  -0.036 &    0.292 &   0.298 &     94.6 &  &   0.011 &    0.308 &   0.325 &                      93.3 \\
				&            &                                       \multicolumn{9}{c}{$s=0.5$}                                 \\
				$\beta_{1}$ & 200  &   0.003 &    0.351 &   0.371 &     92.4 &  &  -0.005 &    0.367 &   0.397 &                      92.0 \\
				& 400 &   0.004 &    0.279 &   0.293 &     93.1 &  &  -0.009 &    0.294 &   0.319 &                      92.0 \\
				$\beta_{2}$ & 200 &  -0.011 &    0.396 &   0.411 &     93.7 &  &  -0.043 &    0.415 &   0.443 &                      93.2 \\
				& 400 &  -0.036 &    0.317 &   0.327 &     93.3 &  &  -0.041 &    0.333 &   0.356 &                      92.7 \\
				&            &   \multicolumn{9}{c}{$s=0.75$}                                 \\
				$\beta_{1}$  & 200 &   0.003 &    0.381 &   0.412 &     93.0 &  &  -0.015 &    0.407 &   0.417 &                      94.4 \\
				& 400  &   0.008 &    0.304 &   0.323 &     93.3 &  &  -0.015 &    0.318 &   0.333 &                      93.7 \\
				$\beta_{2}$  & 200 &  -0.026 &    0.425 &   0.468 &     93.0 &  &   0.023 &    0.452 &   0.492 &                      92.9 \\
				& 400 &  -0.035 &    0.341 &   0.372 &     93.9 &  &   0.021 &    0.358 &   0.386 &                      93.1 \\
				&            &   \multicolumn{9}{c}{$s=1$}                                 \\
				$\beta_{1}$ & 200 &   0.007 &    0.473 &   0.467 &     93.5 &  &  -0.018 &    0.511 &   0.489 &                      92.8 \\
				& 400 &  -0.003 &    0.355 &   0.365 &     94.3 &  &  -0.021 &    0.382 &   0.358 &                      95.3 \\
				$\beta_{2}$ & 200 &   0.040 &    0.500 &   0.507 &     95.8 &  &   0.023 &    0.545 &   0.540 &                      95.1 \\
				& 400 &   0.015 &    0.385 &   0.406 &     92.9 &  &  -0.010 &    0.414 &   0.419 &                      94.6 \\
				\hline
			\end{tabular}
		\end{table} 
	}
	
	\newpage
	
	\section{The proofs of Theorem~\ref{prop:wrate} and Corollary \ref{coro:wconver}}
	\label{supp:wrateproof}
	
	\subsection{Additional notation}
	We develop the uniform-in-bandwidth convergence rate of $\left\Vert \mathbb{G}_{n}\right\Vert _{\mathcal{W}_{nh}^{k}}$ for $k=1,2$.
	The main technical challenge is that the uniform-in-bandwidth convergence allows the bandwidth to range in an interval whose length decreases with the sample size.
	Consequently, the functions in $\mathcal{W}_{nh}^{k}$, for $k=1,2$, are
	not bounded and square integrable for arbitrarily small $h$; it is difficult to compute the bracketing numbers of the underlying classes.
	To sidestep this technical issue, inspired by \citet{Einmahl2005}, we adopt on the chaining argument that splits both $n$ and $h$ in chains on the classes
	$\bar{\mathcal{W}}_{nh}^{1} \coloneqq \left\{
	h\omega_{1}:\omega_{1}\in\mathcal{W}_{nh}^{1}\right\}$
	and
	$\bar{\mathcal{W}}_{nh}^{2} \coloneqq \left\{
	h\omega_{2}:\omega_{2}\in\mathcal{W}_{nh}^{2}\right\}$, that is, the sets of the functions in $\mathcal{W}_{nh}^{k}$, $k=1,2$, multiplied by the bandwidth.
	Specifically, we consider the following classes
	\[
	\bar{\mathcal{W}}_{j,k}^{1} \coloneqq \{ \omega_{1,j,k}\in\bar{\mathcal{W}}_{n_{k}h}^{1}:h_{j-1,k}< h\le h_{j,k}\}
	\]
	and 
	\[
	\bar{\mathcal{W}}_{j,k}^{2}\coloneqq \{ \omega_{2,j,k}\in\bar{\mathcal{W}}_{n_{k}h}^{2}:h_{j-1,k} < h\le h_{j,k}\}
	\]
	indexed by both $h$ and $\theta \in\Theta_{n}$. 
	Here, $n_{k}=2^{k}$ and $h_{j,k}=\left(2^{j}c_{1}\log n_{k}\right)/n_{k}^{1-\nu}$ for some some nonegnative integers $j,k$ and some constant $c_{1}>0$. 
	Clearly, $h_{j,k}=2h_{j-1,k}$ for each $k$.
	
	Recall that the overall sieve space for $\Theta$ as $\Theta_{n} \coloneqq \mathcal{B}\times\mathcal{A}_{n,l}$ where $\mathcal{A}_{n,l}$ is defined in \eqref{eq:seivespace} with its dimension $q_{n}=O(n^{\nu})$.

	\subsection{Technical lemmas to prove Theorem~\ref{prop:wrate}}
	The following lemmas povide some properties of the classes defined in the above. 
	Lemma \ref{lem:Wbracketing} and Lemma \ref{lem:Wbarhbracketing} provide the respective bracketing numbers of $\bar{\mathcal{W}}_{nh}^{1}$, $\bar{\mathcal{W}}_{nh}^{2}$, $\bar{\mathcal{W}}_{j,k}^{1}$ and $\bar{\mathcal{W}}_{j,k}^{2}$.
	Lemma \ref{lem:WbarhUppers} gives the uniform upper bounds of functions in $\bar{\mathcal{W}}_{nh}^{1}$ and $\bar{\mathcal{W}}_{nh}^{2}$ under the sup and $\mathcal{L}_2$ norms, respectively.
	Lemma \ref{lem:WbarhUppers} then implies Corollary \ref{lem:WhbarhUppers}.
	The detailed proofs of Lemmas \ref{lem:Wbracketing}-\ref{lem:WbarhUppers} are relegated to Section~\ref{supp:wratelemmaproof}.

	\begin{lem}
		\label{lem:Wbracketing}
		Under Conditions \ref{enu:kernel}-\ref{enu:zbounded} and \ref{enu:finiteobs},
		$$N_{[\,]}\left(\epsilon,\bar{\mathcal{W}}_{nh}^{1},\mathcal{L}_{2}\left(\mathbf{P}\right)\right)\lesssim\left(\sqrt{h}/\epsilon\right)^{cq_{n}+p}$$
		and
		$$N_{[\,]}\left(\epsilon,\bar{\mathcal{W}}_{nh}^{2},\mathcal{L}_{2}\left(\mathbf{P}\right)\right)\lesssim\left(\sqrt{h}/\epsilon\right)^{cq_{n}+p}$$.
		Here, $c>0$ is some constant and $c$ appearing in different contexts may represent different constants.
	\end{lem}

	\begin{lem}
		\label{lem:Wbarhbracketing}
		Under Conditions \ref{enu:kernel}-\ref{enu:zbounded} and \ref{enu:finiteobs},
		$$N_{[\,]}(\epsilon,\bar{\mathcal{W}}_{j,k}^{1},\mathcal{L}_{2}(\mathbf{P}))\lesssim(\sqrt{h_{j,k}}/\epsilon)^{cq_{n}+p+1}$$
		and
		$$N_{[\,]}(\epsilon,\bar{\mathcal{W}}_{j,k}^{2},\mathcal{L}_{2}(\mathbf{P}))\lesssim(\sqrt{h_{j,k}}/\epsilon)^{cq_{n}+p+1}$$
		with some constant $c>0$.
		
	\end{lem}

	\begin{lem}
		\label{lem:WbarhUppers}
		Suppose Conditions \ref{enu:kernel}, \ref{enu:alpha_para_space}, \ref{enu:beta_para_space},
		\ref{enu:zbounded}, \ref{enu:H_cond}, \ref{enu:ratefun} and \ref{enu:finiteobs} hold.
		Then, for small enough $h$ and some positive finite constant $D_{11}$, $D_{12}$, $D_{21}$ and $D_{22}$, $$\sup_{f\in\bar{\mathcal{W}}_{nh}^{1}}\left\Vert f\right\Vert _{\infty}\le D_{11}$$ and 
		$$\sup_{f\in\bar{\mathcal{W}}_{nh}^{2}}\left\Vert f\right\Vert _{\infty}\le D_{12}$$ with probability one;
		$$\sup_{f\in\bar{\mathcal{W}}_{nh}^{1}}\mathbf{P}f^{2}\le hD_{21}$$ and
		$$\sup_{f\in\bar{\mathcal{W}}_{nh}^{2}}\mathbf{P}f^{2}\le hD_{22}$$.
	\end{lem}

	\begin{coro}
		\label{lem:WhbarhUppers}
		Assume the conditions of Lemma \ref{lem:WbarhUppers} hold.
		Then, $\sup_{f\in\bar{\mathcal{W}}_{j,k}^{1}}\left\Vert f\right\Vert _{\infty}\le D_{11}$,
		$\sup_{f\in\bar{\mathcal{W}}_{j,k}^{2}}\left\Vert f\right\Vert _{\infty}\le D_{21}$,
		$\sup_{f\in\bar{\mathcal{W}}_{j,k}^{1}}\mathbf{P}f^{2}\le h_{j,k}D_{12}$ and
		$\sup_{f\in\bar{\mathcal{W}}_{j,k}^{2}}\mathbf{P}f^{2}\le h_{j,k}D_{22}$.
	\end{coro}
	
	\begin{proof}
		Following Lemma \ref{lem:WbarhUppers}, it is trivial to show $\sup_{f\in\bar{\mathcal{W}}_{j,k}^{1}}\left\Vert f\right\Vert _{\infty}\le D_{11}$ and $\sup_{f\in\bar{\mathcal{W}}_{j,k}^{1}}\left\Vert f\right\Vert _{\infty}\le D_{21}$.
		It also follows from Lemma \ref{lem:WbarhUppers} that 
		\begin{align*}
			\sup_{f\in\bar{\mathcal{W}}_{j,k}^{1}}\mathbf{P}f^{2} & \le\sup_{h_{j-1,k}<h\le h_{j,k}}\sup_{f\in\bar{\mathcal{W}}_{n_{k}h}^{1}}\mathbf{P}f^{2}\\
			& \le \sup_{h_{j-1,k}<h\le h_{j,k}}hD_{12}\\
			& =h_{j,k}D_{21}.
		\end{align*}
		The same argument also proves $\sup_{f\in\bar{\mathcal{W}}_{j,k}^{2}}\mathbf{P}f^{2}\le h_{j,k}D_{22}$.
	\end{proof}
	
	\subsection{The detailed proof of Theorem~\ref{prop:wrate}}

	
	\begin{proof}
		We mainly prove the uniform-in-bandwidth convergence rate for $\mathcal{W}_{nh}^{1}$ and briefly discuss how the similar arguments can prove that for  $\mathcal{W}_{nh}^{2}$.
		The proof is outlined as follows.
		First, we derive an upper bound of $\sqrt{n_{k}} E\left\Vert \mathbb{G}_{n_{k}}\right\Vert _{\bar{\mathcal{W}}_{j,k}^{1}}$.
		Second, for large enough $k$, we obtain the bound on a tail probability of $\max_{n_{k-1}\le n\le n_{k}}\sqrt{n}\left\Vert \mathbb{G}_{n}\right\Vert _{\bar{\mathcal{W}}_{j,k}^{1}}$.
		Last, we prove the uniform-in-bandwidth convergence rate by the chaining and Borel-Cantelli lemma.
		
		\underline{\textbf{Step 1.}}
		We complete the first step by Lemma \ref{lem:maximial-inequality}.
		Using the notations in Lemma \ref{lem:maximial-inequality}, Corollary \ref{lem:WhbarhUppers} yields that, 
		for $\bar{\mathcal{W}}_{j,k}^{1}$, $U =D_{11}$ and $\xi^{2}=h_{j,k}D_{12}$.
		By Lemma \ref{lem:Wbarhbracketing}, $N_{[\,]}(\epsilon,\bar{\mathcal{W}}_{j,k}^{1},\mathcal{L}_{2}(\mathbf{P}))\le D(\sqrt{h_{j,k}}/\epsilon)^{v_{n_{k}}}$
		where some constant $D>0$ and $v_{n_{k}}=O\left(cq_{n_{k}}+p+1\right)=O(n_{k}^{\nu})$.
		This means that we can choose $\eta=\xi$ up to a constant as $ = O(\sqrt{h_{j,k}}/\left(\sqrt{h_{j,k}}D_{12}\right))=O(1)$.
		Since $ D^{1/v_{n_{k}}}\to1$ as $k\to \infty$, we have 
		 $\log(D_1)\to 1$ for large enough $k$.
		Then, Lemma \ref{lem:maximial-inequality} implies that, for large enough $k$,
		\begin{align}
			\sqrt{n_{k}}E\left\Vert \mathbb{G}_{n_{k}}\right\Vert _{\bar{\mathcal{W}}_{j,k}^{1}} & \lesssim2\sqrt{v_{n_{k}}n_{k}h_{j,k}D_{12}\log\left(D_{1}\right)}+4v_{n_{k}}D_{11}\log\left(D_{1}\right) \nonumber \\
			& \lesssim\sqrt{v_{n_{k}}n_{k}h_{j,k}}+v_{n_{k}}. \label{eq:epbound_raw}
		\end{align}
		We next verify that, for large enough $k$,
		\begin{equation}
			\sqrt{v_{n_{k}}n_{k}h_{j,k}}\gtrsim v_{n_{k}},\label{eq:eporder_ineqaulity}
		\end{equation}
		so that we can only keep $\sqrt{v_{n_{k}}n_{k}h_{j,k}}$ on the RHS of \eqref{eq:epbound_raw}.
		Recall that $v_{n_{k}} = O(n_{k}^{\nu})$,
		$n_{k}=2^{k}$ and $h_{j,k}=\left(2^{j}c_{1}\log n_{k}\right)/n_{k}^{1-\nu}$.
		Then, for $j \ge 0$,
		\begin{align}
			n_{k}h_{j,k} & =n_{k}\left(2^{j}c_{1}\log n_{k}\right)/n_{k}^{1-\nu} \nonumber \\
			& =n_{k}^{\nu}\left(2^{j}c_{1}\log n_{k}\right)\nonumber \\
			& \gtrsim v_{n_{k}}. \label{eq:nhv_order_ineq}
		\end{align}
		Hence, \eqref{eq:nhv_order_ineq} yields that $\sqrt{v_{n_{k}}n_{k}h_{j,k}}\gtrsim\sqrt{v_{n_{k}}^{2}}=v_{n_{k}}$, proving \eqref{eq:eporder_ineqaulity}.
		
		For large enough $k$, it then follows from \eqref{eq:eporder_ineqaulity} that
		\begin{equation}
			\sqrt{n_{k}}E\left\Vert \mathbb{G}_{n_{k}}\right\Vert _{\bar{\mathcal{W}}_{j,k}^{1}}  \le D_{3}a_{j,k}, \label{eq:Gmaximalineq}
		\end{equation}
		where   $D_{3}$ is some positive constant and 
		\[
		a_{j,k}\coloneqq\sqrt{v_{n_{k}}n_{k}h_{j,k}}.	 
		\]
		
		\underline{\textbf{Step 2.}}		
		Based on \eqref{eq:Gmaximalineq}, we proceed to bound the tail probability of
		$$\max_{n_{k-1}\le n\le n_{k}}\sqrt{n}\left\Vert \mathbb{G}_{n}\right\Vert _{\bar{\mathcal{W}}_{j,k}^{1}}$$%
		by the following two inequalities.
		In the context of the Lemma \ref{lem:maximial-inequality}, Theorem
		7.3 in \citet{Bousquet2003} gives a concentration inequality
		of the empirical process 
		\begin{equation}
			\Pr\left(\sqrt{n}\left\Vert \mathbb{G}_{n}\right\Vert _{\mathcal{G}}\ge \sqrt{n}E\left\Vert \mathbb{G}_{n}\right\Vert _{\mathcal{G}}+\sqrt{2s\left(2\sqrt{n}E\left\Vert \mathbb{G}_{n}\right\Vert _{\mathcal{G}}+n\xi^{2}\right)}+s/3\right)\le e^{-s},\label{eq:Tlkagrand1996}
		\end{equation}
		for any $s>0$ where $\sup_{g\in\mathcal{G}}\mathbf{P}g^{2}\le\xi^{2}$. Theorem 1.1.5 in \citet{delaPena1999} provides a maximal
		inequality of the partial sum of the empirical process in the form
		of 
		\begin{equation}
			\Pr\left(\max_{1\le m\le n}\sqrt{m}\left\Vert \mathbb{G}_{m}\right\Vert _{\mathcal{G}}>s\right)\le9\Pr\left(\sqrt{n}\left\Vert \mathbb{G}_{n}\right\Vert _{\mathcal{G}}>s/30\right),\label{eq:Smith1994}
		\end{equation}
		for any $s>0$.
		Combining \eqref{eq:Tlkagrand1996} and \eqref{eq:Smith1994},
		we obtain the following inequality 
		\begin{equation}
			\Pr\left(\max_{1\le m\le n}\sqrt{m}\Vert\mathbb{G}_{m}\Vert_{\mathcal{G}}>30\left\{ \sqrt{n}E\Vert\mathbb{G}_{n}\Vert_{\mathcal{G}}+\sqrt{2s(2\sqrt{n}E\Vert\mathbb{G}_{n}\Vert_{\mathcal{G}}+n\xi^{2})}+\frac{s}{3}\right\} \right)\le9e^{-s}.\label{eq:main_inequality}
		\end{equation}
		
		For large enough $k$, by \eqref{eq:nhv_order_ineq}, $a_{j,k}\lesssim\sqrt{\left(n_{k}h_{j,k}\right)^{2}}\lesssim n_{k}h_{j,k}$.
		It then easily follows that
		\begin{equation}
			\label{eq:2nEGW1up}
			2\sqrt{n_{k}}E\left\Vert \mathbb{G}_{n_{k}}\right\Vert _{\bar{\mathcal{W}}_{j,k}^{1}}+n_{k}h_{j,k}D_{12}\le D_{4}n_{k}h_{j,k},
		\end{equation}
		for some constant $D_{4}>0$.
		Combining \eqref{eq:Gmaximalineq} and \eqref{eq:2nEGW1up} into \eqref{eq:main_inequality} with $\mathcal{G}=\bar{\mathcal{W}}_{j,k}^{1}$, we obtain that
		\[
		\tilde{p}_{j,k}\left(\rho\right)\coloneqq\Pr\left(\max_{n_{k-1}\le n\le n_{k}}\sqrt{n}\left\Vert \mathbb{G}_{n}\right\Vert _{\bar{\mathcal{W}}_{j,k}^{1}}>30\left\{ D_{3}a_{j,k}+\sqrt{2sD_{4}n_{k}h_{j,k}}+\frac{s}{3}\right\} \right)\le9e^{-s},
		\]
		where we can choose $s=\rho v_{n_{k}}$ for any positive $\rho$ and therefore
		\begin{equation}
			\tilde{p}_{j,k}\left(\rho\right) \le 9\exp\left(-\rho v_{n_{k}} \right) = 9\exp\left(-\rho c_{7} 2^{k\nu}\right), \label{eq:pjkupper}
		\end{equation}
		for some constant $c_{7}>0$.
		Furthermore, given $s=\rho v_{n_{k}}$ and $a_{j,k}=\sqrt{v_{n_{k}}n_{k}h_{j,k}}$
		\begin{equation}
			D_{3}a_{j,k}+\sqrt{4sD_{4}n_{k}h_{j,k}}+\frac{s}{3}=a_{j,k}\left\{ D_{3}+\sqrt{2\rho D_{4}}+\frac{\rho a_{j,k}}{3n_{k}h_{j,k}}\right\} ,\label{eq:rhoequation}
		\end{equation}
		where $\rho>0$ is arbitrary as \eqref{eq:Smith1994} holds for any $s$.
		By $a_{j,k}\lesssim n_{k}h_{j,k}$, for large enough $k$, the second and third term on the
		RHS of \eqref{eq:rhoequation} is bounded by $\rho D_{5}$ for some positive constant  $D_{5}$.
		Consequently,
		\[
		\Pr\left(\max_{n_{k-1}\le n\le n_{k}}\sqrt{n}\left\Vert \mathbb{G}_{n}\right\Vert _{\bar{\mathcal{W}}_{j,k}^{1}}>30a_{j,k}\left\{ D_{3}+\sqrt{2\rho D_{4}}+\rho D_{5}\right\} \right)\le\tilde{p}_{j,k}\left(\rho\right).
		\]

		\underline{\textbf{Step 3.}}
		We will show that the summation of $\tilde{p}_{j,k}(\rho)$ with respective to  $j$ and $k$ is finite for some $\rho$ and prove the proposition by the Borel-Cantelli lemma.
		Set  $j^{*}_{k}=\max\left\{ j: h_{j,k}\le 2\right\}$.	
		By the definition of $h_{j,k}$, for large enough
		$k$, $j^{*}_{k}\lesssim\log\left(n_{k}\right)$.
		Therefore, by \eqref{eq:pjkupper} and $n_{k}=2^{k}$,
		\begin{align*}
			\tilde{P}_{k}\left(\rho\right) & \coloneqq\sum_{j=0}^{j_{k}^{*}}\tilde{p}_{j,k}(\rho)\\
			& \le \sum_{j=0}^{j_{k}^{*}} 9\exp\left(-\rho c_{7} 2^{k\nu}\right)\\
			& \lesssim \log\left(n_{k}\right)\exp\left(-\rho c_{7} 2^{k\nu}\right)\\
			& \lesssim k\exp\left(-\rho c_{7} 2^{k\nu}\right).
		\end{align*}	
		For large enough $k$, 
		\begin{align*}
			\sum_{k=1}^{\infty}\tilde{P}_{k}\left(\rho\right) & \lesssim\sum_{k=1}^{\infty}k\exp\left(-\rho c_{7}2^{k\nu}\right)\\
			& \le\sum_{k=1}^{\infty}\frac{k}{2^{\rho c_{7}\nu k}}\\
			& \lesssim\frac{1}{2^{\rho c_{7}\nu}}+\int_{0}^{\infty}\frac{k}{2^{\rho c_{7}\nu k}}dk\\
			& =\frac{1}{2^{\rho c_{7}\nu}}+\frac{1}{\left\{ \rho c_{7}\nu\log(2)\right\} ^{2}}<\infty,
		\end{align*}
		for positive $\rho$, $c_{7}$ and $\nu$.
		
		With $\sum_{k=1}^{\infty}\tilde{P}_{k}\left(\rho\right) < \infty$, we finally provide the proposition.
		By the definition of $j^{*}_{k}$ for large $k$, $h_{j^{*}_{k},k}=\frac{1}{2}h_{j^{*}_{k}+1,k} > 1$.
		Hence, for $n_{k-1}\le n\le n_{k}$, we have 
		\[
		\left[\frac{c_{1}\log\left(n\right)}{n^{1-\nu}},1\right]\subset\left[\frac{c_{1}\log\left(n_{k}\right)}{n_{k}^{1-\nu}},h_{j_{k}^{*},k}\right]=\left[h_{0,k},h_{j_{k}^{*},k}\right],
		\]
		for any $c_{1}>0$.
		Thus, for all large enough $k$ and $n_{k-1}\le n\le n_{k}$,
		\begin{multline*}
			A_{k}\left(\rho\right)=\left\{ \max_{n_{k-1}\le n\le n_{k}}\sup_{c_{1}\log n/n^{1-\nu}\le h\le 1}\frac{n\left\Vert \mathbb{P}_{n}-\mathbf{P}\right\Vert _{\bar{\mathcal{W}}_{nh}^{1}}}{\sqrt{v_{n}nh  }}\right.\\
			\left.\ge30\left(D_{3}+\sqrt{2\rho D_{4}}+\rho D_{5}\right)\right\} \\
			\subset\bigcup_{j=0}^{j^{*}_{k}}\left\{ \max_{n_{k-1}\le n\le n_{k}}\sqrt{n}\left\Vert \mathbb{G}_{n}\right\Vert _{\bar{\mathcal{W}}_{j,k}^{1}}\ge30a_{j,k}\left(D_{3}+\sqrt{2\rho D_{4}}+\rho D_{5}\right)\right\}.
		\end{multline*}
		It follows that $\Pr\left(A_{k}\left(\rho\right)\right)\le\sum_{j=0}^{j^{*}_{k}}\tilde{p}_{j,k}\left(\rho\right)=\tilde{P}_{k}\left(\rho\right)$ and 
		\[
		\sum_{k=1}^{\infty}\Pr\left(A_{k}\left(\rho\right)\right)\le\sum_{k=1}^{\infty}\tilde{P}_{k}\left(\rho\right)<\infty.
		\]
		Since $D_{3}+\sqrt{4\rho D_{4}}+\rho D_{5}$ is finite, by the Borel-Cantelli lemma, there exists some finite constant $d_{1}$ such
		that, with probability one,
		\[
		\limsup_{n\to\infty}\sup_{c_{1}\log n/n^{1-\nu}\le h\le 1}\sqrt{n^{1-\nu}/h}\left\Vert \mathbb{P}_{n}-\mathbf{P}\right\Vert _{\bar{\mathcal{W}}_{nh}^{1}}=d_{1} < \infty.
		\]
		Using the fact that $\left\Vert \mathbb{P}_{n}-\mathbf{P}\right\Vert _{\bar{\mathcal{W}}_{nh}^{1}}=h\left\Vert \mathbb{P}_{n}-\mathbf{P}\right\Vert _{\mathcal{W}_{nh}^{1}}$, the uniform-in-bandwidth convergence on $\mathcal{W}_{nh}^{1}$ holds.
		
		From the proof above, we can see that the upper bounds and bracketing numbers in Lemma \ref{lem:WbarhUppers} and \ref{lem:Wbarhbracketing}
		determine the overall behavior of the empirical process of $\bar{\mathcal{W}}_{nh}^{1}$.
		Hence, the uniform-in-bandwidth convergence rate will be same for $\bar{\mathcal{W}}_{nh}^{2}$ since the $\bar{\mathcal{W}}_{nh}^{2}$ has the same orders of the upper bounds
		as those of $\bar{\mathcal{W}}_{nh}^{1}$ in terms of $h$ and the bracketing numbers of $\bar{\mathcal{W}}_{j,k}^{1}$ and $\bar{\mathcal{W}}_{j,k}^{2}$ are the same in Lemma \ref{lem:Wbarhbracketing}.
		The detailed proof is omitted.
	\end{proof}
	%

	\subsection{The proof of Corollary \ref{coro:wconver}}

	For large enough $n$, the range  of $h_n$ could fall in  $h_n$ falls in $[c_{1}\log n/n^{1-\nu}, 1]$ since $h_{n}\to0$ as $n\to\infty$.
	Theorem~\ref{prop:wrate} immediately implies that, for large enough $n$,  with probability one,
	\[
	\Vert\mathbb{P}_{n}-\mathbf{P}\Vert_{\mathcal{W}_{nh_{n}}^{\iota}}\le O\left(\sqrt{1/(n^{1-\nu}h_{n})}\right).
	\]
	Some algebra shows that 
	\begin{equation}
		1/(n^{1-\nu}h_{n})=\log(n)/(n^{1-\nu}h_{n})/\log(n). \label{eq:Gn_CR}
	\end{equation}	
	It clearly follows from $n^{1-\nu}h_{n}/\log(n)\to\infty$ that  $\log(n)/(n^{1-\nu}h_{n})\to0$.
	Then, the RHS of \eqref{eq:Gn_CR} converges to zero as $n\to\infty$, implying $\Vert \mathbb{P}_{n}-\mathbf{P}\Vert_{\mathcal{W}_{nh_{n}}^{\iota}} \to 0$ almost surely.

	\newpage
	
	\section{The detailed proof of Theorem~\ref{them:consistency}-\ref{them:normality}}
	\label{sec:proof}

	
	
	We sketch the proofs of Theorem~\ref{them:consistency}-\ref{them:normality} in the following three subsection respectively.
	Recall that  $$H(\theta;t,Z(r))\coloneqq
	H(\alpha(t)+\beta^{\top}Z(r)),$$
	$$H^{\prime}(\theta;t,Z(r))\coloneqq
	H^{\prime}(\alpha(t)+\beta^{\top}Z(r)),$$
	$$H^{(1)}(\theta;t,Z(r))\coloneqq \frac{H^{\prime}(\theta;t,Z(r)) }{H(\theta;t,Z(r))}$$
	and
	$$H^{(2)}(\theta;t,Z(r))\coloneqq \frac{H^{\prime}(\theta;t,Z(r))^{2}}{H(\theta;t,Z(r))}$$
	where $H^{\prime}$ is the first order derivative of $H$. These notations involving $H$ and $H^{\prime}$ would be used extensively in the following proof.

	\subsection{The proof of Theorem \ref{them:consistency}}

	Before proceeding further, we introduce the following notation.
	Define $m_{\theta,h} \coloneqq m_{\theta,h,1}-m_{\theta,h,2}$
	where
	$$
	m_{\theta,h,1}(O) \coloneqq \int_{0}^{\infty}K_{h}(X-r)I(r\le
	X)\Delta\cdot\log\left\{ H\left(\theta;X,Z\left(r\right)\right)\right\} dN(r),
	$$
	$$
	m_{\theta,h,2}(O) \coloneqq \int_{0}^{\tau}\int_{0}^{\infty}I(r\le
	t)K_{h}(t-r)I(X\ge t)\allowbreak
	H\left(\theta;t,Z\left(r\right)\right)\allowbreak dN(r)\allowbreak
	dt,$$
	and
	$$
	m_{\theta}(O)\coloneqq \Delta\cdot\log\{ H\left(\theta;X,Z\left(X\right)\right)\}\mu(X)-\int_{0}^{X}H\left(\theta;t,Z\left(t\right)\right)\mu(t)dt.
	$$
	Let
	$\mathbb{M}_{n,h}\left(\theta\right) \coloneqq\mathbb{P}_{n}m_{\theta,h}= \tilde{\ell}_{n}\left(\theta\right)$,
	$\bM_{h}\left(\theta\right) \coloneqq \mathbf{P}m_{\theta,h}$
	and
	$\bM\left(\theta\right) \coloneqq \mathbf{P}m_{\theta}$.

	The outline of the proof of Theorem~\ref{them:consistency} is as follows.
	First, we verify $\bM\left(\theta\right)$ is uniquely maximized at
	$\theta_{0}$ over $\Theta$.
	Then, we show $\sup_{\theta\in\Theta_{n}}|\mathbb{M}_{n,h_{n}}(\theta)-\bM(\theta)|$ converges to zero a.s. as $n\to \infty$, facilitated by Corollary \ref{coro:wconver}.
	Finally, for any $\epsilon>0$, we complete the proof by showing 
	$\Pr(\limsup_{n\to\infty}\{ d(\hat{\theta}_{n},\theta_{0})>\epsilon\} )=0$ for any $\epsilon>0$.
	
	\underline{\textbf{Step 1.}}
	We prove that $\theta_{0}$  is the unique maximizer of $\bM\left(\theta\right)$ over $\Theta$.
	Using the expectation conditional on $X$, $Z$, $\Delta$ and $N$ characterized in
	\ref{enu:ratefun}, after some algebra, we can obtain
	$\bM\left(\theta\right)=\int_{0}^{\tau}\tilde{g}\left(\theta;t,t\right)dt$
	and
	$\bM_{h}\left(\theta\right)=\int_{0}^{\tau}\int_{0}^{\infty}I\left(r\le
	t\right)K_{h}\left(t-r\right)\tilde{g}\left(\theta;t,r\right)drdt$,
	where
	\begin{multline}
		\label{eq:gtileddef}
		\tilde{g}\left(\theta;t,r\right) \coloneqq E\left[S_{T\wedge C|Z}\left(t\right)\mu\left(r\right)\right.
		\\
		\times\left.\left\{ \log\left\{
		H\left(\alpha\left(t\right)+\beta^{\top}Z\left(r\right)\right)\right\}
		H\left(\alpha_{0}\left(t\right)+\beta_{0}^{\top}Z\left(t\right)\right)-H\left(\alpha\left(t\right)+\beta^{\top}Z\left(r\right)\right)\right\}
		\right].
	\end{multline}
	Hence,
	\begin{equation}
		\bM\left(\theta_{0}\right)-\bM\left(\theta\right)=E\left[\int_{0}^{\tau}S_{T\wedge C|Z}\left(t\right)H\left(\theta_{0};t,Z\left(t\right)\right)\mu\left(t\right)q\left(\frac{H\left(\theta;t,Z\left(t\right)\right)}{H\left(\theta_{0};t,Z\left(t\right)\right)}\right)dt\right],\label{eq:M0diff}
	\end{equation}
	where $q\left(x\right)=x-\log\left(x\right)-1$.
	It is easy to check that $\mu$ is positive almost everywhere (a.e.) on
	$\left[0,\tau\right]$ under Condition \ref{enu:ratefun},
	$H\left(\theta_{0};t,Z\left(t\right)\right)$ is positive on
	$\left[0,\tau\right]$ under Condition \ref{enu:H_cond}, and
	$S_{T\wedge C|Z}\left(\cdot\right)=S_{T|Z}\left(\cdot\right)S_{C|Z}\left(\cdot\right)$ is positive on
	$\left[0,\tau\right]$ given Conditions \ref{enu:H_cond} and
	\ref{enu:censor}.
	Consequently,
	$\bM\left(\theta_{0}\right)\ge\bM\left(\theta\right)$
	and
	$\bM\left(\theta_{0}\right)=\bM\left(\theta\right)$
	if and only
	if
	\begin{equation}
		\frac{H\left(\theta;t,Z\left(t\right)\right)}{H\left(\theta_{0};t,Z\left(t\right)\right)}=1\quad \text{a.e. with respect to } \upsilon_{1}, \label{eq:Hequalcon}
	\end{equation}
	because $q\left(x\right)\ge0$ for all $x>0$ with the
	equality holding only at $x=1$.
	Since $H$ is strictly increasing, by
	Condition \ref{enu:identifiability}, \eqref{eq:Hequalcon} implies that
	$\beta=\beta_{0}$ and
	$\alpha_{0}\left(t\right)=\alpha\left(t\right)$
	a.e. with respect to $\upsilon_{1}$.
	It follows that $\theta_{0}$ maximizes
	$\bM\left(\theta\right)$ uniquely.

	\underline{\textbf{Step 2.}} We show the almost sure convergence of $\sup_{\theta\in\Theta_{n}}|\mathbb{M}_{n,h_{n}}(\theta)-\bM(\theta)|$, which is a 
	sufficient condition for the strong consistency of $\hat{\theta}_{n}$.
	We write
	$\mathbb{M}_{n,h}\left(\theta\right)-\bM\left(\theta\right)=(\mathbb{M}_{n,h}\left(\theta\right)-\bM_{h}\left(\theta\right))+(\bM_{h}\left(\theta\right)-\bM\left(\theta\right))$.
	The former term is a stochastic empirical process and the latter is deterministic,
	usually referred to as the ``bias'' term 
	\citep{Einmahl2000,Einmahl2005,Tsybakov2009}.
	We shall next show the uniform convergence of the two terms.

	For the stochastic term, $m_{\theta,h_{n},k}$ belongs to $W^{k}_{nh_n}$, for $k=1,2$, respectively, by setting $\varphi_{1}\left(\cdot\right)=\log\left(H\left(\cdot\right)\right)$
	and
	$\varphi_{2}\left(\cdot\right)=H\left(\cdot\right)$.
	Corollary \ref{coro:wconver} then implies that, if
	$n^{1-\nu}h_{n}/\log(n)\to\infty$ and
	$h_{n}\to0$, $$
	\sup_{\theta\in\Theta_{n}}\left|\left(\mathbb{P}_{n}-\mathbf{P}\right)m_{\theta,h_{n}}\right|\le\sum_{k=1}^{2}\sup_{\theta\in\Theta_{n}}\left|\left(\mathbb{P}_{n}-\mathbf{P}\right)m_{\theta,h_{n},k}\right|\to0$$
	a.s..
	In other words, $\sup_{\theta\in\Theta_{n}}|\mathbb{M}_{n,h_{n}}\left(\theta\right)-\bM_{h_{n}}\left(\theta\right)|\to0$
	a.s. as $n\to\infty$.

	For the the ``bias'' term $\bM_{h_{n}}(\theta)-\bM(\theta)$, we begin with the following lemma regarding the uniform convergence of a general ``bias'' term.
	\begin{lem}
		\label{lem:kernelconverge}
		
		Let $\mathcal{G}$ be a set of bounded function $g(t,r)$, $t,r\in [0,\tau]$, that is uniformly Lipschitz continuous over $r\in[0,\tau]$ and $\mathcal{G}$,
		that is, there exists a positive
		constant $L^{*}$,
		$\sup_{g\in\mathcal{G}}|g(t,r_{1})-g(t,r_{2})|<L^{*}|r_{1}-r_{2}|$,
		for all $t,r_{1},r_{2}\in[0,\tau]$.
		Suppose Condition \ref{enu:kernel} holds. Then,
		\begin{multline*}
			\left|2\int_{0}^{\tau}\int_{0}^{\infty}\frac{1}{h}K\left(\frac{t-r}{h}\right)I\left(r\le t\right)g\left(t,r\right)drdt-\int_{0}^{\tau}g\left(t,t\right)dt\right| \\
			\le 2 h\tau\left(2\sup_{g\in\mathcal{G},t\in[0,\tau]}\left|g\left(t,t\right)\right|+L^{*}\right)\int_{0}^{\infty}K\left(u\right)udu.
		\end{multline*}
	\end{lem}
	\begin{rem}
		Lemma \ref{lem:kernelconverge} requires $h\to0$, leading to the condition $h_{n} \to 0$.
		If the Lipschitz continuity is relaxed to continuity, we can still show the convergence but the convergence rate in $h$ cannot be explicitly determined if the smoothness of $g$ is unspecified.
		The rate in Lemma \ref{lem:kernelconverge} is similar to
		Proposition 1.2 in \citet{Tsybakov2009} for kernel density estimation, which, however, assumed a stronger condition that $g$ is	differential to some degree.
	\end{rem}
	\begin{lem}
		\label{lem:BnLg}
		Suppose that Conditions \ref{enu:kernel}, \ref{enu:alpha_para_space}-\ref{enu:zbounded}, \ref{enu:H_cond} and \ref{enu:ratefun} hold. Then,  $\sup_{t\in[0,\tau],\theta\in\Theta_{n}}\left|\tilde{g}\left(\theta;t,t\right)\right|<\infty$ and $\sup_{\theta\in\Theta_{n}}\left|\tilde{g}\left(\theta;t,t\right)-\tilde{g}\left(\theta;t,r\right)\right|\le L_{1}^{*}\left|t-r\right|$ where $L_{1}^{*}$ is some finite constant independent of $t$, $r$ and $\theta$.
	\end{lem}
	Lemma \ref{lem:BnLg} shows that $\sup_{\theta\in\Theta_{n},t\in\left[0,\tau\right]}\left|\tilde{g}\left(\theta;t,t\right)\right|$ and $L_1^{*}$ are finite and Condition \ref{enu:kernel} entails that $\int_{0}^{\infty}K\left(u\right)udu<\infty$.
	Combining Lemma \ref{lem:kernelconverge} and \ref{lem:BnLg} for $\tilde{g}\left(\theta;t,r\right)$ in $\bM\left(\theta\right)$, we have $\sup_{\theta\in\Theta_{n}}|\bM_{h}(\theta)-\bM(\theta)|\lesssim h$.
	It 
	follows from $h_n\to0$ as $n\to\infty$ that $\sup_{\theta\in\Theta_{n}}|\bM_{h_{n}}(\theta)-\bM(\theta)| \to 0$ 
	as $n \to \infty$.
	
	
	
	
	
	Since we have shown the convergence of both stochastic and deterministic terms, we conclude that, if $n^{1-\nu}h_{n}/\log(n)\to\infty$ and
	$h_{n}\to0$, with probability one, $\sup_{\theta\in\Theta_{n}}|\mathbb{M}_{n,h_{n}}(\theta)-\bM(\theta)|$
	converges to $0$ as $n\to\infty$.
	
	\underline{\textbf{Step 3.}} We establish the strong consistency by showing $\Pr(\limsup_{n\to\infty}\{ d(\hat{\theta}_{n},\theta_{0})>\epsilon\} )=0$ for any $\epsilon>0$.
	We begin with the property of the sieve space $\mathcal{A}_{n,l}$ defined in \eqref{eq:seivespace}.
	By Conditions \ref{enu:sieveoder} and \ref{enu:alpha_para_space}, Lemma 2
	in
	\citet{Zhao2017a} implies that there exists a
	function $\alpha_{0n}\in\mathcal{A}_{n,l}$ defined in the sieve space $\mathcal{A}_{n,l}$ such that
	$\left\Vert
	\alpha_{0n}-\alpha_{0}\right\Vert _{\infty}=O\left(n^{-\kappa\nu}\right)$.
	
	We next find a space that contains $\{d(\hat{\theta}_{n},\theta_{0})>\epsilon\}$.
	Since $\hat{\theta}_{n}$ maximizes $\mathbb{M}_{n,h_{n}}$ over $\Theta_{n}=\mathcal{B}\times \mathcal{A}_{n,l}$, $d(\hat{\theta}_{n},\theta_{0})>\epsilon$ implies that 
	$$\sup_{\{ d(\theta,\theta_{0})>\epsilon,\theta\in\Theta_{n}\} }\mathbb{M}_{n,h_{n}}(\theta)\ge\mathbb{M}_{n,h_{n}}(\hat{\theta}_{n})\ge\mathbb{M}_{n,h_{n}}(\theta_{0n}),$$
	for any $\epsilon>0$, where $\theta_{0n} \coloneqq (\beta_{0},\alpha_{0n})\in \Theta_{n}$.
	In other words, for any $\epsilon>0$,
	\begin{equation}
		\label{eq:depsineq}
		\left\{ d(\hat{\theta}_{n},\theta_{0})>\epsilon\right\}  \subseteq \left\{ \sup_{\{ d(\theta,\theta_{0})>\epsilon,\theta\in\Theta_{n}\} }\mathbb{M}_{n,h_{n}}(\theta)-\bM(\theta_{0n})\ge\mathbb{M}_{n,h_{n}}(\theta_{0n})-\bM(\theta_{0n})\right\}.
	\end{equation}
	Furthermore, since  $\Theta_{n_{1}} \subseteq \Theta_{n_{2}} \subseteq \Theta$ for any $n_{1} \le n_{2}$ with $q_{n}=O(n^{\nu})\to\infty$  as $n\to\infty$ \citep{Schumaker2007}, it follows that   
	\begin{align}
		\sup_{d\left(\theta,\theta_{0}\right)>\epsilon,\theta\in\Theta_{n}}\mathbb{M}_{n,h_{n}}\left(\theta\right) & \le\sup_{d\left(\theta,\theta_{0}\right)>\epsilon,\theta\in\Theta_{n}}\left|\mathbb{M}_{n,h_{n}}\left(\theta\right)-\bM\left(\theta\right)\right|+\sup_{d\left(\theta,\theta_{0}\right)>\epsilon,\theta\in\Theta_{n}}\bM\left(\theta\right)\nonumber\\
		& \le\zeta_{n}+\sup_{d\left(\theta,\theta_{0}\right)>\epsilon,\theta\in\Theta}\bM\left(\theta\right),\label{eq:Mtechineqal1}
	\end{align}
	where $\zeta_{n} \coloneqq \sup_{\theta\in\Theta_{n}}|\mathbb{M}_{n,h_{n}}(\theta)-\bM(\theta)|$.
	Besides, 
	\begin{equation}
		\mathbb{M}_{n,h_{n}}\left(\theta_{0n}\right)-\bM\left(\theta_{0n}\right)\ge-\zeta_{n}, \label{eq:Mtechineqal2}
	\end{equation}
	because $\theta_{0n} \in \Theta_{n}$.
	Thus, combining the inequalities  \eqref{eq:Mtechineqal1} and \eqref{eq:Mtechineqal2}  into the RHS of \eqref{eq:depsineq}s, we have
	$$
	\left\{ d(\hat{\theta}_{n},\theta_{0})>\epsilon\right\}  \subseteq \left\{ \sup_{d\left(\theta,\theta_{0}\right)>\epsilon,\theta\in\Theta}\bM\left(\theta\right)-\bM\left(\theta_{0n}\right)+\zeta_{n}\ge-\zeta_{n}\right\} =\left\{ 2\zeta_{n} +\delta_{2n}\ge\delta_{1\epsilon}\right\},
	$$
	where $\delta_{1\epsilon} \coloneqq \bM\left(\theta_{0}\right)-\sup_{\left\{ d\left(\theta,\theta_{0}\right)>\epsilon,\theta\in\Theta\right\} }\bM\left(\theta\right)$
	and $\delta_{2n} \coloneqq \bM\left(\theta_{0}\right)-\bM\left(\theta_{0n}\right)$.
	It then follows that, for any $\epsilon>0$, 
	$$
	\Pr(\limsup_{n\to\infty}\{ d(\hat{\theta}_{n},\theta_{0})>\epsilon\} )\le\Pr(\limsup_{n\to\infty}\{ 2\zeta_{n}+\delta_{2n} \ge\delta_{1\epsilon}\} ).
	$$

	It only remains to show $\Pr(\limsup_{n\to\infty}\{ 2\zeta_{n}+\delta_{2n} \ge\delta_{1\epsilon}\} )=0$ by proving $2\zeta_{n}+\delta_{2n}\to0$ a.s. as $n\to0$.
	As proved in Step 2, $\zeta_{n}$ converges to 0 a.s..
	Under Conditions \ref{enu:alpha_para_space}, \ref{enu:beta_para_space} and \ref{enu:H_cond}, $\left\Vert
	\alpha_{0n}-\alpha_{0}\right\Vert _{\infty}=O\left(n^{-\kappa\nu}\right)$ implies that  $\lim_{n\to\infty}\delta_{2n}=\lim_{n\to\infty}O\left(n^{-\kappa\nu}\right)=0$.
	Note that $\delta_{2n}$ is deterministic and its convergence happens with probability one.
	Thus,  $2\zeta_{n}+\delta_{2n}\to0$ a.s. as $n\to0$. 
	Since $\theta_{0}$ is the unique maximizer
	of $\bM$ as shown in Step 1, $\delta_{1\epsilon}>0$ for any $\epsilon>0$.
	By the definition of the almost sure convergence, for any $\epsilon>0$, 
	$
	\Pr(\limsup_{n\to\infty}\{2\zeta_{n}+\delta_{2n}\ge\delta_{1\epsilon}\})=0
	$.
	Therefore,  $\Pr(\limsup_{n\to\infty}\{ d(\hat{\theta}_{n},\theta_{0})>\epsilon\} )=0$, for any $\epsilon>0$,
	and Theorem~\ref{them:consistency} follows.
	
	\subsection{The proof of Theorem \ref{them:rate}} 
	
	The proof of Theorem \ref{them:rate} will be divided  into three steps
	to
	verify three conditions of Theorem 3.2.5 of \citet{Vaart1996}.
	We will first show
	$\bM(\theta_{0})-\bM(\theta)\apprge
	d^{2}(\theta,\theta_{0})$.
	We then derive the ``modulus of continuity'' $\phi_{n}(\eta)$ of the
	centered processes $\sqrt{n}(\mathbb{M}_{n,h_{n}}(\theta)-\bM(\theta))$ over $\Theta_{n}$.
	Finally, we determine the convergence rate $r_{n}$ from $\phi_{n}(\eta)$ and $\mathbb{M}_{n,h_{n}}(\hat{\theta}_{n})\ge\mathbb{M}_{n,h_{n}}(\theta)-O_{p}(r_{n}^{-2})$ given that $\hat{\theta}_{n}$ is consistent.
	
	\underline{\textbf{Step 1.}}
	We will show that $\bM(\theta_{0})-\bM(\theta)\apprge d^{2}(\theta,\theta_{0})$ for any $\theta$ in a neighborhood of $\theta_{0}$.
	Note that, in \eqref{eq:M0diff}, $q\left(x\right)=x-\log(x)-1\ge\frac{1}{4}\left(x-1\right)^{2}$ for $0<x<2$.
	Thus, for any $\theta$ in a neighborhood of $\theta_{0}$, some algebra on \eqref{eq:M0diff} yields that
	\begin{equation*}
		\bM(\theta_{0})-\bM(\theta)  
		\ge \frac{1}{4}E\left\{\int_{0}^{\tau}\frac{S_{T\wedge C|Z}(t)\mu(t)}{H(\theta_{0};t,Z(t))}(H(\theta;t,Z(t))-H(\theta_{0};t,Z(t)))^{2}dt\right\}.
	\end{equation*}
	Let $\tilde{f}(\xi)=H\left(\theta_{\xi};t,Z\left(t\right)\right)$ where
	$\theta_{\xi}=\xi\theta+\left(1-\xi\right)\theta_{0}$.
	Then,
	$H\left(\theta;t,Z\left(t\right)\right) - H\left(\theta_{0};t,Z\left(t\right)\right)=\tilde{f}(1)-\tilde{f}(0)$.
	The mean value theorem implies that there exists
	$0\le\xi^{*}\le1$ such that 
	$$\tilde{f}(1)-\tilde{f}(0)=\tilde{f}^{\prime}(\xi^{*})
	= H^{\prime}(\theta_{\xi^{*}};t,Z(t))(\alpha(t)-\alpha_{0}(t)+(\beta-\beta_{0})^{\top}Z(t)).
	$$
	Here,  $\xi^{*}$ may be dependent on $\theta$, $t$ and $Z$.
	Under Conditions \ref{enu:alpha_para_space}-\ref{enu:zbounded} and \ref{enu:H_cond}-\ref{enu:censor}, for any $t\in[0,\tau]$,
	$S_{T\wedge C|Z}(t)\mu(t)$
	is bounded below by
	a positive constant and $H\left(\theta;t,Z\left(t\right)\right)$ is uniformly bounded for $\theta\in\Theta$ a.e. on $\upsilon_{1}$.
	Thus,
	\begin{equation}
		\bM(\theta_{0})-\bM(\theta) \gtrsim E\left\{\int_{0}^{\tau}H^{\prime}(\theta_{\xi^{*}};t,Z(t))^{2}(\alpha(t)-\alpha_{0}(t)+(\beta-\beta_{0})^{\top}Z(t))^{2}dt\right\}.\label{eq:M0diff_lb}
	\end{equation}
	By Condition~\ref{enu:H_cond}, $H(\cdot)$ is strictly increasing and hence $H^{\prime}(\cdot)$ is positive. 
	It then follows from Conditions \ref{enu:alpha_para_space}-\ref{enu:zbounded} that
	$H^{\prime}\left(\theta_{\xi^{*}};t,Z\left(t\right)\right)^{2}$ is bounded below by a positive constant for any $\theta \in \Theta$ and $\xi^{*}\in[0,1]$ a.e. on $\upsilon_{1}$.
	Therefore, \eqref{eq:M0diff_lb} gives that
	\begin{align}
		\bM\left(\theta_{0}\right)-\bM\left(\theta\right) & \ge c_{4}\int_{\mathbb{R}^{+}\times\mathcal{Z}}\left(\alpha\left(u\right)-\alpha_{0}\left(u\right)+\left(\beta-\beta_{0}\right)^{\top}z\left(u\right)\right)^{2}d\upsilon_{1}\left(u,z\right),\label{eq:dinequal1}
	\end{align}
	for some constant $c_{4}>0$.
	For convenience, we let $E_{1}$ denote the
	expectation under $\upsilon_{1}$.
	Then, \eqref{eq:dinequal1} could be rewritten as
	$$\bM\left(\theta_{0}\right)-\bM\left(\theta\right)\gtrsim
	E_{1}\left[\left\{
	f_{1}\left(U\right)+f_{2}\left(Z,U\right)\right\}
	^{2}\right],$$
	where
	$f_{1}\left(U\right) \coloneqq \alpha\left(U\right)-\alpha_{0}\left(U\right)$,
	$f_{2}\left(Z,U\right) \coloneqq \left(\beta-\beta_{0}\right)^{\top}Z\left(U\right)$
	and
	$U$ is the random variable defined in Condition \ref{enu:UZinequality}.
	
	To find the lower bound of the right-hand side (RHS) of \eqref{eq:dinequal1}, we
	utilize Lemma 25.86 in \citet{Vaart1998} that requires to bound
	$[E_{1}\left\{f_{1}\left(U\right)f_{2}\left(Z,U\right)\right\}]^{2}$.
	By the Cauchy-Schwarz inequality and Condition \ref{enu:UZinequality},
	\begin{align*}
		\left[E_{1}\left\{ f_{1}\left(U\right)f_{2}\left(Z,U\right)\right\} \right]^{2} & \le E_{1}\left\{ f_{1}^{2}\left(U\right)\right\} E_{1}\left\{ \left[E_{1}\{f_{2}\left(Z,U\right)|U\}\right]^{2}\right\} \\
		& =E_{1}\left\{ f_{1}^{2}\left(U\right)\right\} E_{1}\left[\left(\beta-\beta_{0}\right)^{\top}E_{1}\{Z\left(U\right)|U\}^{\otimes2}\left(\beta-\beta_{0}\right)\right]\\
		& \le E_{1}\left\{ f_{1}^{2}\left(U\right)\right\} E_{1}\left[\left(\beta-\beta_{0}\right)^{\top}\left(1-\eta^{*}\right)E_{1}\left\{ Z\left(U\right)^{\otimes2}|U\right\} \left(\beta-\beta_{0}\right)\right]\\
		& =\left(1-\eta^{*}\right)E_{1}\left\{ f_{1}^{2}\left(U\right)\right\} E_{1}\left\{ f_{2}^{2}\left(Z,U\right)\right\} .
	\end{align*}
	%
	The second inequality holds because 
	\begin{multline*}
		E_{1}\left(\left(\beta-\beta_{0}\right)^{\top}\left[E_{1}\left\{ Z\left(U\right)^{\otimes2}|U\right\} -E_{1}\left\{ Z\left(U\right)|U\right\} ^{\otimes2}\right]\left(\beta-\beta_{0}\right)\right)\\
		\ge\eta^{*}E_{1}\left\{ \left(\beta-\beta_{0}\right)^{\top}E_{1}\left\{ Z\left(U\right)^{\otimes2}|U\right\} \left(\beta-\beta_{0}\right)\right\} ,
	\end{multline*} implied by Condition \ref{enu:UZinequality} and some algebra.
	Since $1-\eta^{*}<1$ from Condition \ref{enu:UZinequality}, Lemma
	25.86 in
	\citet{Vaart1998} yields that $ E_{1}[\{
	f_{1}(U)+f_{2}(Z,U)\} ^{2}] \ge
	c_{5}E_{1}[f_{1}^{2}(U)+f_{2}^{2}(Z,U)]
	$ for a constant $c_{5}>0$.
	Since $E_{1}[Z(U)Z(U)^{\top}]$ is
	positive definite by Condition
	\ref{enu:identifiability}, some algebra gives that
	$$
	E_{1}[f_{1}^{2}(U)+f_{2}^{2}(Z,U)]
	\apprge\Vert
	\alpha-\alpha_{0}\Vert_{2}^{2}+\Vert
	\beta-\beta_{0}\Vert _{2}^{2}=d^{2}(\theta,\theta_{0}).
	$$
	We thus obtain $\bM\left(\theta_{0}\right)-\bM\left(\theta\right) \apprge
	d^{2}\left(\theta,\theta_{0}\right)$.
	
	\underline{\textbf{Step 2.}}
	We will derive $\phi_{n}\left(\eta\right)$ such that, for small enough $\eta$,
	\begin{equation}
		E\sqrt{n}\sup_{\theta \in \Theta_{n,\eta}}|\{
		\mathbb{M}_{n,h_{n}}(\theta)-\bM(\theta)\} -\{
		\mathbb{M}_{n,h_{n}}(\theta_{0})-\bM(\theta_{0})\}
		|\apprle\phi_{n}(\eta),\label{eq:contmodule}
	\end{equation}
	where $\Theta_{n,\eta}\coloneqq\left\{ \theta\in\Theta_{n},d\left(\theta,\theta_{0}\right)<\eta\right\} $ and $\phi_{n}(\eta)/\eta$ is decreasing with
	respect to $\eta$.
	Similar to the Step 2 of the proof of Theorem~\ref{them:consistency}, we have
	\begin{multline}	    
		E\sqrt{n}\sup_{\theta \in \Theta_{n,\eta}}|(\mathbb{M}_{n,h_{n}}(\theta)-\bM(\theta))-(\mathbb{M}_{n,h_{n}}(\theta_{0})-\bM(\theta_{0}))|\\
		\le  E\left\Vert \mathbb{G}_{n}\right\Vert_{\mathcal{F}_{n,\eta}}+\sqrt{n}\sup_{\theta \in \Theta_{n,\eta}}|\bM_{h_{n}}(\theta)-\bM_{h_{n}}(\theta_{0})-(\bM(\theta)-\bM(\theta_{0}))|,
		\label{eq:empricaldiffsup}
	\end{multline}
	where $\mathcal{F}_{n,\eta}\coloneqq \{
	m_{\theta,h_{n}}\left(O\right)-m_{\theta_{0},h_{n}}\left(O\right):\theta\in\Theta_{n,\eta}\}
	$.
	We now proceed to find the respective upper bounds of the two terms on the RHS of \eqref{eq:empricaldiffsup}, of which the latter represents the uniform kernel-based approximation error to $\bM(\theta)-\bM(\theta_{0})$.
	
	For $E\left\Vert\mathbb{G}_{n}\right\Vert_{\mathcal{F}_{n,\eta}}$, for technical simplicity, we consider
	$E\Vert
	\mathbb{G}_{n}\Vert _{\bar{\mathcal{F}}_{n,\eta}}$ where $\bar{\mathcal{F}}_{n,\eta} \coloneqq \left\{h_{n}f:f\in\mathcal{F}_{n,\eta}\right\} $ so that
	$E\Vert
	\mathbb{G}_{n}\Vert _{\mathcal{F}_{n,\eta}}=E\Vert
	\mathbb{G}_{n}\Vert _{\bar{\mathcal{F}}_{n,\eta}}/h_{n}$.
	\begin{lem}
		\label{lem:hempridiffprop}
		Under Conditions \ref{enu:kernel},
		\ref{enu:alpha_para_space}-\ref{enu:zbounded}, \ref{enu:H_cond}, \ref{enu:ratefun} and
		\ref{enu:finiteobs}, 
		$\sup_{f\in\mathcal{\bar{F}}_{n,\eta}}\left\Vert f\right\Vert _{\infty}\le\bar{K}J_{0}\left(\tau+1\right)$ a.s.,
		$\sup_{f\in\mathcal{\bar{F}}_{n,\eta}}\mathbf{P}f^{2}\lesssim h_{n}\eta^{2}$ and
		$N_{[\,]}\left(\epsilon,\mathcal{\bar{F}}_{n,\eta},\mathcal{L}_{2}\left(\mathbf{P}\right)\right)\lesssim\left(\eta\sqrt{h_{n}}/\epsilon\right)^{cq_{n}+p}$, for some constant $c$ and $0 < \epsilon < \eta\sqrt{h_{n}}$.
	\end{lem}
	By Lemma \ref{lem:hempridiffprop} and Lemma 19.36 in \citet{Vaart1998}, some algebra gives
	$E\Vert \mathbb{G}_{n}\Vert_{\bar{\mathcal{F}}_{n,\eta}}\lesssim\eta\sqrt{q_{n}h_{n}}+q_{n}/\sqrt{n}$
	and hence
	\begin{equation}
		E\Vert \mathbb{G}_{n}\Vert_{\mathcal{F}_{n,\eta}}\lesssim\eta\sqrt{q_{n}/h_{n}}+q_{n}/(\sqrt{n}h_{n}).
		\label{eq:empiricalmaximal}
	\end{equation}


	
	
	For the second term on the RHS of \eqref{eq:empricaldiffsup}, some algebra shows that
	\[
	\bM_{h_{n}}(\theta)-\bM_{h_{n}}(\theta_{0})-(\bM(\theta)-\bM(\theta_{0}))=\int_{0}^{\tau} \left\{\int_{0}^{t}K_{h_{n}}\left(t-r\right)\tilde{w}\left(\theta,t,r\right)dr-\tilde{w}\left(\theta,t,t\right) \right\}dt,
	\]
	where 
	\begin{equation}
		\label{eq:wtildedef}
		\tilde{w}\left(\theta,t,r\right) \coloneqq E\left[S_{T\wedge C|Z}\left(t\right)w\left(\theta,t,Z\left(r\right)\right)\right]\mu\left(r\right)
	\end{equation}
	and 
	\[
	w\left(\theta,t,Z\left(r\right)\right) \coloneqq H\left(\theta_{0};t,Z\left(t\right)\right)\left\{\log\left( \frac{H\left(\theta;t,Z\left(r\right)\right)}{H\left(\theta_{0};t,Z\left(r\right)\right)}\right) -\left(\frac{H\left(\theta;t,Z\left(r\right)\right)}{H\left(\theta_{0};t,Z\left(r\right)\right)}-1\right)\right\}.
	\]
	
	\begin{lem}
		\label{lem:wtiproperty}
		Suppose Conditions \ref{enu:alpha_para_space}, \ref{enu:beta_para_space},
		\ref{enu:zbounded}, \ref{enu:H_cond}, \ref{enu:ratefun} and \ref{enu:censor} hold. Then,  $\sup_{t\in[0,\tau],\,\theta\in\Theta_{n,\eta}}\left|\tilde{w}\left(\theta;t,t\right)\right|\lesssim \eta^{2/3}$ and $\sup_{\theta\in\Theta_{n,\eta}}\left|\tilde{w}\left(\theta;t,t\right)-\tilde{w}\left(\theta;t,r\right)\right|\le L_{2}^{*}\eta^{2/3}|t-r|$ for all $t$ and $r\in[0,\tau]$ where $L_{2}^{*}$ is some finite constant independent of $t$, $r$ and $\theta$.
	\end{lem}
	The term $\eta^{2/3}$ in Lemma \ref{lem:wtiproperty} comes from the following lemma which shares the same spirits with Lemma 7.1 in \citet{Wellner2007}.
	\begin{lem}
		\label{lem:supL2}
		Under Condition \ref{enu:alpha_para_space}, for any
		$\alpha\in\mathcal{\mathcal{G}}_{n,\eta}=\{\alpha\in\mathcal{A}_{n,l}:\Vert\alpha-\alpha_{0}\Vert_{2}<\eta\}$,
		$\sup_{\alpha\in\mathcal{G}_{n,\eta},t\in\left[0,\tau\right]}|\alpha(t)-\alpha_{0}(t)|\le
		O(\eta^{2/3})$.
	\end{lem}
	Applying Lemma \ref{lem:kernelconverge} to the conclusion of Lemma \ref{lem:wtiproperty}, we have 
	\begin{equation}
		\sqrt{n}\sup_{\alpha\in\mathcal{A}_{n,\eta},d(\theta,\theta_{0})<\eta}|\bM_{h_{n}}(\theta)-\bM_{h_{n}}(\theta_{0})-(\bM(\theta)-\bM(\theta_{0}))|\lesssim\sqrt{n}h_{n}\eta^{2/3}.
		\label{eq:kernelbiasorder}
	\end{equation}
	Combing \eqref{eq:empiricalmaximal} and \eqref{eq:kernelbiasorder}, we obtain
	$$
	\phi_{n}\left(\eta\right)=\eta\sqrt{q_{n}/h_{n}}+q_{n}/\left(h_{n}\sqrt{n}\right)+\sqrt{n}h_{n}\eta^{2/3}.
	$$
	It is not hard to show that $\phi_{n}\left(\eta\right)/\eta$ is
	decreasing in $\eta$.
	
	\underline{\textbf{Step 3.}}
	We shall find $r_{n}$ such that
	$r_{n}^{2}\phi_{n}\left(1/r_{n}\right)\le\sqrt{n}$
	for every $n$ and
	$\mathbb{M}_{n,h_{n}}(\hat{\theta}_{n})\ge\mathbb{M}_{n,h_{n}}(\theta)-O_{p}(r_{n}^{-2})$.
	Since $h_{n}=O(n^{-a})$ and $0.4\left(1-\nu\right)\le a$, we see that $n^{2\left(1-\nu\right)}h^{5}\to
	c_{6}$ as
	$n\to\infty$ for some nonnegative constant $c_{6}$.
	Consequently,
	$r_{n}=\min\{
	\sqrt{nh_{n}/q_{n}},n^{\kappa\nu}\} =n^{\min\{ (1-\nu-a)/2,\kappa\nu\}  }$ satisfies that
	$$r_{n}^{2}\phi_{n}\left(1/r_{n}\right) \le \sqrt{n}+\sqrt{n}+\sqrt{n}\left(n^{2\left(1-\nu\right)}h_{n}^{5}\right)^{1/3}\lesssim\sqrt{n}.$$
	
	Since $1-a-\nu >0$ and $a>0$, we have $O(n^{-1+\nu}\log n) \le O(n^{-a}) = h_{n}$ and $h_{n} \to 0$ as $n\to\infty$. 
	Thus, requirement of $h_{n}$ by Theorem~\ref{them:consistency} is satisfied and $\hat{\theta}_{n}$ is strongly consistent under the conditions of Theorem~\ref{them:rate}.
	Then, the next lemma completes the third step.
	\begin{lem}
		\label{lem:CR_order_inequal}
		Suppose that the regularity conditions of Theorem~\ref{them:rate} hold
		and $0.4(1-\nu)\le a < (1-\nu)$,
		$\mathbb{M}_{n,h_{n}}(\hat{\theta}_{n})-\mathbb{M}_{n,h_{n}}\left(\theta_{0}\right)\ge-O_{P}\left(r_{n}^{-2}\right)$.
	\end{lem}
	
	Finally, by Theorem 3.2.5 in \citet{Vaart1996}, the convergence rate of
	$\hat{\theta}_{n}$ is in the form of
	$r_{n}d(\hat{\theta}_{n},\theta_{0})=O_{p}(1)$.
	It is easily to check that $\min\{ (1-\nu-a)/2,\kappa\nu\} \le(1-a)\kappa/(2\kappa+1)$ achieves its maximum when $\nu=(1-a)/(2\kappa+1)$ and $a=0.4(1-\nu)$.
	In other words, when $a=4\kappa/(10\kappa+3)$ and $\nu=3/(3+10\kappa)$, the rate of convergence is optimal as $n^{3\kappa/(10\kappa+3)}$.
	\begin{rem}
		The kernel-based approximation error, or ``bias'' term, in \eqref{eq:empricaldiffsup} does not influence the form of $r_{n}$.
		However, the bias term requires the constrain $0.4\left(1-\nu\right)\le a$ to bound $r_{n}^{2}\phi_{n}\left(1/r_{n}\right)/\sqrt{n}$ and consequently determines the optimal
		convergence rate in Theorem \ref{them:rate}.
	\end{rem}
	
	\subsection{The proof of Theorem \ref{them:normality}} 
	The three steps of the proof are as follows. 
	First, we develop a general asymptotic normality theory as Proposition \ref{prop:genasym} for general kernel-weighted M-estimation regardless the model parametric setting.
	Second, we verify the sufficient conditions of Proposition \ref{prop:genasym} for SMKLE.
	Third, we apply Proposition \ref{prop:genasym} to prove Theorem \ref{them:normality}.
	
	\underline{\textbf{Step 1.}}
	We define a sequence of maps $\mathbb{S}_{n,h}$ mapping a neighborhood
	of
	$\theta_{0}$, denoted by $\mathcal{U}$, in the
	parameter space $\Theta$ into
	$l^{\infty}\left(\mathcal{G}_{1}\times\mathcal{G}_{2}\right)$  as
	\[
	\mathbb{S}_{n,h}\left(\theta\right)\left[g_{1},g_{2}\right]\coloneqq\frac{d}{d\epsilon}\left.\tilde{\ell}_{n}\left(\beta+\epsilon g_{1},\alpha+\epsilon g_{2}\right)\right|_{\epsilon=0}=\mathbb{P}_{n}\left(m_{\theta,h}^{\prime}\left[g_{1},g_{2}\right]\right),
	\]
	where $\tilde{\ell}_{n}$ is defined in \eqref{eq:smoothlogll}, 
	$\mathcal{G}_{1}=\left\{
	g_{1}:g_{1}\in\mathcal{B}\right\} $ and
	$\mathcal{G}_{2}=\left\{ g_{2}:g_{2}\in\mathcal{A}\right.
	$ is a bounded function on
	$\left[0,\tau\right]\}$.
	Here,
	\begin{multline*}
		m_{\theta,h}^{\prime}[g_{1},g_{2}]\coloneqq\int_{0}^{X}K_{h_{}}(X-r)\Delta\cdot H^{(1)}(\theta;X,Z(r))(g_{2}(X)+g_{1}^{\top}Z(r))dN(r)\\
		-\int_{0}^{X}\int_{0}^{t}K_{h}(t-r)H^{\prime}(\alpha(t)+\beta^{\top}Z(r))(g_{2}(t)+g_{1}^{\top}Z(r))dN(r)dt.
	\end{multline*}
	We then define the limit mapping
	$\mathbf{S}_{h}:\mathcal{U}\to
	l^{\infty}(\mathcal{G}_{1}\times\mathcal{G}_{2})$ as $\mathbf{S}_{h}(\theta)[g_{1},g_{2}]\coloneqq\mathbf{P}m_{\theta,h}^{\prime}[g_{1},g_{2}]$ and the limit mapping $\mathbf{S}:\mathcal{U}\to
	l^{\infty}\left(\mathcal{G}_{1}\times\mathcal{G}_{2}\right)$ as $
	\mathbf{S}(\theta)[g_{1},g_{2}]\coloneqq\mathbf{P}m_{\theta}^{\prime}[g_{1},g_{2}]$,
	where
	\begin{multline*}
		m_{\theta}^{\prime}[g_{1},g_{2}]\coloneqq\Delta \cdot H^{(1)}(\theta;X,Z(X))(g_{2}(X)+g_{1}^{\top}Z(X))\mu(X)\\
		-\int_{0}^{X}H^{\prime}(\theta;t,Z(t))(g_{2}(t)+g_{1}^{\top}Z(t))\mu(t)dt.
	\end{multline*}
	By the conditional arguments used to derive $\bM_{h}(\theta)$ and $\bM(\theta)$, we have 
	\begin{multline*}
		\mathbf{S}_{h}(\theta)[g_{1},g_{2}]=E\left[\int_{0}^{\tau}\int_{0}^{t}K_{h}\left(t-r\right)S_{T\wedge C|Z}\left(t\right)H^{\prime}\left(\theta,t,Z(r)\right)\right.\\
		\times\bigg.\left(H\left(\theta_{0};t,Z(r)\right)/H\left(\theta;t,Z(r)\right)-1\right)\left(g_{2}\left(t\right)+g_{1}^{\top}Z\left(r\right)\right)\mu\left(r\right)drdt\bigg],
	\end{multline*}
	and
	\begin{multline*}
		\mathbf{S}(\theta)[g_{1},g_{2}]=E\left[\int_{0}^{\tau}S_{T\wedge C|Z}\left(t\right)H^{\prime}\left(\theta,t,Z(r)\right)\right.\\
		\times\bigg.\left(H\left(\theta_{0};t,Z(t)\right)/H\left(\theta;t,Z(t)\right)-1\right)\left(g_{2}\left(t\right)+g_{1}^{\top}Z\left(t\right)\right)\mu\left(t\right)dt\bigg].
	\end{multline*}
	
	Inspired by \citet{He2016} and \citet{Zhao2017}, to derive
	the asymptotic
	normality of $\hat{\beta}$, we need to check the following
	conditions:
	\begin{enumerate}[label=(A\arabic*)]
		\item $\sqrt{nh_{n}}(\mathbb{S}_{n,h_{n}}-\mathbf{S})(\hat{\theta}_{n})[g_{1},g_{2}]-\sqrt{nh_{n}}(\mathbb{S}_{n,h_{n}}-\mathbf{S})(\theta_{0})[g_{1},g_{2}]=o_{p}(1)$.
		\item
		$\mathbf{S}(\theta_{0})[g_{1},g_{2}]=0$
		and
		$\mathbb{S}_{n,h_{n}}(\hat{\theta}_{n})[g_{1},g_{2}]=o_{p}((nh_{n})^{-1/2})$.
		\item
		$\sqrt{nh_{n}}(\mathbb{S}_{n,h_{n}}-\mathbf{S})(\theta_{0})[g_{1},g_{2}]$
		converges in distribution to a tight zero-mean
		Gaussian process on
		$l^{\infty}(\mathcal{G}_{1}\times\mathcal{G}_{2})$ provided the variance function $\sigma^{2}[g_{1},g_{2}]$ exists.
		\item
		$\mathbf{S}(\theta)[g_{1},g_{2}]$
		is
		Fr\'{e}chet-differentiable at $\theta_{0}$ with a continuous derivative
		$\dot{\mathbf{S}}(\theta_{0})[g_{1},g_{2}]$.
		\item
		$\sqrt{nh_{n}}\{
		\mathbf{S}(\hat{\theta}_{n})[g_{1},g_{2}]-\mathbf{S}(\theta_{0})[g_{1},g_{2}]-\dot{\mathbf{S}}(\theta_{0})(\hat{\theta}_{n}-\theta_{0})[g_{1},g_{2}]\}
		=o_{p}(1)$.
	\end{enumerate}
	The following proposition provides a general asymptotic
	distribution of $\hat{\theta}_{n}$ that is applicable to general kernel-weighted sieve M-estimators.
	\begin{prop}
		\label{prop:genasym}
		Assume that Conditions (A1)-(A5) hold.
		Then, for any $(g_{1},g_{2})\in\mathcal{G}_{1}\times\mathcal{G}_{2}$,
		\[
		-\sqrt{nh_{n}}\dot{\mathbf{S}}(\theta_{0})(\hat{\theta}_{n}-\theta_{0})[g_{1},g_{2}]=\sqrt{nh_{n}}(\mathbb{S}_{n,h_{n}}-\mathbf{S})(\theta_{0})[g_{1},g_{2}]+o_{p}(1)  \to_{d} N(0,\sigma^{2}[g_{1},g_{2}]).
		\]
	\end{prop}
	Proposition \ref{prop:genasym} requires weaker conditions than Theorem 6.1 in \citet{Wellner2007} and gives the asymptotic normality of both Euclidean parameters and infinite-dimensional parameters.
	
	\underline{\textbf{Step 2.}}
	To verify (A1)-(A5), we need the following notation and lemmas.
	Define the following classes
	\begin{multline*}
		\Psi_{n}(\eta)\coloneqq\{
		\psi_{n}(\theta)[g_{1},g_{2}]=\sqrt{h_{n}}(m_{\theta,h_{n}}^{\prime}-m_{\theta_{0},h_{n}}^{\prime})[g_{1},g_{2}]:
		\theta\in\Theta_{n,\eta},(g_{1},g_{2})\in\mathcal{G}_{1}\times\mathcal{G}_{2}\},
	\end{multline*}
	$$
	\Gamma_{n}\coloneqq\{\sqrt{h_{n}}m_{\theta_{0},h_{n}}^{\prime}[g_{1},g_{2}]:(g_{1},g_{2})\in\mathcal{G}_{1}\times\mathcal{G}_{2}\}
	$$ and 
	$$
	\Gamma_{n}(\eta) \coloneqq\{\sqrt{h_{n}}m_{\theta_{0},h_{n}}^{\prime}[g_{1},g_{2}-g_{2n}]:(g_{1},g_{2})\in\mathcal{G}_{1}\times\mathcal{G}_{2},\left\Vert g_{2}-g_{2n}\right\Vert _{\infty}\le\eta\},
	$$ 
	for $\Theta_{n,\eta}\coloneqq\left\{ \theta\in\Theta_{n},d\left(\theta,\theta_{0}\right)<\eta\right\}$.
	\begin{lem}
		\label{lem:PsiSidiffprop}
		Under Conditions \ref{enu:kernel}, \ref{enu:alpha_para_space}-\ref{enu:zbounded},
		\ref{enu:H_cond}, \ref{enu:ratefun} and \ref{enu:finiteobs}, for each $n$,
		\begin{eqnarray*}
			\sup_{f\in\Psi_{n}(\eta)}\Vert f\Vert_{\infty}\lesssim1/\sqrt{h_{n}}, & \sup_{f\in\Psi_{n}(\eta)}\mathbf{P}f^{2}\lesssim\eta^{2}, & \log N_{[\,]}(\epsilon,\Psi_{n}(\eta),\mathcal{L}_{2}(\mathbf{P}))\lesssim\eta/\epsilon;\\
			\sup_{f\in\Gamma_{n}(\eta)}\Vert f\Vert_{\infty}\lesssim1/\sqrt{h_{n}}, & \sup_{f\in\Gamma_{n}(\eta)}\mathbf{P}f^{2}<\eta, & \log N_{[\,]}(\epsilon,\Gamma_{n}(\eta),\mathcal{L}_{2}(\mathbf{P}))\lesssim\eta/\epsilon.
		\end{eqnarray*}
		Furthermore, $\log N_{[\,]}(\epsilon,\Gamma_{n},\mathcal{L}_{2}(\mathbf{P}))\lesssim1/\epsilon$ and there exists an envelope function for $\Gamma_{n}$, $F_{n}$, such that $\mathbf{P}F_{n}^{2}=O(1)$ and $\mathbf{P}F_{n}^{2}I(F_{n}>\delta\sqrt{n})\to 0 $ as $n\to\infty$ for any positive $\delta$.
	\end{lem}

	To verify Condition (A1), similar to \eqref{eq:empricaldiffsup}, we can decompose
	\begin{multline}
		|\sqrt{nh_{n}}(\mathbb{S}_{n,h_{n}}-\mathbf{S})(\hat{\theta}_{n})-\sqrt{nh_{n}}(\mathbb{S}_{n,h_{n}}-\mathbf{S})(\theta_{0})|\\
		= |\mathbb{G}_{n}\psi_{n}(\hat{\theta}_{n})|+|\sqrt{nh_{n}}\{(\mathbf{S}_{h_{n}}-\mathbf{S})(\hat{\theta}_{n})-(\mathbf{S}_{h_{n}}-\mathbf{S})(\theta_{0})\}|,\label{eq:SA1decom}
	\end{multline}
	where $g_1$ and $g_2$ are omitted for simplicity.
	We first show $|\mathbb{G}_{n}\psi_{n}(\hat{\theta}_{n})|=o_p(1)$.
	By Theorem \ref{them:rate}, with probability tending to one,
	$\psi_{n}(\hat{\theta}_{n})\in \Psi_{n}(\eta)$ with $\eta = \eta_{n}=O(r_{n}^{-1})$ for $r_{n}$ defined in the proof of Theorem \ref{them:rate}.
	Plugging in $\eta=O(r_{n}^{-1})$ in Lemma \ref{lem:PsiSidiffprop} and applying Lemma 19.36 in \citet{Vaart1998},
	$$E\Vert \mathbb{G}_{n}\Vert_{\Psi_{n}(\eta_{n})}	\lesssim \eta_{n}+1/(\sqrt{nh_{n}})= O(r_{n}^{-1})+O(n^{-\frac{1-a}{2}})=o(1),$$
	for $0<a<1$.
	Therefore, employing the Markov inequality, $\Vert\mathbb{G}_{n}\Vert_{\Psi_{n}(\eta_{n})}  =o_p(1)$ and hence $|\mathbb{G}_{n}\psi_{n}(\hat{\theta}_{n})|  =o_p(1)$.

	For the second term on the RHS of \eqref{eq:SA1decom}, similar to deriving \eqref{eq:kernelbiasorder}, we obtain that
	\begin{align*}
		\sqrt{nh_{n}}|(\mathbf{S}_{h_{n}}-\mathbf{S})(\hat{\theta}_{n})[g_{1},g_{2}]-(\mathbf{S}_{h_{n}}-\mathbf{S})(\theta_{0})[g_{1},g_{2}]|
		& \lesssim O(\sqrt{nh_{n}^{3}}d(\hat{\theta}_{n},\theta_{0})^{2/3}).
	\end{align*}
	We can show that
	$O(\sqrt{nh_{n}^{3}}d(\hat{\theta}_{n},\theta_{0})^{2/3})\lesssim
	O_{p}(n^{\frac{1}{6}\max\{1+2\nu-7a,3-9a-4\kappa\nu\}})=o_{p}(1)$ because $1+2\nu-7a$ and $3-9a-4\kappa\nu$ are both
	negative given $0.4\left(1-\nu\right)\le a$,
	$\left(1-a\right)/\left(4\kappa\right)<\nu<\left(1-a\right)/2$
	and $\kappa>1$.
	Therefore, (A1) follows.
	
	For (A2), clearly,
	$\mathbf{S}\left(\theta_{0}\right)\left[g_{1},g_{2}\right]=0$
	in (A2) for any $g_{1}$ and $g_{2}$.
	We next show
	$\mathbb{S}_{n,h_{n}}(\hat{\theta}_{n})=o_{p}(\left(nh_{n}\right)^{-1/2})$.
	As shown in the proof of Theorem \ref{them:consistency}, for any $g_2\in\mathcal{G}_2$,  there exists
	$g_{2n}\in\mathcal{A}_{n,l}$ such that
	$\left\Vert
	g_{2n}-g_{2}\right\Vert _{\infty}=O\left(n^{-\kappa\nu}\right)$.
	Clearly, 
	$\mathbb{S}_{n,h_{n}}(\hat{\theta}_{n})\left[g_{1},g_{2n}\right]=0$ given $g_{2n}\in\mathcal{A}_{n,l}$.
	Note that $m_{\theta,h}^{\prime}[g_{1},g_{2}]$ and $m_{\theta}^{\prime}[g_{1},g_{2}]$ is linear in $(g_1,g_2)$ as well as $\mathbf{S}_{h}(\theta_{0})[g_{1},g_{2}]=0$ for any $h$, $g_{1}$ and $g_{2}$.
	Thus, we can decompose
	$$\sqrt{nh_{n}}\mathbb{S}_{n,h_{n}}(\hat{\theta}_{n})[g_{1},g_{2}]=\mathbf{I}_{n1}+\mathbf{I}_{n2}+\mathbf{I}_{n3}$$
	where $$
	\mathbf{I}_{n1}=\sqrt{h_{n}}\mathbb{G}_{n}\{m_{\hat{\theta}_{n},h_{n}}^{\prime}[g_{1},g_{2}-g_{2n}]-m_{\theta_{0},h_{n}}^{\prime}[g_{1},g_{2}-g_{2n}]\},
	$$
	$$
	\mathbf{I}_{n2}=\sqrt{h_{n}}\mathbb{G}_{n}\{m_{\theta_{0},h_{n}}^{\prime}[g_{1},g_{2}-g_{2n}]\}
	$$ and
	$$
	\mathbf{I}_{n3}=\sqrt{nh_{n}}\mathbf{P}\{ (m_{\hat{\theta}_{n},h_{n}}^{\prime}-m_{\theta_{0},h_{n}}^{\prime})[g_{1},g_{2}-g_{2n}]\}.$$
	
	Since $g_{2n}-g_{2} \in \mathcal{G}_2$ and $\left\Vert
	g_{2n}-g_{2}\right\Vert _{\infty}=O\left(n^{-\kappa\nu}\right)$, with probability tending to one, $\mathbf{I}_{n1} \in \Psi_{n}(O(r_{n}^{-1}))$. By the arguments similar to proving (A1), we have
	$\mathbf{I}_{n1}=o_{p}\left(1\right)$.
	Similarly, $\mathbf{I}_{n2}\in \Gamma_{n}(\eta) $ with  $\eta=O(n^{-\kappa\nu})$ and thus $$E\Vert\mathbb{G}_{n}\Vert_{\Gamma_{n}(O(n^{-\kappa\nu}))}\lesssim O(n^{-\kappa\nu})+1/(\sqrt{nh_{n}}) = o(1).$$
	Following the proof of (A1), we have $\mathbf{I}_{n2}=o_{p}\left(1\right)$.
	By Conditions \ref{enu:alpha_para_space}, \ref{enu:beta_para_space}, 
	\ref{enu:zbounded} and \ref{enu:H_cond},
	$$
	\left|\mathbf{I}_{n3}\right|\lesssim\sqrt{nh_{n}}d(\hat{\theta}_{n},\theta_{0})\left\Vert
	g_{2n}-g_{2}\right\Vert _{\infty}=O_{p}\left(\max\left\{
	n^{\left(1/2-\kappa\right)\nu},n^{\left(1-a\right)/2-2\kappa\nu}\right\}
	\right)=o_{p}\left(1\right),
	$$
	since $\nu>\left(1-a\right)/\left(4\kappa\right)$
	and $\kappa>1$.
	
	We now verify (A3) mainly by Lemma 2.11.23 in \citet{Vaart1996}.
	Note that $$\sqrt{nh_{n}}\left(\mathbb{S}_{n,h_{n}}-\mathbf{S}\right)\left(\theta_{0}\right)=\sqrt{nh_{n}}\left(\mathbb{S}_{n,h_{n}}-\mathbf{S}_{h_{n}}\right)\left(\theta_{0}\right)+\sqrt{nh_{n}}\left(\mathbf{S}_{h_{n}}-\mathbf{S}\right)\left(\theta_{0}\right).$$
	We can write
	$\sqrt{nh_{n}}\left(\mathbb{S}_{n,h_{n}}-\mathbf{S}_{h_{n}}\right)\left(\theta_{0}\right)\left[g_{1},g_{2}\right]=\mathbb{G}_{n}f_{n}$
	for $f_{n}\in\Gamma_{n}$.
	From Lemma \ref{lem:PsiSidiffprop}, the envelope function of $\Gamma_{n}$ satisfies the first two conditions of (2.11.21) in \citet{Vaart1996} and 
	$\int_{0}^{\delta_{n}}\sqrt{\log
		N_{[\,]}\left(\epsilon,\Gamma_{n},\mathcal{L}_{2}\left(\mathbf{P}\right)\right)}d\epsilon\lesssim\sqrt{\delta_{n}}\to0$
	for any $\delta_n\to0$.
	Recall that $\sqrt{h_{n}}m_{\theta_{0},h_{n}}^{\prime}[g_1,g_2]$ is linear in $g_1$ and $g_2$.
	Then, it easily follows that $\sqrt{h_{n}}m_{\theta_{0},h_{n}}^{\prime}[g_1,g_2]$ is uniformly continuous in $(g_1,g_2)$ for each $n$, verifying the third condition of (2.11.21) in \citet{Vaart1996}.
	By the calculation in Section~\ref{supp:sigmaderive}, the variance function exists as 
	\begin{multline*}
		\sigma^{2}[g_{1},g_{2}] \coloneqq\lim_{h\to0}hE\left[m_{\theta_{0},h}^{\prime}\left[g_{1},g_{2}\right]^{2}\right]\\
		=           
		4\int_{0}^{\infty}K^{2}\left(u\right)duE\left[\int_{0}^{\tau}S_{C\wedge T|Z}(t)H^{(2)}(\theta_{0};t,Z(t))(g_{2}(t)+g_{1}^{\top}Z(t))^{2}\mu(t)\right]dt.
	\end{multline*}
	Then, by Lemma 2.11.23 in \citet{Vaart1996},
	$\sqrt{nh_{n}}(\mathbb{S}_{n,h_{n}}-\mathbf{S}_{h_{n}})(\theta_{0})[g_{1},g_{2}]$
	converges in distribution to a tight Gaussian process on
	$l^{\infty}\left(\mathcal{G}_{1}\times\mathcal{G}_{2}\right)$.
	Since
	$\mathbf{S}_{h_{n}}\left(\theta_{0}\right)\left[g_{1},g_{2}\right]=\mathbf{S}\left(\theta_{0}\right)\left[g_{1},g_{2}\right]=0$
	for any $h_{n}$, $g_1$ and $g_2$,
	$\sqrt{nh_{n}}\left(\mathbf{S}_{h_{n}}-\mathbf{S}\right)\left(\theta_{0}\right)$
	converges to 0 as $n\to\infty$.
	By Slutsky's theorem, (A3) holds.
	
	It is easy to check that the Fr\'{e}chet differentiability in (A4) holds by the continuity of
	$\mathbf{S}\left(\theta\right)$ in $\theta$.
	Denote the derivative of $\mathbf{S}\left(\theta\right)$ at
	$\theta_{0}$ by
	\[
	\dot{\mathbf{S}}\left(\theta_{0}\right)\left(\theta-\theta_{0}\right)\left[g_{1},g_{2}\right]=\left.\frac{d\mathbf{S}\left(\theta_{0}+\epsilon\left(\theta-\theta_{0}\right)\right)\left[g_{1},g_{2}\right]}{d\epsilon}\right|_{\epsilon=0}
	\]
	that is a map from the space $\left\{
	\theta-\theta_{0}:\theta\in\mathcal{U}\right\} $ to
	$l^{\infty}\left(\mathcal{G}_{1}\times\mathcal{G}_{2}\right)$.
	Recall that 
	\begin{multline*}
		\mathbf{S}(\theta)[g_{1},g_{2}]=E\left[\int_{0}^{\tau}S_{T\wedge C|Z}\left(t\right)H^{\prime}\left(\theta,t,Z(r)\right)\right.\\
		\times\bigg.\left(H\left(\theta_{0};t,Z(t)\right)/H\left(\theta;t,Z(t)\right)-1\right)\left(g_{2}\left(t\right)+g_{1}^{\top}Z\left(t\right)\right)\mu\left(t\right)dt\bigg].
	\end{multline*}
	Then, some algebra shows that 
	\[\dot{\mathbf{S}}\left(\theta_{0}\right)\left(\theta-\theta_{0}\right)\left[g_{1},g_{2}\right]=-\left(\beta-\beta_{0}\right)^{\top}Q_{1}\left(g_{1},g_{2}\right)-\int_{0}^{\tau}\left(\alpha\left(t\right)-\alpha_{0}\left(t\right)\right)dQ_{2}\left(g_{1},g_{2}\right)\left(t\right),
	\]
	where
	\[
	Q_{1}\left(g_{1},g_{2}\right) \coloneqq E\left[\int_{0}^{\infty}Z\left(t\right)S_{T\wedge C|Z}\left(t\right)H^{(2)}\left(\theta_{0};t,Z\left(t\right)\right)\left(g_{2}\left(t\right)+g_{1}^{\top}Z\left(t\right)\right)\mu\left(t\right)dt\right],
	\]
	and $$dQ_{2}(g_{1},g_{2})(t)\coloneqq (s^{(0)}(\theta_{0},t)g_{2}(t)+g_{1}^{\top}s^{(1)}(\theta_{0},t))dt.$$
	Here, \[ 
	s^{\left(k\right)}\left(\theta,t\right)\coloneqq E\left[Z^{\otimes k}\left(t\right)S_{T\wedge C|Z}\left(t\right)H^{(2)}\left(\theta;t,Z(t)\right)\right],
	\]
	as defined in the main paper.
	
	Last, for (A5), under Conditions \ref{enu:alpha_para_space},
	\ref{enu:beta_para_space}, \ref{enu:zbounded}, \ref{enu:H_cond}
	and
	\ref{enu:ratefun}, by the mean value theorem, we can show
	$$|\mathbf{S}(\hat{\theta}_{n})[g_{1},g_{2}]-\mathbf{S}(\theta_{0})[g_{1},g_{2}]-\dot{\mathbf{S}}(\theta_{0})(\hat{\theta}_{n}-\theta_{0})[g_{1},g_{2}]|\lesssim
	d^{2}(\hat{\theta}_{n},\theta_{0})=O_{p}(r_{n}^{-2}).$$
	It follows that 
	$$r_{n}^{-2}(nh_{n})^{1/2}\le\max\{n^{\nu-(1-a)/2},\allowbreak n^{-2\kappa\nu+(1-a)/2}\}\to0$$
	as $n\to\infty$ since $\left(1-a\right)/\left(4\kappa\right)<\nu<\left(1-a\right)/2$.
	Therefore, we have (A5).
	
	\underline{\textbf{Step 3 .}} After verifying (A1)-(A5), Proposition
	\ref{prop:genasym} implies
	$\sqrt{nh_{n}}\left(\mathbb{S}_{n,h_{n}}-\mathbf{S}\right)\left(\theta_{0}\right)\to
	N\left(0,\sigma^{2}\right)$, for any $g_{1}$ and
	$g_{2}$.
	It remains to find a specific relationship between
	$g_{1}$ and $g_{2}$ to
	derive the explicit form of the asymptotic normality of
	$\hat{\beta}$.
	Setting
	$$g^{*}_{2}\left(t\right)=-g_{1}^{\top}s^{\left(1\right)}\left(\theta_{0},t\right)/s^{\left(0\right)}\left(\theta_{0},t\right),$$
	it easily follows that
	$\int_{0}^{\tau}\left(\hat{\alpha}_{n}\left(t\right)-\alpha_{0}\left(t\right)\right)dQ_{2}\left(g_{1},g^{*}_{2}\right)\left(t\right)$
	vanishes.
	Thus, after some algebra,
	$\dot{\mathbf{S}}(\theta_{0})(\theta-\theta_{0})[g_{1},g^{*}_{2}]=-g_{1}^{\top}\Xi(\hat{\beta}_{n}-\beta_{0})$
	where $\Xi=\Sigma$.
	By Condition \ref{enu:identifiability}, $\Sigma$ is positive-definite and
	invertible.
	Besides,
	we have that $\sigma^{2}[g_1,g_2^{*}]= g_{1}^{\top}\Omega g_{1}$.
	Finally, since $g_{1}$ is arbitrary in
	$\mathcal{G}_{1}$, by Cram\'er--Wold
	theorem \citep[][Proposition 18.5]{Kosorok2008}, 
	$\sqrt{nh_{n}}(\hat{\beta}-\beta_{0})\to N\left(0,\Xi^{-1}\Omega
	\Xi^{-1}\right)$
	in distribution.
	With arguments similar to proving the uniform convergence of
	$\mathbb{M}_{n,h_{n}}\left(\theta\right)$, we can show $\Xi$
	and $\Omega$
	can be consistently estimated by $\hat{\Xi}$ and
	$\hat{\Omega}$, respectively.	
		\begin{rem}
			In Step 3, we select a particular $g_2$ as $g_2^{*}$ so that $\alpha$ vanishes in $\dot{\mathbf{S}}(\beta-\beta_{0},\alpha-\alpha_{0})[g_{1},g_{2}^{*}]$.
			This parallels the process of identifying a least favorable direction so that the score function becomes an efficient score function of $\beta$ that is orthogonal to nuisance parameter space regarding $\alpha$ \citep[Chapter 25.5.1]{Vaart1998}.
			However, this does not imply that the proposed estimator is semiparametrically efficient as we explained in the main manuscript.
			Specifically, as shown in the proof of Theorem 3.3, compared with the limit of the log-likelihood with fully observed covariates 
			\[
			E\left[\Delta\cdot\log\{ H\left(\theta;X,Z\left(X\right)\right)\}-\int_{0}^{X}H\left(\theta;t,Z\left(t\right)\right)dt\right],
			\]
			the limit of (4) is 
			\[
			E\left[\Delta\cdot\log\{ H\left(\theta;X,Z\left(X\right)\right)\}\mu(X)-\int_{0}^{X}H\left(\theta;t,Z\left(t\right)\right)\mu(t)dt\right]\, ,
			\]
			which contains additional terms involving $\mu(\cdot)$ due to smoothing of sparse longitudinal covariates.
			Both \citet{Wellner2007} and \citet{Sun2024} recognized that if the limit of the loss function differs from that of the log-likelihood function, the resulting M-estimator may not be asymptotically efficient even if the score of the Euclidean parameter is orthogonal to the nuisance parameter space.
		\end{rem}
	
	\newpage
	
	\section{The detailed proofs of the lemmas to prove Theorem~~\ref{prop:wrate}}
	\label{supp:wratelemmaproof}
	\subsection{The proof of Lemma~\ref{lem:Wbracketing}}

	\begin{proof}
		We first derive the bracketing number of $\bar{\mathcal{W}}_{nh}^{1}$.
		Denote the ceiling of $x$ by $\left\lceil x\right\rceil $, which is the largest integer smaller than $x$.
		By Lemma \ref{lem:splinebracket} and the argument in \citet[p. 2127]{Wellner2007},
		for any $0<\epsilon\le 1$, there exists a set of brackets 
		\[
		\left\{ \left[\alpha_{nj}^{L},\alpha_{nj}^{U}\right]:j=1,\ldots,\left\lceil \left(1/\epsilon\right)^{cq_{n}}\right\rceil \right\},
		\]
		such that, for any $\alpha\in\mathcal{A}_{n,l}$, $\alpha_{nj}^{L}(t)<\alpha(t)<\alpha_{nj}^{U}(t)$
		over $t\in\left[0,\tau\right]$ as well as $\Vert\alpha_{nj}^{L}-\alpha_{nj}^{U}\Vert_{2}<\epsilon$ for some $1<j<\left\lceil \left(1/\epsilon\right)^{cq_{n}}\right\rceil $,
		and some constant $c$. Since $\mathcal{B}$ is compact under Condition
		\ref{enu:beta_para_space}, $\mathcal{B}$ can be covered by $\left\lceil c_{2}\left(1/\epsilon\right)^{p}\right\rceil $ number of $\mathcal{L}_2$ balls
		with radius $\epsilon$. Then, for any $\beta\in\mathcal{B}$, the
		exists $\beta_{k}$ for $1\le k\le\left\lceil c_{2}\left(1/\epsilon\right)^{p}\right\rceil $
		such that $\left\Vert \beta-\beta_{k}\right\Vert _{2}\le\epsilon$.
		Therefore, under Condition \ref{enu:zbounded}, $\left|\beta^{\top}Z\left(t\right)-\beta_{k}^{\top}Z\left(t\right)\right|\le c_{3}\epsilon$
		for some positive constant $c_{3}$ with probability one.
		
		Then, by the montonicity of $\varphi_{1}$ in $\mathcal{W}_{nh}^{1} $, we can construct a set of brackets for $\bar{\mathcal{W}}_{nh}^{1}$
		as $\left[\omega_{1,j,k}^{L},\omega_{1,j,k}^{U}\right]$ with $1<j<\left\lceil \left(1/\epsilon\right)^{cq_{n}}\right\rceil $
		and $1\le k\le\left\lceil c_{2}\left(1/\epsilon\right)^{p}\right\rceil $
		where
		\[
		\omega_{1,j,k}^{L} \coloneqq 2\int_{0}^{\infty}I\left(X>r\right)K\left(\frac{X-r}{h}\right)\Delta\cdot\varphi_{1}\left(\alpha_{nj}^{L}\left(X\right)+\beta_{k}^{\top}Z\left(r\right)-c_{3}\epsilon\right)dN\left(r\right),
		\]
		and 
		\[
		\omega_{1,j,k}^{U} \coloneqq 2\int_{0}^{\infty}I\left(X>r\right)K\left(\frac{X-r}{h}\right)\Delta\cdot\varphi_{1}\left(\alpha_{nj}^{U}\left(X\right)+\beta_{k}^{\top}Z\left(r\right)+c_{3}\epsilon\right)dN\left(r\right).
		\]
		Next we need to show $\left\Vert \omega_{1,j,k}^{D}\right\Vert _{\mathcal{L}_{2}\left(\mathbf{P}\right)}\lesssim\epsilon$
		where $\omega_{1,j,k}^{D}\coloneqq\omega_{1,j,k}^{U}-\omega_{1,j,k}^{L}$. 
		
		With probability one, write out $\left|\omega_{1,j,k}^{D}\right|$ as %
		\begin{align*}
			\left|\omega_{1,j,k}^{D}\right|= & \left|2\int_{0}^{\infty}I\left(X>r\right)K\left(\frac{X-r}{h}\right)\right.\Delta\\
			& \times\left.\left\{ \varphi_{1}\left(\alpha_{nj}^{U}\left(X\right)+\beta_{k}^{\top}Z\left(r\right)-c_{3}\epsilon\right)-\varphi_{1}\left(\alpha_{nj}^{L}\left(X\right)+\beta_{k}^{\top}Z\left(r\right)+c_{3}\epsilon\right)\right\} dN\left(r\right)\right|\\
			\le & 2\int_{0}^{\infty}I\left(X>r\right)K\left(\frac{X-r}{h}\right)\Delta\\
			& \times\left|\varphi_{1}\left(\alpha_{nj}^{U}\left(X\right)+\beta_{k}^{\top}Z\left(r\right)-c_{3}\epsilon\right)-\varphi_{1}\left(\alpha_{nj}^{L}\left(X\right)+\beta_{k}^{\top}Z\left(r\right)+c_{3}\epsilon\right)\right|dN\left(r\right).
		\end{align*}
		With probability one, by the continuity of $\varphi_{1}\left(\cdot\right)$ and the mean value
		theorem, 
		\begin{multline*}
			\left|\omega_{1,j,k}^{D}\right|\le2\int_{0}^{\infty}I\left(X>r\right)K\left(\frac{X-r}{h}\right)\Delta\cdot\left|\varphi_{1}^{\prime}\left(\alpha_{nj}^{*}\left(X\right)+\beta_{k}^{\top}Z\left(r\right)+c_{3}^{*}\epsilon\right)\right|\\
			\times\left|\alpha_{nj}^{U}\left(X\right)-\alpha_{nj}^{L}\left(X\right)+2c_{3}\epsilon\right|dN\left(r\right),
		\end{multline*}
		where $\alpha_{nj}^{*}\left(\cdot\right)$ is on a line segment between $\alpha_{nj}^{L}\left(\cdot\right)$
		and $a_{nj}^{U}\left(\cdot\right)$, and $c_{3}^{*}\in\left(-c_{3},c_{3}\right)$.
		Since $\varphi_{1}^{\prime}$ is also continuous, by Conditions \ref{enu:alpha_para_space},
		\ref{enu:beta_para_space} and \ref{enu:zbounded}, we can conclude that
		$$\varphi_{1}^{\prime}\left(\alpha_{nj}^{*}\left(t\right)+\beta_{k}^{\prime}Z\left(r\right)+c_{3}^{*}\epsilon\right)$$
		is also bounded for $t,r\in\left[0,\tau\right]$ with probability one. Hence, according
		to Lemma \ref{lem:integralvar},
		\begin{align*}
			\mathbf{P}\left|\omega_{1,j,k}^{D}\right|^{2}\le & E\left[4\int_{0}^{\infty}I\left(X>r\right)K^{2}\left(\frac{X-r}{h}\right)\left|\varphi_{1}^{\prime}\left(\alpha_{nj}^{*}\left(X\right)+\beta_{k}^{\top}Z\left(r\right)+c_{3}^{*}\epsilon\right)\right|^{2}\right.\\
			& \times\left.\left(\left|\alpha_{nj}^{U}\left(X\right)-\alpha_{nj}^{L}\left(X\right)\right|+2c_{3}\epsilon\right)^{2}\mu\left(r\right)dr\right]\\
			& +E\left[\left(2\int_{0}^{\infty}I\left(X>r\right)K\left(\frac{X-r}{h}\right)\left|\varphi_{1}^{\prime}\left(\alpha_{nj}^{*}\left(X\right)+\beta_{k}^{\top}Z\left(r\right)+c_{3}^{*}\epsilon\right)\right|\right.\right.\\
			& \times\left.\left.\left|\alpha_{nj}^{U}\left(X\right)-\alpha_{nj}^{L}\left(X\right)+2c_{3}\epsilon\right|\mu\left(r\right)dr\right)^{2}\right]\\
			\lesssim & E\left[4\epsilon^{2}\int_{0}^{\infty}I\left(X>r\right)K^{2}\left(\frac{X-r}{h}\right)\mu\left(r\right)dr\right]\\
			& +E\left[4\epsilon^{2}\left(\int_{0}^{\infty}I\left(X>r\right)K\left(\frac{X-r}{h}\right)\mu\left(r\right)dr\right)^{2}\right]\\
			\lesssim & \epsilon^{2}\left\{ 2h\sup_{r\in\left[0,\tau\right]}\left(\mu\left(r\right)\right)+h^{2}\left(\sup_{r\in\left[0,\tau\right]}\left(\mu\left(r\right)\right)\right)^{2}\right\} \\
			\lesssim & h \epsilon^{2}.
		\end{align*}
		Therefore, $\left\Vert \omega_{1,j,k}^{D}\right\Vert _{\mathcal{L}_{2}\left(\mathbf{P}\right)}\lesssim\epsilon$.
		It then follows that
		\[
		N_{[\,]}\left(\sqrt{h}\epsilon,\mathcal{\bar{W}}_{nh}^{1},\mathcal{L}_{2}\left(\mathbf{P}\right)\right)\lesssim\left(1/\epsilon\right)^{cq_{n}}\left(1/\epsilon\right)^{p}\lesssim\left(1/\epsilon\right)^{cq_{n}+p},
		\]
		for some constant $c$.
		Setting $\epsilon^{*} = \sqrt{h} \epsilon$, we have $N_{[\,]}\left(\epsilon^{*},\mathcal{\bar{W}}_{nh}^{1},\mathcal{L}_{2}\left(\mathbf{P}\right)\right)\lesssim\left(\sqrt{h} /\epsilon^{*}\right)^{cq_{n}+p}$ 
		and the bracketing number of $\mathcal{W}_{nh}^{1}$ follows.
		
		We can now proceed analogously to the proof of $N_{[\,]}\left(\epsilon,\mathcal{\bar{W}}_{nh}^{2},\mathcal{L}_{2}\left(\mathbf{P}\right)\right)$.
		It follows similarly that, 
		\[
		\omega_{2,j,k}^{L} \coloneqq \int_{0}^{\infty}\int_{0}^{\tau}I\left(t\le X\right)I\left(r\le t\right)2K\left(\frac{t-r}{h}\right)\varphi_{2}\left(\alpha_{nj}^{L}\left(X\right)+\beta_{k}^{\top}Z\left(r\right)-c_{3}\epsilon\right)dtdN\left(r\right),
		\]
		\[
		\omega_{2,j,k}^{U} \coloneqq \int_{0}^{\infty}\int_{0}^{\tau}I\left(t\le X\right)I\left(r\le t\right)2K\left(\frac{t-r}{h}\right)\varphi_{2}\left(\alpha_{nj}^{U}\left(X\right)+\beta_{k}^{\top}Z\left(r\right)+c_{3}\epsilon\right)dtdN\left(r\right),
		\]
		and
		\begin{align*}
			& \mathbf{P}\left|\omega_{2,j,k}^{U}-\omega_{2,j,k}^{L}\right|^{2}\\
			\lesssim & E\int_{0}^{\infty}\left(\int_{0}^{\tau}I\left(t\le X\right)I\left(r\le t\right)2K\left(\frac{t-r}{h}\right)\left|\alpha_{nj}^{U}\left(t\right)-\alpha_{nj}^{L}\left(t\right)+2c_{3}\epsilon\right|dt\right)^{2}\mu\left(r\right)dr\\
			& +E\left(\int_{0}^{\infty}\int_{0}^{\tau}I\left(t\le X\right)I\left(r\le t\right)2K\left(\frac{t-r}{h}\right)\left|\alpha_{nj}^{U}\left(t\right)-\alpha_{nj}^{L}\left(t\right)+2c_{3}\epsilon\right|\mu\left(r\right)dtdr\right)^{2}\\
			\lesssim & \epsilon^{2}\left\{ 2h\sup_{r\in\left[0,\tau\right]}\mu\left(r\right)+h^{2}\left(\sup_{r\in\left[0,\tau\right]}\mu\left(r\right)\right)^{2}\right\} \\
			\lesssim & h \epsilon^{2}.
		\end{align*}
		Therefore, the bracketing number of $\bar{\mathcal{W}}_{nh}^{2}$ follows.
	\end{proof}
	
	\subsection{The proof of Lemma~\ref{lem:Wbarhbracketing}}

	\begin{proof}
		Since $h_{j-1,k}\le h\le h_{j,k}$ for $\bar{\mathcal{W}}_{j,k}^{1}$ and  $h_{j,k}-h_{j-1,k}=h_{j-1,k}$, 
		we can construct a set of brackets
		\[
		\left\{ \left(h_{l_{1}}^{L},h_{l_{1}}^{U}\right):l_{1}=1,\ldots,\left\lceil 1/\epsilon\right\rceil \right\}  ,
		\]
		such that, for any $h\in\left[h_{j-1,k},h_{j,k}\right]$, there exists
		$1\le l_{1}\le\left\lceil 1/\epsilon\right\rceil  $
		such that $h_{l_{1}}^{L}\le h\le h_{l_{1}}^{U}$ and $\left|h_{l_{1}}^{U}-h_{l_{1}}^{L}\right|\le h_{j-1,k}\epsilon$.
		By the monotonicity property of $K$ under Condition \ref{enu:kernel} on either $\left[-1,0\right]$
		or $\left[0,1\right]$, $K\left(t/h_{l_{1}}^{L}\right)\le K\left(t/h\right)\le K\left(t/h_{l_{1}}^{U}\right)$
		for all $t\in\mathbb{R}$.
		Then, we can have that
		\[
		\left\{ \left[K\left(t/h_{l_{1}}^{L}\right),K\left(t/h_{l_{1}}^{U}\right)\right]:l_{1}=1,\ldots,\left\lceil 1/\epsilon\right\rceil \right\}
		\]
		is a set of brackets for $\mathcal{K}=\left\{ K\left(\cdot/h\right):h_{j-1,k}\le h \le h_{j,k}\right\}$. 
		Furthermore,
		since the support of $K$  is $[-1,1]$ and $K$ is Lipschitz continuous under Condition \ref{enu:kernel},
		\begin{align*}
			\int_{0}^{\infty}\left\{ K\left(\frac{t}{h_{l_{1}}^{L}}\right)-K\left(\frac{t}{h_{l_{1}}^{U}}\right)\right\} ^{2}dt & \lesssim\int_{0}^{h_{l_{1}}^{U}}t^{2}\left(\frac{1}{h_{l_{1}}^{L}}-\frac{1}{h_{l_{1}}^{U}}\right)^{2}dt\\
			& \lesssim\frac{h_{l_{1}}^{U}}{\left(h_{l_{1}}^{L}\right)^{2}}\left(h_{l_{1}}^{U}-h_{l_{1}}^{L}\right)^{2}\\
			& =\frac{h_{l_{1}}^{U}}{\left(h_{l_{1}}^{L}\right)^{2}}h_{j-1,k}^{2}\epsilon^{2}\\
			& \le\frac{h_{j,k}}{h_{j-1,k}^{2}}h_{j-1,k}^{2}\epsilon^{2}\\
			& =h_{j,k}\epsilon^{2}.
		\end{align*}
		In other words, 
		\begin{equation}
			\int_{0}^{\infty}\left\{ K\left(\frac{t}{h_{l_{1}}^{L}}\right)-K\left(\frac{t}{h_{l_{1}}^{U}}\right)\right\} ^{2}dt\lesssim h_{j,k} \epsilon^{2}, \label{eq:Keps}
		\end{equation}
		for all $1\le l_{1}\le\left\lceil 1/\epsilon\right\rceil $.

		
		Following the proof of Lemma \ref{lem:Wbracketing}, we can similarly
		construct a set of brackets for $\bar{\mathcal{W}}_{j,k}^{1}$
		as $\left[\omega_{1,j_{1},k_{1}}^{L}\left(h_{l_{1}}^{L}\right),\omega_{1,j_{1},k_{1}}^{U}\left(h_{l_{1}}^{U}\right)\right]$
		with $1<j_{1}<\left\lceil \left(1/\epsilon\right)^{c\left(m_{n}+l\right)}\right\rceil $,
		$1\le k_{1}\le\left\lceil c_{2}\left(1/\epsilon\right)^{p}\right\rceil $
		and $1\le l_{1}\le\left\lceil 1/\epsilon\right\rceil$
		where 
		\[
		\omega_{1,j_{1},k_{1}}^{L}\left(h_{l_{1}}^{L}\right) \coloneqq 2\int_{0}^{\infty}I\left(X>r\right)K\left(\frac{X-r}{h_{l_{1}}^{L}}\right)\Delta\cdot\varphi_{1}\left(\alpha_{nj_{1}}^{L}\left(X\right)+\beta_{k_{1}}^{\top}Z\left(r\right)-c_{3}\epsilon\right)dN\left(r\right)
		\]
		and 
		\[
		\omega_{1,j_{1},k_{1}}^{U}\left(h_{l_{1}}^{U}\right)\coloneqq 2\int_{0}^{\infty}I\left(X>r\right)K\left(\frac{X-r}{h_{l_{1}}^{U}}\right)\Delta\cdot\varphi_{1}\left(\alpha_{nj_{1}}^{U}\left(X\right)+\beta_{k_{1}}^{\top}Z\left(r\right)+c_{3}\epsilon\right)dN\left(r\right).
		\]
		Then,
		\begin{multline}
			\mathbf{P}\left|\omega_{1,j_{1},k_{1}}^{U}\left(h_{l_{1}}^{U}\right)-\omega_{1,j_{1},k_{1}}^{L}\left(h_{l_{1}}^{L}\right)\right|^{2}\le\mathbf{P}\left|\omega_{1,j_{1},k_{1}}^{U}\left(h_{l_{1}}^{U}\right)-\omega_{1,j_{1},k_{1}}^{U}\left(h_{l_{1}}^{L}\right)\right|^{2}\\
			+\mathbf{P}\left|\omega_{1,j_{1},k_{1}}^{U}\left(h_{l_{1}}^{L}\right)-\omega_{1,j_{1},k_{1}}^{L}\left(h_{l_{1}}^{L}\right)\right|^{2}.\label{eq:wULHdiff}
		\end{multline}
		For the first term on the RHS of \eqref{eq:wULHdiff}, by Conditions \ref{enu:alpha_para_space},
		\ref{enu:beta_para_space} and \ref{enu:zbounded}, Lemma \ref{lem:integralvar}
		yields that
		\begin{alignat*}{1}
			& \mathbf{P}\left|\omega_{1,j_{1},k_{1}}^{U}\left(h_{l_{1}}^{U}\right)-\omega_{1,j_{1},k_{1}}^{U}\left(h_{l_{1}}^{L}\right)\right|^{2}\\
			\lesssim & E\left[\int_{0}^{\infty}I\left(X>r\right)\left\{ K\left(\frac{X-r}{h_{l_{1}}^{U}}\right)-K\left(\frac{X-r}{h_{l_{1}}^{L}}\right)\right\} ^{2}dr\right]\\
			& +E\left[\left\{ \int_{0}^{\infty}I\left(X>r\right)\left|K\left(\frac{X-r}{h_{l_{1}}^{U}}\right)-K\left(\frac{X-r}{h_{l_{1}}^{L}}\right)\right|dr\right\} ^{2}\right].
		\end{alignat*}
		By the change of variable, $u=X-r$, and similar arguments of showing \eqref{eq:Keps}, 
		\begin{align*}
			& E\left[\left\{ \int_{0}^{\infty}I\left(X>r\right)\left|K\left(\frac{X-r}{h_{l_{1}}^{U}}\right)-K\left(\frac{X-r}{h_{l_{1}}^{L}}\right)\right|dr\right\} ^{2}\right]\\
			\le & E\left[\left\{ \int_{0}^{\infty}\left|K\left(\frac{u}{h_{l_{1}}^{U}}\right)-K\left(\frac{u}{h_{l_{1}}^{L}}\right)\right|du\right\} ^{2}\right]\\
			\lesssim & E\left[\left\{ \int_{0}^{h_{l_{1}}^{U}}u\left|\frac{1}{h_{l_{1}}^{U}}-\frac{1}{h_{l_{1}}^{L}}\right|du\right\} ^{2}\right]\\
			= & E\left[\left\{ \frac{\left(h_{l_{1}}^{U}\right)^{2}}{2}\left|\frac{1}{h_{l_{1}}^{U}}-\frac{1}{h_{l_{1}}^{L}}\right|\right\} ^{2}\right]\\
			= & E\left[\left\{ \frac{h_{l_{1}}^{U}}{h_{l_{1}}^{L}}\left|h_{l_{1}}^{U}-h_{l_{1}}^{L}\right|\right\} ^{2}\right]\\
			\le & E\left[\left\{ \frac{h_{j,k}}{h_{j-1,k}}\epsilon h_{j-1,k}\right\} ^{2}\right]\\
			= & h^{2}_{j,k}\epsilon^{2}.
		\end{align*}
		Thus, by \eqref{eq:Keps}, we find the upper bound of the first term on RHS of \eqref{eq:wULHdiff} as
		\begin{align*}
			\mathbf{P}\left|\omega_{1,j_{1},k_{1}}^{U}\left(h_{l_{1}}^{U}\right)-\omega_{1,j_{1},k_{1}}^{U}\left(h_{l_{1}}^{L}\right)\right|^{2} 
			& \lesssim h_{j,k}^{2}\epsilon^{2}+h_{j,k}\epsilon^{2}\\
			& \lesssim h_{j,k}\epsilon^{2}.
		\end{align*} 	
		From the proof of Lemma \ref{lem:Wbracketing}, the second term on the right hand side of \eqref{eq:wULHdiff},
		$$
		\mathbf{P}\left|\omega_{1,j_{1},k_{1}}^{U}\left(h_{l_{1}}^{L}\right)-\omega_{1,j_{1},k_{1}}^{L}\left(h_{l_{1}}^{L}\right)\right|^{2}\lesssim h_{l_{1}}^{L}\epsilon^{2}\lesssim h_{j,k}\epsilon^{2},
		$$
		for $h_{l_{1}}^{L}\le h_{j,k}$.
		Thus, \eqref{eq:wULHdiff} yields that $\mathbf{P}\left|\omega_{1,j_{1},k_{1}}^{U}\left(h_{l_{1}}^{U}\right)-\omega_{1,j_{1},k_{1}}^{L}\left(h_{l_{1}}^{L}\right)\right|^{2}\lesssim  h_{j,k}\epsilon^{2}$.
		Finally, we obtain
		\[
		N_{[\,]}(\sqrt{h_{j,k}}\epsilon,\bar{\mathcal{W}}_{j,k}^{1},\mathcal{L}_{2}(\mathbf{P}))\lesssim\left(1/\epsilon\right)^{cq_{n}+p+1},
		\]
		and hence 
		\[
		N_{[\,]}(\epsilon,\bar{\mathcal{W}}_{j,k}^{1},\mathcal{L}_{2}(\mathbf{P}))\lesssim\left(\sqrt{h_{j,k}}/\epsilon\right)^{cq_{n}+p+1},
		\]
		
		The similar arguments to derive $N_{[\,]}(\epsilon,\bar{\mathcal{W}}_{j,k}^{1},\mathcal{L}_{2}(\mathbf{P}))$
		can apply to $N_{[\,]}(\epsilon,\bar{\mathcal{W}}_{j,k}^{2},\mathcal{L}_{2}(\mathbf{P}))$
		and the proof is omitted.
	\end{proof}
	
	\subsection{The proof of Lemma \ref{lem:WbarhUppers}}
	
	\begin{proof}
		We first derive the uniform upper bound of $\mathbf{P}\left(\omega_{1}\right)^{2}$
		for $\omega_{1}\in\bar{\mathcal{W}}_{nh}^{1}$.
		By Conditions \ref{enu:alpha_para_space}-\ref{enu:zbounded} and continuity
		of $\varphi_{1}$, $\varphi_{1}\left(\alpha_{n}\left(t\right)+\beta^{\top}Z\left(r\right)\right)$
		is bounded for $t,r\in\left[0,\tau\right]$. 
		Using the conditional expectation
		argument based on \ref{enu:ratefun} and Lemma \ref{lem:integralvar},
		we have $\mathbf{P}\left(\omega_{1}\right)^{2}=\mathbf{I}+\mathbf{II}$
		where
		\[
		\mathbf{I} \coloneqq 4E\left[\int_{0}^{\infty}I\left(r\le X\right)K^{2}\left(\frac{X-r}{h}\right)\Delta\cdot\left\{ \varphi_{1}\left(\alpha_{n}\left(X\right)+\beta^{\top}Z\left(r\right)\right)\right\} ^{2}\mu\left(r\right)dr\right],
		\]
		and 
		\[
		\mathbf{II} \coloneqq 4E\left[\left\{ \int_{0}^{\infty}I\left(r\le X\right)K\left(\frac{X-r}{h}\right)\Delta\cdot\varphi_{1}\left(\alpha_{n}\left(X\right)+\beta^{\top}Z\left(r\right)\right)\mu\left(r\right)dr\right\} ^{2}\right].
		\]
		
		Letting $r=t-hu$ and by change of variables in the integral, under
		Conditions \ref{enu:kernel}, \ref{enu:alpha_para_space}, \ref{enu:beta_para_space},
		\ref{enu:zbounded}, \ref{enu:H_cond} and \ref{enu:ratefun},
		\begin{align*}
			\mathbf{I} & =4hE\left[\int_{0}^{\infty}I\left(u\le\frac{X}{h}\right)K^{2}\left(u\right)\Delta\cdot\left\{ \varphi_{1}\left(\alpha_{n}\left(X\right)+\beta^{\top}Z\left(X-hu\right)\right)\right\} ^{2}\mu\left(X-hu\right)du\right]\\
			& \le 4h\left(\sup_{t,r\in\left[0,\tau\right],Z\in\mathcal{Z}}\left|\varphi_{1}\left(\alpha_{n}\left(t\right)+\beta^{\top}Z\left(r\right)\right)\right|^{2}\mu\left(r\right)\right)E\left[\int_{0}^{\infty}I\left(u<\frac{X}{h}\right)\Delta \cdot K^{2}\left(u\right)du\right]\\
			& \lesssim 4h\bar{K}E\left[\int_{0}^{\tau}K\left(u\right)du\right]\lesssim h.
		\end{align*}
		The similar argument applies to $\mathbf{II}$ as
		\begin{align*}
			\mathbf{II}= & 4h^{2}E\left[\left\{ \int_{0}^{\infty}I\left(u\le\frac{X}{h}\right)K\left(u\right)\Delta\cdot\varphi_{1}\left(\alpha_{n}\left(X\right)+\beta^{\top}Z\left(X-hu\right)\right)\mu\left(X-hu\right)du\right\} ^{2}\right]\\
			\le & 4h^{2}\left(\sup_{t,r\in\left[0,\tau\right],Z\in\mathcal{Z}}|\varphi_{1}(\alpha_{n}(t)+\beta^{\top}Z(r))|\mu(r)\right)^{2}E\left[\left\{ \int_{0}^{\infty}I\left(u\le\frac{X}{h}\right)K\left(u\right)\Delta \cdot du\right\} ^{2}\right]\\
			\lesssim & h^{2}.
		\end{align*}
		Hence, since $0<h\le1$, there exists some constant $D_{12}$ such that $\mathbf{P}\left(\omega_{1}\right)^{2}\lesssim h+h^{2}\le hD_{12}$.
		Clearly, this upper bound is uniform in $\bar{\mathcal{W}}_{nh}^{1}$ since it is independent of $\theta$ and the observed data $O$. Thus, $\sup_{f\in\bar{\mathcal{W}}_{nh}^{1}}\mathbf{P}f^{2}\le hD_{21}$.
		
		In the same manner, we can calculate the uniform upper bound of $\left|\omega_{1}\right|$.
		Under Conditions \ref{enu:alpha_para_space}, \ref{enu:beta_para_space},
		\ref{enu:zbounded} and \ref{enu:finiteobs}, with probability one,
		\begin{align*}
			\left|\omega_{1}\right|\le & 2\int_{0}^{\infty}I\left(r\le X\right)K\left(\frac{X-r}{h}\right)\Delta\cdot\left|\varphi_{1}\left(\alpha_{n}\left(X\right)+\beta^{\top}Z\left(r\right)\right)\right|dN\left(r\right)\\
			= & 2\bar{K}\left(\sup_{t,r\in\left[0,\tau\right],Z\in\mathcal{Z}}\left|\varphi_{1}\left(\alpha_{n}\left(t\right)+\beta^{\top}Z\left(r\right)\right)\right|\right)J_{0},
		\end{align*}
		which is smaller than some constant independent of $O$, $\alpha_{n}$ and $\beta$.
		It then follows that there exists some positive constant $D_{11}$, such that $\sup_{f\in\bar{\mathcal{W}}_{nh}^{1}}\left\Vert f\right\Vert _{\infty}\le D_{11}$ almost surely.
		
		For
		$\omega_{2}\in\bar{\mathcal{W}}_{nh}^{2}$, we similarly can show the uniform upper bound as
		\begin{align*}
			\mathbf{P}\left(\omega_{2}\right)^{2}= & \mathbf{P}\int_{0}^{\infty}\left\{ \int_{0}^{\tau}I\left(t\le X\right)I\left(r\le t\right)2K\left(\frac{t-r}{h}\right)\varphi_{2}\left(\alpha_{n}\left(t\right)+\beta^{\top}Z\left(r\right)\right)dt\right\} ^{2}\mu\left(r\right)dr\\
			& +\mathbf{P}\left\{ \int_{0}^{\infty}\int_{0}^{\tau}I\left(r\le t\le X\right)2K\left(\frac{t-r}{h}\right)\varphi_{2}\left(\alpha_{n}\left(t\right)+\beta^{\top}Z\left(r\right)\right)dt\mu\left(r\right)dr\right\} ^{2}\\
			\lesssim & 2\bar{K}\left(\tau h\wedge\bar{K}\tau^{2}\right)\le D_{22}h,
		\end{align*}
		and, with probability one,
		\begin{align*}
			\left|\omega_{2}\right|\le & \int_{0}^{\infty}\int_{0}^{\tau}I\left(t\le X\right)I\left(r\le t\right)2K\left(\frac{t-r}{h}\right)\left|\varphi_{2}\left(\alpha_{n}\left(t\right)+\beta^{\top}Z\left(r\right)\right)\right|dtdN\left(r\right)\\
			\le & 2\bar{K}\left(\sup_{t,r\in\left[0,\tau\right],Z\in\mathcal{Z}}\left|\varphi_{2}\left(\alpha_{n}\left(t\right)+\beta^{\top}Z\left(r\right)\right)\right|\right)\int_{0}^{\infty}\int_{0}^{\tau}I\left(t\le X\right)I\left(r\le t\right)dtdN\left(r\right)\\
			\le & \tau J_{0}\le D_{21}.
		\end{align*}
	\end{proof}

	\newpage
	
	\section{The detailed derivation and proofs of lemmas and propositions to prove Theorem~\ref{them:consistency}-\ref{them:normality}}

	\label{supp:thmproof}
	
	\subsection{The derivation of the closed forms of $\bM_{h}\left(\theta\right)$ and $\bM\left(\theta\right)$}
	We first derive the expectation of $m_{\theta,h,1}(O)$, i.e.,
	$\bM_{h,1}\left(\theta\right)$. By the definition of $m_{\theta,h,1}(O)$,
	using the conditional expectation on $\{X,\Delta,Z\}$ and Condition~\ref{enu:ratefun},
	\begin{align*}
		& E\left[m_{\theta,h,1}(O)\right]\\
		= & E\left[\int_{0}^{\infty}K_{h}(X-r)I(r\le X)\Delta\cdot\log\{H(\theta;X,Z(r))\}dN(r)\right]\\
		= & E\left[\int_{0}^{\infty}K_{h}(X-r)I(r\le X)\Delta\cdot\log\{H(\theta;X,Z(r))\}E\left[dN(r)|X,\Delta,Z(u),0\le0\le r\right]\right]\\
		= & E\left[\int_{0}^{\infty}K_{h}(X-r)I(r\le X)\Delta\cdot\log\{H(\theta;X,Z(r))\}\mu(r)dr\right].
	\end{align*}
	Since $\Delta=I(T\le C)$ and $X=\min\left\{ T,C\right\} $, 
	\begin{align*}
		& E\left[m_{\theta,h,1}(O)\right]\\
		= & E\left[\int_{0}^{\infty}K_{h}(T-r)I(r\le T)I\left(T\le C\right)\log\{H(\theta;T,Z(r))\}\mu(r)dr\right]\\
		= & E\left[E\left[\int_{0}^{\infty}K_{h}(T-r)I(r\le T)I\left(T\le C\right)\log\{H(\theta;T,Z(r))\}|Z,C\right]\mu(r)dr\right].
	\end{align*}
	Since $T$ and $C$ are independent given $Z$ under \ref{enu:censor}, it not
	hard to show that the probability density function of $T$ given $Z$
	and $C$ is $H(\theta_{0};t,Z(t))S_{T|Z}(t)$, which is independent
	of $C$. Then, taking expectation with respect to $T$ given $Z$
	and $C$, we have 
	\begin{align*}
		& E\left[m_{\theta,h,1}(O)\right]\\
		= & \int_{0}^{\infty}\int_{0}^{\infty}K_{h}(t-r)I(r\le t)\\
		& \times E\left[I\left(t\le C\right)\log\{H(\theta;t,Z(r))\}H(\theta_{0};t,Z(t))S_{T|Z}(t)\right]dt\mu(r)dr\\
		= & \int_{0}^{\infty}\int_{0}^{\infty}K_{h}(t-r)I(r\le t)\\
		& \times E\left[S_{T|Z}(t)E\left[I\left(t\le C\right)|Z\right]\log\{H(\theta;t,Z(r))\}H(\theta_{0};t,Z(t))\right]dt\mu(r)dr\\
		= & \int_{0}^{\infty}\int_{0}^{\infty}K_{h}(t-r)I(r\le t)\\
		& \times E\left[S_{T|Z}(t)S_{C|Z}(t)\log\{H(\theta;t,Z(r))\}H(\theta_{0};t,Z(t))\right]dt\mu(r)dr\\
		= & \int_{0}^{\tau}\int_{0}^{\infty}K_{h}(t-r)I(r\le t)\\
		& \times E\left[S_{T \wedge C|Z}(t)\log\{H(\theta;t,Z(r))\}H(\theta_{0};t,Z(t))\right]\mu(r)drdt.
	\end{align*}
	The last equality holds because we can then replace the integral upper bound
	of $t$ in $E\left[m_{\theta,h,1}(O)\right]$ by $\tau$ since
	$S_{C|Z}(t)=0$ for $t>\tau$ under Condition~\ref{enu:censor}.
	
	For $\bM_{h,2}\left(\theta\right)=E\left[m_{\theta,h,2}(O)\right]$,
	by using the similar conditional expectation argument, 
	\begin{align*}
		& E\left[m_{\theta,h,2}(O)\right]\\
		=& E\left[\int_{0}^{\tau}\int_{0}^{\infty}I(r\le t)K_{h}(t-r)I(X\ge t)H(\theta;t,Z(r))\mu(r)drdt\right]\\
		=& E\left[\int_{0}^{\tau}\int_{0}^{\infty}I(r\le t)K_{h}(t-r)E\left[I(T\wedge C\ge t)|Z\right]H(\theta;t,Z(r))\mu(r)drdt\right].
	\end{align*}
	Since $T$ and $C$ are independent given $Z$, 
	\begin{align*}
		E\left[I(T\wedge C\ge t)|Z\right] & =E\left[I(T\ge t)I(C\ge t)|Z\right]\\
		& =E\left[I(T\ge t)|Z\right]E\left[I(C\ge t)|Z\right]\\
		& =S_{T|Z}(t)S_{C|Z}(t).
	\end{align*}
	Thus, 
	\[
	E\left[m_{\theta,h,2}(O)\right]=\int_{0}^{\tau}\int_{0}^{\infty}I(r\le t)K_{h}(t-r)E\left[S_{T|Z}(t)S_{C|Z}(t)H(\theta;t,Z(r))\right]\mu(r)drdt.
	\]
	Then, combining $E\left[m_{\theta,h,1}(O)\right]$ and $E\left[m_{\theta,h,2}(O)\right]$,
	we have 
	\begin{align*}
		\bM_{h}\left(\theta\right)= & E\left[m_{\theta,h,1}(O)\right]-E\left[m_{\theta,h,2}(O)\right]\\
		= & \int_{0}^{\tau}\int_{0}^{\infty}K_{h}(t-r)I(r\le t)E\left[S_{T|Z}(t)S_{C|Z}(t)\right.\\
		& \left.\times \left\{ \log\{H(\theta;t,Z(r))\}H(\theta_{0};t,Z(t))-H(\theta;t,Z(r))\right\} \right]\mu(r)drdt\\
		= & \int_{0}^{\tau}\int_{0}^{\infty}\tilde{g}\left(\theta,t,r\right)drdt,
	\end{align*}
	where $\tilde{g}\left(\theta,t,r\right)$ is defined in \eqref{eq:gtileddef}.
	
	Next, we give a heuristic proof that $\bM_{h}\left(\theta\right)$
	converges to $\bM\left(\theta\right)$ as $h\to0$ for each
	$\theta \in \Theta$. The rigorous proof is established by Lemma~\ref{lem:kernelconverge} in the Step 2 of the proof of Theorem \ref{them:consistency}. Let $u=(t-r)/h$.
	By the change of variable in $\bM_{h}\left(\theta\right)$,
	we have 
	\begin{align*}
		\bM_{h}\left(\theta\right)  = & 2\int_{0}^{\tau}\int_{0}^{\infty}K(u)I(u\le t/h)E\left[S_{T|Z}(t)S_{C|Z}(t)\right.\\
		& \times\big.\left\{ \log\{H(\theta;t,Z(t-uh))\}H(\theta_{0};t,Z(t))-H(\theta;t,Z(t-uh)\right\} \big]\mu(t-uh)dudt\\
		= & 2\int_{0}^{\tau}\int_{0}^{\infty}K(u)I(u\le t/h)\tilde{g}\left(\theta,t,t-uh\right)dudt.
	\end{align*}
	By Conditions \ref{enu:sieveoder}-\ref{enu:zbounded} and \ref{enu:H_cond}-\ref{enu:censor} as well as Lemma~\ref{lem:GZLip}, we can show $\tilde{g}\left(\theta,t,r\right)$
	is bounded and continuous in $r$ for any $\theta$.
	The upper bound of $u$, $t/h$, also goes to infinity as $h\to\infty$ sine $t\le\tau<\infty$.
	Then, for every $\theta$, by the dominated convergence theorem, 
	\begin{align*}
		\lim_{h\to0}\bM_{h}\left(\theta\right) & =2\int_{0}^{\tau}\int_{0}^{\infty}K(u)\tilde{g}\left(\theta,t,t\right)dudt\\
		& =2\int_{0}^{\infty}K(u)du\int_{0}^{\tau}\tilde{g}\left(\theta,t,t\right)dt.
	\end{align*}
	By Condition \ref{enu:kernel} and the definition of a kernel, $2\int_{0}^{\infty}K(u)du=1$.
	Hence, $\lim_{h\to0}\bM_{h}\left(\theta\right)=\bM\left(\theta\right)$ for every $\theta \in \Theta$.

	\subsection{The proof of Lemma \ref{lem:kernelconverge}}
	\begin{proof}
		We first need to find an upper bound of $\int_{0}^{\infty}K\left(u\right)I\left(u>b\right)du$
		for some constant $b>0$. By the definition of the kernel $K$,
		$2K\left(u\right)$ can be seen as a probability density function
		for a positive random variable since $\int_{0}^{\infty}2K\left(u\right)du=1$.
		Markov's inequality implies that
		\begin{equation}
			\int_{0}^{\infty}K\left(u\right)I\left(u>b\right)du \le\left(1/b\right)\int_{0}^{\infty}K\left(u\right)udu.\label{eq:Kerneltailprob}
		\end{equation}
		
		It is easy to see
		\begin{align}
			& \sup_{g\in\mathcal{G}}\left|2\int_{0}^{\tau}\int_{0}^{\infty}\frac{1}{h}K\left(\frac{t-r}{h}\right)I\left(r\le t\right)g\left(t,r\right)drdt-\int_{0}^{\tau}g\left(t,t\right)dt\right|\nonumber \\
			\le &  \mathbf{A}_{1} +\mathbf{A}_{2},
		\end{align}
		where 
		\[
		\mathbf{A}_{1} \coloneqq \sup_{g\in\mathcal{G}}\left|2\int_{0}^{\tau}\int_{0}^{\infty}\frac{1}{h}K\left(\frac{t-r}{h}\right)I(r\le t)(g(t,r)-g(t,t))drdt\right|	
		\]
		and
		\[
		\mathbf{A}_{2} \coloneqq \sup_{g\in\mathcal{G}}\left|2\int_{0}^{\tau}\int_{0}^{\infty}\frac{1}{h}K\left(\frac{t-r}{h}\right)I(r\le t)g(t,t)drdt-\int_{0}^{\tau}g(t,t)dt\right|.	
		\]
		We next find the upper bounds of $\mathbf{A}_{1}$ and $\mathbf{A}_{2}$.
		
		For $\mathbf{A}_{1}$, by the Lipschitz continuity of $g(t,r)$ in $r$, 
		\begin{align*}
			\mathbf{A}_{1} \le & 2\int_{0}^{\tau}\int_{0}^{\infty}\frac{1}{h}K\left(\frac{t-r}{h}\right)I(r\le t)\sup_{g\in\mathcal{G}}|g(t,r)-g(t,t)|drdt\nonumber\\
			\le & 2L^{*}\int_{0}^{\tau}\int_{0}^{\infty}\frac{1}{h}K\left(\frac{t-r}{h}\right)I\left(r\le t\right)\left|t-r\right|drdt.
		\end{align*}
		Letting $u=(t-r)/h$, the change of variable  yields that 
		\begin{multline*}
			2L^{*}\int_{0}^{\tau}\int_{0}^{\infty}\frac{1}{h}K\left(\frac{t-r}{h}\right)I\left(r\le t\right)\left|t-r\right|drdt= 2L^{*}h\int_{0}^{\tau}\int_{0}^{\infty}uK\left(u\right)I\left(uh\le t\right)dudt.
		\end{multline*}
		Since $I\left(uh\le t\right)\le1$, by Tonelli's theorem, it follows that 
		\begin{align*}
			\int_{0}^{\tau}\int_{0}^{\infty}uK\left(u\right)I\left(uh\le t\right)dudt & =\int_{0}^{\infty}uK\left(u\right)\int_{0}^{\tau}I\left(uh\le t\right)dtdu\\
			& \le\tau\int_{0}^{\infty}uK\left(u\right)du.
		\end{align*}
		Thus,
		\begin{equation*}
			\mathbf{A}_{1} \le 2L^{*}h\tau\int_{0}^{\infty}uK\left(u\right)du.
		\end{equation*}
		
		For $\mathbf{A}_{2}$, by the change of variable as $u=\left(t-r\right)/h$,
		\begin{align*}
			\mathbf{A}_{2} & =\sup_{g\in\mathcal{G}}\left|2\int_{0}^{\tau}\int_{0}^{\infty}\frac{1}{h}K\left(\frac{t-r}{h}\right)I\left(r\le t\right)g\left(t,t\right)drdt-\int_{0}^{\tau}g\left(t,t\right)dt\right|\\
			& =\sup_{g\in\mathcal{G}}\left|2\int_{0}^{\tau}\int_{0}^{\infty}K\left(u\right)I\left(uh\le t\right)g\left(t,t\right)dudt-2\int_{0}^{\infty}K\left(u\right)du\int_{0}^{\tau}g\left(t,t\right)dt\right|\\
			& =\sup_{g\in\mathcal{G}}\left|2\int_{0}^{\tau}\int_{0}^{\infty}K\left(u\right)I\left(uh\le t\right)g\left(t,t\right)dudt-2\int_{0}^{\tau}\int_{0}^{\infty}K\left(u\right)g\left(t,t\right)dudt\right|\\
			& \le 2\sup_{g\in\mathcal{G},t\in[0,\tau]}\int_{0}^{\tau}\int_{0}^{\infty}K\left(u\right)I\left(uh>t\right)dudt.
		\end{align*}
		Besides, 
		\begin{align*}
			\int_{0}^{\tau}\int_{0}^{\infty}K\left(u\right)I\left(uh>t\right)dudt  =&\int_{0}^{\tau/h}K\left(u\right)\int_{0}^{\tau}I\left(uh>t\right)dtdu\\
			&+\int_{\tau/h}^{\infty}K\left(u\right)\int_{0}^{\tau}I\left(uh>t\right)dtdu\\
			=&h\int_{0}^{\tau/h}K\left(u\right)udu+\int_{\tau/h}^{\infty}K\left(u\right)\tau du\\
			\le &  h\int_{0}^{\infty}K\left(u\right)udu+h\int_{0}^{\infty}K\left(u\right)udu \\
			= & 2h\int_{0}^{\infty}K\left(u\right)udu.
		\end{align*}
		The last inequality  is due to the application of \eqref{eq:Kerneltailprob} to $\int_{\tau/h}^{\infty}K\left(u\right)\tau du$. 
		Thus,
		\begin{align*}
			\mathbf{A}_{2} & \le 2\sup_{g\in\mathcal{G},t\in[0,\tau]}\int_{0}^{\tau}\int_{0}^{\infty}K\left(u\right)I\left(uh>t\right)dudt\\
			& \le4h\sup_{g\in\mathcal{G},t\in[0,\tau]}\left|g\left(t,t\right)\right|\int_{0}^{\infty}K\left(u\right)udu.
		\end{align*}

		Combining the upper bounds of $\mathbf{A}_{1}$ and $\mathbf{A}_{2}$, we have
		\begin{multline*}
			\left|2\int_{0}^{\tau}\int_{0}^{\infty}\frac{1}{h}K\left(\frac{t-r}{h}\right)I\left(r\le t\right)g\left(t,r\right)drdt-\int_{0}^{\tau}g\left(t,t\right)dt\right| \\
			\le 2 h\tau\left(2\sup_{g\in\mathcal{G},t\in[0,\tau]}\left|g\left(t,t\right)\right|+L^{*}\right)\int_{0}^{\infty}K\left(u\right)udu.
		\end{multline*}
		Then, this lemma follows.

	\end{proof}
	
	\subsection{The proof of Lemma \ref{lem:BnLg}}
	
	We first prove $\sup_{t\in[0,\tau],\theta\in\Theta_{n}}\left|\tilde{g}\left(\theta;t,t\right)\right|<\infty$.
	Define a function
	\[
	g\left(x;t,\alpha\right) \coloneqq \log\left\{ H\left(\alpha\left(t\right)+x\right)\right\} H\left(\alpha_{0}\left(t\right)+\beta_{0}^{\top}Z\left(t\right)\right)-H\left(\alpha\left(t\right)+x\right),
	\]
	and clearly, for $\tilde{g}\left(\theta,t,r\right)$ defined in \eqref{eq:gtileddef},
	\[
	\tilde{g}\left(\theta;t,r\right)=E\left[S_{T\wedge C|Z}\left(t\right)\mu\left(r\right)g\left(\beta^{\top}Z\left(r\right);t,\alpha\right)\right].
	\]
	Since $0\le S_{T\wedge C|Z}\left(t\right)\le 1$ for $t\in[0,\tau]$, it is not hard to see 
	\[
	\sup_{t\in[0,\tau],\theta\in\Theta_{n}}\left|\tilde{g}\left(\theta;t,t\right)\right|\le E\left[\sup_{t\in[0,\tau],\theta\in\Theta_{n}}\mu\left(t\right)\left|g\left(\beta^{\top}Z\left(t\right);t,\alpha\right)\right|\right].	
	\]
	Conditions \ref{enu:alpha_para_space}, \ref{enu:beta_para_space}, \ref{enu:zbounded} and \ref{enu:H_cond} ensure that
	$H\left(\alpha\left(t\right)+\beta^{\top}Z\left(r\right)\right)$ is positive and bounded
	for all $\theta\in\Theta$ and $t,r\in\left[0,\tau\right]$; Condition \ref{enu:ratefun} implies that $\mu(t)$
	is bounded for $t\in\left[0,\tau\right]$.
	It then follows that $\sup_{t\in[0,\tau],\theta\in\Theta_{n}}\mu\left(t\right)\left|g\left(\beta^{\top}Z\left(t\right);t,\alpha\right)\right|<\infty$.
	Hence,
	$\sup_{t\in[0,\tau],\theta\in\Theta_{n}}\left|\tilde{g}\left(\theta;t,t\right)\right|$
	is finite.
	
	For the uniform Lipschitz continuity, we write
	\[
	\sup_{\theta\in\Theta_{n}}\left|\tilde{g}\left(\theta;t,t\right)-\tilde{g}\left(\theta;t,r\right)\right|=\mathbf{A}_{3}+\mathbf{A}_{4},
	\]
	where
	\[
	\mathbf{A}_{3} \coloneqq \sup_{\theta\in\Theta_{n}}\left|E\left[S_{T\wedge C|Z}\left(t\right)\left\{ \mu\left(t\right)-\mu\left(r\right)\right\} g\left(\beta^{\top}Z\left(t\right);t,\alpha\right)\right]\right|,
	\]
	and 
	\[
	\mathbf{A}_{4} \coloneqq \sup_{\theta\in\Theta_{n}}\left|E\left[S_{T\wedge C|Z}\left(t\right)\mu\left(r\right)\left\{ g\left(\beta^{\top}Z\left(t\right);t,\alpha\right)-g\left(\beta^{\top}Z\left(r\right);t,\alpha\right)\right\} \right]\right|.
	\]
	We next find the upper bounds of $\mathbf{A}_{3}$ and $\mathbf{A}_{4}$ in terms of $|t-r|$.
	
	For $\mathbf{A}_{3},$ by Condition \ref{enu:ratefun} and $S_{T\wedge C|Z}\left(t\right)\le 1$,
	\begin{align*}
		\mathbf{A}_{3} & \le E\left[S_{T\wedge C|Z}\left(t\right)\left|\mu\left(t\right)-\mu\left(r\right)\right|\sup_{\theta\in\Theta_{n}}\left|g\left(\beta^{\top}Z\left(t\right);t,\alpha\right)\right|\right]\\
		& \le L_{1}E\left[\sup_{t\in[0,\tau],\theta\in\Theta_{n}}\left|g\left(\beta^{\top}Z\left(t\right);t,\alpha\right)\right|\right]\left|t-r\right|,
	\end{align*}
	where $L_{1}$ is the Lipschitz constant of $\mu\left(\cdot\right)$
	in view of Condition \ref{enu:ratefun}.
	
	For $\mathbf{A}_{4}$, applying Lemma \ref{lem:GZLip}, for $L$ defined in \ref{enu:zbounded},	we have
	\begin{align*}
		\mathbf{A}_{4} & \le L\sup_{\beta\in\mathcal{B}}\left\Vert \beta\right\Vert _{2}\sup_{\theta\in\Theta_{n}}\sqrt{E\left[\left|S_{T\wedge C|Z}\left(t\right)g^{\prime}\left(\beta^{\top}\left(\xi^{*}Z\left(t\right)+\left(1-\xi^{*}\right)Z\left(r\right)\right);t,\alpha\right)\right|^{2}\right]}\left|t-r\right|\\
		& \le L\sup_{\beta\in\mathcal{B}}\left\Vert \beta\right\Vert _{2}\mathbf{A}_{5}\left|t-r\right|,
	\end{align*}
	where the derivative of $g\left(x;t,\alpha\right)$ with respective to $x$,
	\[
	g^{\prime}\left(x;t,\alpha\right) \coloneqq H^{\prime}\left(\alpha\left(t\right)+x\right)\left\{ \frac{H\left(\alpha_{0}\left(t\right)+\beta_{0}^{\top}Z\left(t\right)\right)}{H\left(\alpha\left(t\right)+x\right)}-1\right\} ,
	\]
	and 
	\[
	\mathbf{A}_{5} \coloneqq \sup_{t,r\in\left[0,\tau\right],\theta\in\Theta_{n}}\sqrt{E\left[\left|g^{\prime}\left(\beta^{\top}\left(\xi^{*}Z\left(t\right)+\left(1-\xi^{*}\right)Z\left(r\right)\right);t,\alpha\right)\right|^{2}\right]}.
	\]
	The last inequality is due to that $0\le S_{T\wedge C|Z}\left(t\right)\le1$ for $t\in[0,\tau]$. 
	Thus, 
	\[
	\sup_{\theta\in\Theta_{n}}\left|\tilde{g}\left(\theta;t,t\right)-\tilde{g}\left(\theta;t,r\right)\right|\le L_{1}^{*}\left|t-r\right|
	\]
	where
	\[
	L_{1}^{*} \coloneqq L_{1}E\left[\sup_{t\in[0,\tau]\theta\in\Theta_{n}}\left|g\left(\beta^{\top}Z\left(t\right);t,\alpha\right)\right|\right]+L\sup_{\beta\in\mathcal{B}}\left\Vert \beta\right\Vert _{2}\mathbf{A}_{5}.
	\]
	Clearly, $L_{1}^{*}$ does not depend on $t$, $r$ and $\theta$. 
	
	We next show $L_{1}^{*}$
	is finite. By Conditions \ref{enu:alpha_para_space}, \ref{enu:beta_para_space}, \ref{enu:zbounded} and \ref{enu:H_cond}, $L\sup_{\beta\in\mathcal{B}}\left\Vert \beta\right\Vert _{2}$
	and $L_{1}E\left[\sup_{t \in [0,\tau],\theta\in\Theta_{n}}\left|g\left(\beta^{\top}Z\left(t\right);t,\alpha\right)\right|\right]$
	are both finite.
	It remains to show $\mathbf{A}_5$ is finite.
	Under Condition \ref{enu:H_cond}, $H$ is strictly increasing and hence
	\begin{multline*}
		H\left(\alpha\left(t\right)+\beta^{\top}\left(\xi^{*}Z\left(t\right)+\left(1-\xi^{*}\right)Z\left(r\right)\right)\right)\\
		\ge\min\left\{ H\left(\alpha\left(t\right)+\beta^{\top}Z\left(t\right)\right),H\left(\alpha\left(t\right)+\beta^{\top}Z\left(r\right)\right)\right\} >0.
	\end{multline*}
	In other words, the denominator in $g^{\prime}\left(\beta^{\top}\left(\xi^{*}Z\left(t\right)+\left(1-\xi^{*}\right)Z\left(r\right)\right);t,\alpha\right)$ is nonzero.
	Then, under Conditions \ref{enu:alpha_para_space}, \ref{enu:beta_para_space}, \ref{enu:zbounded} and \ref{enu:H_cond}, we can show 
	\[
	\left|g^{\prime}\left(\beta^{\top}\left(\xi^{*}Z\left(t\right)+\left(1-\xi^{*}\right)Z\left(r\right)\right);t,\alpha\right)\right|	
	\]
	is finite for all $t,r\in[0,\tau]$ and $\theta\in \Theta_n$ by the similar arguments for proving
	$$
	\sup_{t\in[0,\tau],\theta\in\Theta_{n}}\mu\left(r\right)\left|g\left(\beta^{\top}Z\left(r\right);t,\alpha\right)\right|<\infty.
	$$
	Thus, $\mathbf{A}_5$ is also finite and hence $L_{1}^{*}<\infty$.
	Now, the lemma follows.

	\subsection{The proof of Lemma \ref{lem:hempridiffprop}}
	
	\begin{proof}

		We first find the upper bounds of $\left\Vert f\right\Vert _{\infty}$ and $\mathbf{P}\left(f\right)^{2}$ for $f\in\mathcal{\bar{F}}_{n,\eta}$, respectively.
		For simplicity, without loss of generality, we drop the $n$ subscript
		for $h_{n}$ in the following derivation.
		
		Write out $f=\bU_{1}-\bU_{2}$ for $f\in\mathcal{\bar{F}}_{n,\eta}$
		where
		\[
		\bU_{1} \coloneqq \int_{0}^{\infty}2K\left(\frac{X-r}{h}\right)I\left(r\le X\right)\Delta\cdot\log\left\{ \frac{H\left(\theta;X,Z\left(r\right)\right)}{H\left(\theta_{0};X,Z\left(r\right)\right)}\right\} dN\left(r\right),
		\]
		and
		\begin{multline*}
			\bU_{2} \coloneqq \int_{0}^{\infty}\int_{0}^{\infty}I\left(r\le t\right)2K\left(\frac{X-r}{h}\right)I\left(X\ge t\right)\\
			\times\left\{ H\left(\theta;t,Z\left(r\right)\right)-H\left(\theta_{0};t,Z\left(r\right)\right)\right\} dN\left(r\right)dt.
		\end{multline*}
		Then, by the fact that $\left\Vert f\right\Vert _{\infty}\le\left\Vert \bU_{1}\right\Vert _{\infty}+\left\Vert \bU_{2}\right\Vert _{\infty}$,
		we calculate the upper bounds of $\left\Vert \bU_{1}\right\Vert _{\infty}$
		and $\left\Vert \bU_{2}\right\Vert _{\infty}$, respectively.
		Then, by Conditions \ref{enu:alpha_para_space}, \ref{enu:beta_para_space},
		\ref{enu:zbounded}, \ref{enu:H_cond} and \ref{enu:finiteobs}, with probability one,
		\begin{align*}
			\left|\bU_{1}\right| & =\left|\int_{0}^{\infty}2K\left(\frac{X-r}{h}\right)I\left(r\le X\right)\Delta\cdot\log\left\{ \frac{H\left(\theta;X,Z\left(r\right)\right)}{H\left(\theta_{0};X,Z\left(r\right)\right)}\right\} dN\left(r\right)\right|\\
			& \le2\bar{K}\int_{0}^{\infty}I\left(r\le X\right)\Delta\cdot\left|\log\left\{ \frac{H\left(\theta;X,Z\left(r\right)\right)}{H\left(\theta_{0};X,Z\left(r\right)\right)}\right\} \right|dN\left(r\right)\\
			& \lesssim2\bar{K}\int_{0}^{\infty}I\left(r\le X\right)dN\left(r\right)\\
			& \lesssim2\bar{K}N\left(X\right)\\
			& \lesssim2\bar{K}J_{0}.
		\end{align*}
		Similarly, 
		\begin{align*}
			\left|\bU_{2}\right|  =&\left|\int_{0}^{\infty}\int_{0}^{\infty}I\left(r\le t\right)2K\left(\frac{t-r}{h}\right)I\left(X\ge t\right)\right.\\
			&\times\left.\left\{ H\left(\theta;t,Z\left(r\right)\right)-H\left(\theta_{0};t,Z\left(r\right)\right)\right\} dN\left(r\right)dt\right|\\
			\le & 2\bar{K}\int_{0}^{\infty}\int_{0}^{\infty}I\left(r\le t\right)I\left(X\ge t\right)\left|H\left(\theta;t,Z\left(r\right)\right)-H\left(\theta_{0};t,Z\left(r\right)\right)\right|dN\left(r\right)dt\\
			\lesssim & 2\bar{K}\int_{0}^{\infty}\int_{0}^{\infty}I\left(r\le t\right)I\left(X\ge t\right)dN\left(r\right)dt\\
			\le & 2\bar{K}\int_{0}^{\infty}N\left(t\right)I\left(X\ge t\right)dt\\
			\le & 2\bar{K}J_{0}X\\
			\le & 2\bar{K}J_{0}\tau.
		\end{align*}
		The above upper bounds are independent of $\theta$.
		Therefore, $\sup_{f\in\mathcal{\bar{F}}_{n,\eta}}\left\Vert f\right\Vert _{\infty}\le\bar{K}J_{0}\left(\tau+1\right)$.
		
		For $\sup_{f\in\mathcal{\bar{F}}_{n,\eta}}\mathbf{P}f^{2}$, it follows that $\mathbf{P}\left(f\right)^{2}=E\left[f^{2}\right]\le2\left(E\left[\bU_{1}^{2}\right]+E\left[\bU_{2}^{2}\right]\right)$.
		For the first term, by Lemma \ref{lem:integralvar},
		$
		E\left[\bU_{1}^{2}\right]=E\left[\bU_{11}\right]+E\left[\bU_{12}\right],
		$
		where 
		\[
		E\left[\bU_{11}\right]=4E\left[\int_{0}^{\infty}K^{2}\left(\frac{X-r}{h}\right)I\left(r\le X\right)\Delta\cdot\left[\log\left\{ \frac{H\left(\theta;X,Z\left(r\right)\right)}{H\left(\theta_{0};X,Z\left(r\right)\right)}\right\} \right]^{2}\mu\left(r\right)dr\right],
		\]
		and 
		\[
		E\left[\bU_{12}\right]=4E\left[\left(\int_{0}^{\infty}K\left(\frac{X-r}{h}\right)I\left(r\le X\right)\Delta\cdot\log\left\{ \frac{H\left(\theta;X,Z\left(r\right)\right)}{H\left(\theta_{0};X,Z\left(r\right)\right)}\right\} \mu\left(r\right)dr\right)^{2}\right].
		\]
		For $E\left[\bU_{11}\right]$, by the conditional expectation argument
		and the change of variable as $u=t-r$ and the boundedness of $\mu(\dot)$ on $[0,\tau]$, we have that
		\begin{align*}
			E\left[\bU_{11}\right]= & 4E\left[\int_{0}^{\tau}K^{2}\left(\frac{X-r}{h}\right)I\left(r\le X\right)\Delta\cdot\left[\log\left\{ \frac{H\left(\theta;X,Z\left(r\right)\right)}{H\left(\theta_{0};X,Z\left(r\right)\right)}\right\} \right]^{2}\mu\left(r\right)dr\right]\\
			\le & 4E\left[\int_{0}^{\tau}K^{2}\left(\frac{X-r}{h}\right)I\left(r\le X\right)\Delta\cdot\left[\log\left\{ \frac{H\left(\theta;X,Z\left(r\right)\right)}{H\left(\theta_{0};X,Z\left(r\right)\right)}\right\} \right]^{2}dr\right]\\
			= & 4E\left[\int_{0}^{\tau}\int_{0}^{\infty}K^{2}\left(\frac{t-r}{h}\right)I\left(r\le t\right)S_{T\wedge C|Z}\left(t\right)\right.\\
			& \times\left.\left[\log\left\{ \frac{H\left(\theta;t,Z\left(r\right)\right)}{H\left(\theta_{0};t,Z\left(r\right)\right)}\right\} \right]^{2}H\left(\theta_{0};t,Z\left(t\right)\right)drdt\right]\\
			\le & 4E\left[\int_{0}^{\tau}\int_{0}^{\infty}K^{2}\left(\frac{t-r}{h}\right)I\left(r\le t\right)\left[\log\left\{ \frac{H\left(\theta;t,Z\left(r\right)\right)}{H\left(\theta_{0};t,Z\left(r\right)\right)}\right\} \right]^{2}drdt\right]\\
			= & 4E\left[\int_{0}^{\tau}\int_{0}^{\infty}K^{2}\left(u/h\right)I\left(u\le t\right)\left[\log\left\{ \frac{H\left(\theta;t,Z\left(t-u\right)\right)}{H\left(\theta_{0};t,Z\left(t-u\right)\right)}\right\} \right]^{2}dudt\right].
		\end{align*}
		By the mean value theorem and Conditions \ref{enu:kernel},
		\ref{enu:alpha_para_space}, \ref{enu:beta_para_space}, \ref{enu:zbounded}
		and \ref{enu:H_cond},
		\begin{align*}
			& E\left[\bU_{11}\right]\\
			= & 4E\left[\int_{0}^{\tau}\int_{0}^{\infty}K^{2}\left(u/h\right)I\left(u\le t\right)\left[\frac{H^{\prime}\left(\theta_{\xi};t,Z\left(t-u\right)\right)}{H\left(\theta_{\xi};t,Z\left(t-u\right)\right)}\right]^{2}\right.\\
			& \times\left\{ \alpha\left(t\right)-\alpha_{0}\left(t\right)+\left(\beta-\beta_{0}\right)^{T}Z\left(t-u\right)\right\} ^{2}\left.\mu\left(t-u\right)dudt\right]\\
			\le & 4\int_{0}^{\tau}\int_{0}^{\infty}K^{2}\left(u/h\right)I\left(u\le t\right)E\left[\left\{ \alpha\left(t\right)-\alpha_{0}\left(t\right)+\left(\beta-\beta_{0}\right)^{T}Z\left(t-u\right)\right\} ^{2}\right]dudt\\
			\lesssim & \int_{0}^{\tau}\int_{0}^{\infty}hK^{2}\left(u\right)I\left(u\le t/h\right)duE\left[\left(\alpha\left(t\right)-\alpha_{0}\left(t\right)\right)^{2}+\left(\beta-\beta_{0}\right)^{T}\left(\beta-\beta_{0}\right)\right]dt\\
			= & \left\{ \sup_{t\in\left[0,\tau\right]}h\int_{0}^{\infty}K^{2}\left(u\right)I\left(u\le t/h\right)du\right\} \left\{ \left\Vert \alpha-\alpha_{0}\right\Vert _{2}^{2}+\left\Vert \beta-\beta_{0}\right\Vert _{2}^{2}\right\} \\
			\lesssim & hd^{2}\left(\theta,\theta_{0}\right).
		\end{align*}
		Applying the similar arguments to calculate $E\left[\bU_{12}\right]$ and letting $u=\left(X-r\right)/h$,
		we have that
		\begin{align*}
			E\left[\bU_{12}\right]= & 4E\left[\left(\int_{0}^{\tau}K\left(\frac{X-r}{h}\right)I\left(r\le X\right)\Delta\cdot\log\left\{ \frac{H\left(\theta;X,Z\left(r\right)\right)}{H\left(\theta_{0};X,Z\left(r\right)\right)}\right\} \mu\left(r\right)dr\right)^{2}\right]\\
			\le & 4E\left[\int_{0}^{\tau}K^{2}\left(\frac{X-r}{h}\right)I\left(r\le X\right)\Delta\cdot\left[\log\left\{ \frac{H\left(\theta;X,Z\left(r\right)\right)}{H\left(\theta_{0};X,Z\left(r\right)\right)}\right\} \right]^{2}\mu^{2}\left(r\right)dr\right]\\
			\le & 4E\left[\int_{0}^{\tau}K^{2}\left(\frac{X-r}{h}\right)I\left(r\le X\right)\Delta\cdot\left[\log\left\{ \frac{H\left(\theta;X,Z\left(r\right)\right)}{H\left(\theta_{0};X,Z\left(r\right)\right)}\right\} \right]^{2}dr\right]\\
			\lesssim & hd^{2}\left(\theta,\theta_{0}\right).
		\end{align*}
		Therefore, $\sup_{f\in\mathcal{\bar{F}}_{n,\eta}}E\left[\bU_{1}^{2}\right]\lesssim h\eta^{2}$.
		
		For $\bU_{2}$, we similarly have that 
		$
		E\left[\bU_{2}^{2}\right]=E\left[\bU_{21}\right]+E\left[\bU_{22}\right]
		$
		where
		\begin{multline*}
			E\left[\bU_{21}\right]=4E\left[\int_{0}^{\infty}\left(\int_{0}^{\tau}I\left(r\le t\right)K\left(\frac{t-r}{h}\right)I\left(X\ge t\right)\right.\right.\\
			\times\left.\left\{ H\left(\theta;t,Z\left(r\right)\right)-H\left(\theta_{0};t,Z\left(r\right)\right)\right\} dt\right)^{2}\left.\mu\left(r\right)dr\right],
		\end{multline*}
		and
		\begin{multline*}
			E\left[\bU_{22}\right]=4E\left[\left\{ \int_{0}^{\infty}\int_{0}^{\tau}I\left(r\le t\right)K\left(\frac{t-r}{h}\right)I\left(X\ge t\right)\right.\right.\\
			\times\left.\left.\left\{ H\left(\alpha\left(t\right)+\beta^{T}Z\left(r\right)\right)-H\left(\alpha_{0}\left(t\right)+\beta_{0}^{T}Z\left(r\right)\right)\right\} dt\mu\left(r\right)dr\right\} ^{2}\right].
		\end{multline*}
		Then, by the Cauchy--Schwarz inequality and the similar arguments to calculating $E\left[\bU_{11}\right]$ and $E\left[\bU_{12}\right]$, 
		\begin{align*}
			& E\left[\bU_{21}\right] \\
			\le & 4\int_{0}^{\infty}E\left[\int_{0}^{\tau}I\left(r\le t\right)K^{2}\left(\frac{t-r}{h}\right)\left\{ H\left(\theta;t,Z\left(r\right)\right)-H\left(\theta_{0};t,Z\left(r\right)\right)\right\} ^{2}dt\right]\mu\left(r\right)dr\\
			\le & 4\int_{0}^{\infty}E\left[\int_{0}^{\tau}I\left(u\le t\right)K^{2}\left(\frac{u}{h}\right)\left\{ H\left(\theta;t,Z\left(t-u\right)\right)-H\left(\theta_{0};t,Z\left(t-u\right)\right)\right\} ^{2}dt\right]\\
			& \times\mu\left(t-u\right)du\\
			= & 4E\left[\int_{0}^{\tau}\int_{0}^{\infty}I\left(u\le t\right)K^{2}\left(\frac{u}{h}\right)\mu\left(t-u\right)\right.\\
			& \times\left.\left\{ H^{\prime}\left(\theta_{\xi};t,Z\left(t-u\right)\right)\left\{ \alpha\left(t\right)-\alpha_{0}\left(t\right)+\left(\beta-\beta_{0}\right)^{T}Z\left(t-u\right)\right\} \right\} ^{2}dudt\right]\\
			\le & \sup_{t\in\left[0,\tau\right]}\mu\left(t\right)4\int_{0}^{\infty}I\left(u\le t\right)K^{2}\left(\frac{u}{h}\right)duE\left[\int_{0}^{\tau}\left(\alpha\left(t\right)-\alpha_{0}\left(t\right)\right)^{2}dt+\left(\beta-\beta_{0}\right)^{T}\left(\beta-\beta_{0}\right)\right]\\
			\le & \sup_{t\in\left[0,\tau\right]}\mu\left(t\right)4h\int_{0}^{\infty}K^{2}\left(u\right)duE\left[\int_{0}^{\tau}\left(\alpha\left(t\right)-\alpha_{0}\left(t\right)\right)^{2}dt+\left(\beta-\beta_{0}\right)^{T}\left(\beta-\beta_{0}\right)\right]\\
			\lesssim & hd^{2}\left(\theta,\theta_{0}\right),
		\end{align*}
		and similarly,
		\begin{align*}
			E\left[\bU_{22}\right]\le&  4E\left[\int_{0}^{\tau}\int_{0}^{\infty}I\left(r\le t\right)K^{2}\left(\frac{t-r}{h}\right)I\left(X\ge t\right)\right.\\
			& \times\left.\left\{ H\left(\alpha\left(t\right)+\beta^{T}Z\left(r\right)\right)-H\left(\alpha_{0}\left(t\right)+\beta_{0}^{T}Z\left(r\right)\right)\right\} ^{2}\mu^{2}\left(r\right)drdt\right]\\
			= & 4E\left[\int_{0}^{\tau}\int_{0}^{\infty}I\left(r\le t\right)K^{2}\left(\frac{t-r}{h}\right)I\left(X\ge t\right)I\left(r\le t\right)\right.\\
			& \left.\left\{ H\left(\alpha\left(t\right)+\beta^{T}Z\left(r\right)\right)-H\left(\alpha_{0}\left(t\right)+\beta_{0}^{T}Z\left(r\right)\right)\right\} ^{2}\mu^{2}\left(r\right)drdt\right]\\
			\le & \sup_{t\in\left[0,\tau\right]}\mu^{2}\left(t\right)4E\left[\int_{0}^{\tau}\int_{0}^{\infty}I\left(r\le t\right)K^{2}\left(\frac{t-r}{h}\right)\right.\\
			& \times \left.\left\{ H\left(\alpha\left(t\right)+\beta^{T}Z\left(r\right)\right)-H\left(\alpha_{0}\left(t\right)+\beta_{0}^{T}Z\left(r\right)\right)\right\} ^{2}drdt\right]\\
			\lesssim & hd^{2}\left(\theta,\theta_{0}\right).
		\end{align*}
		It follows that $\sup_{f\in\mathcal{\bar{F}}_{n,\eta}}\mathbf{P}\left(f\right)^{2}\lesssim \sup_{f\in\mathcal{\bar{F}}_{n,\eta}}(E\left[\bU_{1}^{2}\right]+E\left[\bU_{2}^{2}\right])\lesssim h\eta^{2}$.
		
		Last, we obtain the  bracketing number of $\mathcal{\bar{F}}_{n,\eta}$.
		Note that the bracketing number of $\bar{\mathcal{F}}_{n,\eta}$ is determined
		by $\theta$ in $h(m_{\theta,h}\left(O\right)-m_{\theta_{0},h}\left(O\right))$.
		In view of  $\sup_{f\in\mathcal{\bar{F}}_{n,\eta}}\mathbf{P}\left(f\right)^{2}\lesssim h\eta^{2}$, the similar arguments to the proof of
		Lemma \ref{lem:Wbracketing} yield that $N_{[\,]}\left(\epsilon,\mathcal{\bar{F}}_{n,\eta},\mathcal{L}_{2}\left(\mathbf{P}\right)\right)\lesssim\left(\eta\sqrt{h_{n}}/\epsilon\right)^{cq_{n}+p}$, for
		some $0<\epsilon<\eta\sqrt{h_{n}}$. 
		
	\end{proof}

	\subsection{The proof of Lemma \ref{lem:wtiproperty}}
	\begin{proof}
		By the mean value theorem, $\exists\xi\in\left[0,1\right]$, such
		that, for $w\left(\theta,t,Z\left(r\right)\right) $ defined in \eqref{eq:wtildedef},
		\begin{equation*}
			w\left(\theta,t,Z\left(r\right)\right)  =w_{1}\left(\theta_{\xi},t,Z\left(r\right)\right)\left\{ \alpha\left(t\right)-\alpha_{0}\left(t\right)+\left(\beta-\beta_{0}\right)^{\top}Z\left(r\right)\right\} ,
		\end{equation*}
		where 
		\[
		w_{1}\left(\theta,t,Z(r)\right) \coloneqq \frac{H^{\prime}\left(\theta;t,Z\left(r\right)\right)}{H\left(\theta;t,Z\left(r\right)\right)}H\left(\theta_{0};t,Z\left(t\right)\right)-H^{\prime}\left(\theta;t,Z\left(r\right)\right).
		\]
		Under Conditions \ref{enu:alpha_para_space}, \ref{enu:beta_para_space},
		\ref{enu:zbounded} and \ref{enu:H_cond}, $w_{1}\left(\theta,t,Z\left(t\right)\right)$
		is uniformly bounded and continuous in $\beta^{\top}Z(t)$ in a neighborhood of $\theta_{0}$. 
		Under Conditions \ref{enu:ratefun} and \ref{enu:censor}, $S_{T\wedge C|Z}\left(t\right)\mu\left(t\right)$
		is bounded for $t\in\left[0,\tau\right]$. Thus,
		\begin{align*}
			& \sup_{t\in\left[0,\tau\right]}\left|E[|S_{T\wedge C|Z}(t)w(\theta,t,Z(t))|]\mu\left(t\right)\right|\\
			\le & \sup_{t\in\left[0,\tau\right]}E[|w_{1}(\theta_{\xi},t,Z(t))\{\alpha(t)-\alpha_{0}(t)-(\beta-\beta_{0})^{\top}Z(t)\}|]\\
			\lesssim & \sup_{t\in\left[0,\tau\right]}E[|\alpha(t)-\alpha_{0}(t)-(\beta-\beta_{0})^{\top}Z(t)|]\\
			\le & \sup_{t\in[0,\tau]}|\alpha(t)-\alpha_{0}(t)|+\Vert\beta-\beta_{0}\Vert_{2}.
		\end{align*}
		The last inequality is due to H\"older's inequality and the boundedness of $Z$ under \ref{enu:zbounded}.
		By Lemma \ref{lem:supL2}, for small enough $\eta$,
		\begin{equation}
			\sup_{\theta\in\Theta_{n,\eta},t\in\left[0,\tau\right]}\left|\alpha\left(t\right)-\alpha_{0}\left(t\right)\right|+\left\Vert \beta-\beta_{0}\right\Vert _{2}
			\lesssim  \eta^{2/3}+\eta   \lesssim  \eta^{2/3}.\label{eq:supeta}
		\end{equation}
		Hence,
		\[
		\sup_{\theta\in\Theta_{n,\eta},t\in\left[0,\tau\right]}\left|E[|S_{T\wedge C|Z}(t)w(\theta,t,Z(t))|]\mu\left(t\right)\right|\lesssim \eta^{2/3},
		\]
		that is, $\sup_{t\in[0,\tau],\,\theta\in\Theta_{n,\eta}}\left|\tilde{w}\left(\theta;t,t\right)\right|\lesssim \eta^{2/3}$.
		
		For the Lipschitz continuity of $\tilde{w}\left(\theta;t,r\right)$ in $r$, 
		$\sup_{\theta\in\Theta_{n,\eta}}\left|\tilde{w}\left(\theta;t,t\right)-\tilde{w}\left(\theta;t,r\right)\right|\le \mathbf{A}_{6}+\mathbf{A}_{7}$
		where
		\[
		\mathbf{A}_{6} \coloneqq \sup_{\theta\in\Theta_{n,\eta}}E[S_{T\wedge C|Z}(t)|w(\theta,t,Z(t))-w(\theta,t,Z(r))|]\mu(t),
		\]
		and
		\[
		\mathbf{A}_{7} \coloneqq \sup_{\theta\in\Theta_{n,\eta}}E[S_{T\wedge C|Z}(t)|w(\theta,t,Z(r))|]|\mu(t)-\mu(r)|.
		\] 
		We next find the upper bounds of $\mathbf{A}_{6}$ and $\mathbf{A}_{7}$ in terms of $|t-r|$.
		
		We first note that
		\begin{align*}
			& w\left(\theta,t,Z\left(t\right)\right)-w\left(\theta,t,Z\left(r\right)\right)\\
			= & w_{1}\left(\theta_{\xi},t,Z\left(t\right)\right)\left\{ \alpha\left(t\right)-\alpha_{0}\left(t\right)+\left(\beta-\beta_{0}\right)^{\top}Z\left(t\right)\right\} \\
			& -w_{1}\left(\theta_{\xi},t,Z\left(r\right)\right)\left\{ \alpha\left(t\right)-\alpha_{0}\left(t\right)+\left(\beta-\beta_{0}\right)^{\top}Z\left(r\right)\right\} \\
			= & \left\{ w_{1}\left(\theta_{\xi},t,Z\left(t\right)\right)-w_{1}\left(\theta_{\xi},t,Z\left(r\right)\right)\right\} \left\{ \alpha\left(t\right)-\alpha_{0}\left(t\right)\right\} \\
			& +\left(\beta-\beta_{0}\right)^{\top}\left\{ Z\left(t\right)w_{1}\left(\theta_{\xi},t,Z\left(t\right)\right)-Z\left(r\right)w_{1}\left(\theta_{\xi},t,Z\left(r\right)\right)\right\} \\
			= & \left\{ w_{1}\left(\theta_{\xi},t,Z\left(t\right)\right)-w_{1}\left(\theta_{\xi},t,Z\left(r\right)\right)\right\} \left(\alpha\left(t\right)-\alpha_{0}\left(t\right)\right)\\
			& +\left(\beta-\beta_{0}\right)^{\top}Z\left(t\right)\left\{ w_{1}\left(\theta_{\xi},t,Z\left(t\right)\right)-w_{1}\left(\theta_{\xi},t,Z\left(r\right)\right)\right\} \\
			& +\left(\beta-\beta_{0}\right)^{\top}\left(Z\left(t\right)-Z\left(r\right)\right)w_{1}\left(\theta_{\xi},t,Z\left(r\right)\right)\\
			= & \left\{ w_{1}\left(\theta_{\xi},t,Z\left(t\right)\right)-w_{1}\left(\theta_{\xi},t,Z\left(r\right)\right)\right\} \left\{\alpha\left(t\right)-\alpha_{0}\left(t\right)+\left(\beta-\beta_{0}\right)^{\top}Z\left(t\right)\right\}\\
			& +\left(\beta-\beta_{0}\right)^{\top}\left(Z\left(t\right)-Z\left(r\right)\right)w_{1}\left(\theta_{\xi},t,Z\left(r\right)\right).
		\end{align*}
		Then, since $w_{1}\left(\theta,t,Z\left(t\right)\right)$
		is continuous in $\beta^{\top}Z(t)$ in a neighborhood of $\theta_{0}$, it follows from Lemma \ref{lem:GZLip} that
		\begin{align*}
			& E\left[\left|w\left(\theta,t,Z\left(t\right)\right)-w\left(\theta,t,Z\left(r\right)\right)\right|\right]\\
			\lesssim & \left|t-r\right|E\left[\left|\alpha\left(t\right)-\alpha_{0}\left(t\right)+\left|\left(\beta-\beta_{0}\right)^{\top}Z\left(t\right)\right|+\left\Vert \beta-\beta_{0}\right\Vert _{2}\right|\right]\\
			\le & \left|t-r\right|E\left[\left|\alpha\left(t\right)-\alpha_{0}\left(t\right)\right|+\left\Vert \beta-\beta_{0}\right\Vert _{2}\left\Vert Z\left(t\right)\right\Vert _{2}+\left\Vert \beta-\beta_{0}\right\Vert _{2}\right]\\
			\lesssim & \left|t-r\right|\left(\left|\alpha\left(t\right)-\alpha_{0}\left(t\right)\right|+\left\Vert \beta-\beta_{0}\right\Vert _{2}\right).
		\end{align*}
		The last inequality is due to H\"older's inequality and the boundedness of $Z$ under \ref{enu:zbounded}.
		
		Then, for $\mathbf{A}_{6}$, since $S_{T\wedge C|Z}\left(t\right) \le 1$ and $\mu\left(t\right)$
		is bounded for $t\in\left[0,\tau\right]$ by Condition \ref{enu:ratefun},
		\begin{align*}
			\mathbf{A}_{6} & \lesssim\sup_{\theta\in\Theta_{n,\eta}}E[|w(\theta,t,Z(t))-w(\theta,t,Z(r))|]\\
			& \lesssim\sup_{\theta\in\Theta_{n,\eta}}\left|t-r\right|\left(\left|\alpha\left(t\right)-\alpha_{0}\left(t\right)\right|+\left\Vert \beta-\beta_{0}\right\Vert _{2}\right)\\
			& \lesssim \eta^{2/3} \left|t-r\right|.
		\end{align*}
		The last inequality is from \eqref{eq:supeta}.

		Similarly, for $\mathbf{A}_{7}$, by the Lipschitz continuity of $\mu(t)$ in Condition \ref{enu:ratefun},
		\begin{align*}
			\mathbf{A}_{7} & \lesssim \sup_{\theta\in\Theta_{n,\eta}}(\left|\alpha\left(t\right)-\alpha_{0}\left(t\right)\right|+\left\Vert \beta-\beta_{0}\right\Vert _{2})|\mu(t)-\mu(r)|\\
			& \lesssim \eta^{2/3}\left|t-r\right|.
		\end{align*}
		Note that the upper bounds of $\mathbf{A}_{6}$ and $\mathbf{A}_{7}$ only depend $t$ and $r$ through $|t-r|$. 
		Therefore, there exists a constant $L_{2}^{*}$ independent of $t$, $r$ and $\theta$, such that  
		\[
		\sup_{\theta\in\Theta_{n,\eta}}\left|\tilde{w}\left(\theta;t,t\right)-\tilde{w}\left(\theta;t,r\right)\right|\le L_{2}^{*}\eta^{2/3}|t-r|.
		\]

	\end{proof}

	\subsection{The proof of Lemma \ref{lem:supL2}}
	\begin{proof}
		From some small enough $\xi>0$, we define $\delta$ as some positive constant such that $\xi\delta^{2}=\eta^{2}$.
		For any
		arbitrary $t\in\left[0,\tau\right]$ and any $\alpha\in\mathcal{G}_{n,\eta}$, there must be at least one point $r^{*}\in\left[0,\tau\right]$ satisfying
		that $\left|t-r^{*}\right|<\xi$ and $\left|\alpha\left(r^{*}\right)-\alpha_{0}\left(r^{*}\right)\right|<\delta$.
		Otherwise, there would exist $\alpha\in \mathcal{G}_{n,\eta}$ such that $\left|\alpha\left(r\right)-\alpha_{0}\left(r\right)\right|\ge\delta$
		for all $r$ satisfying $\left|t-r\right|<\xi$.
		If so, we have 
		\begin{align*}
			\int_{0}^{\tau}\left|\alpha\left(r\right)-\alpha_{0}\left(r\right)\right|^{2}dr & \ge\int_{\max\left(t-\xi,0\right)}^{\min\left(t+\xi,\tau\right)}\left|\alpha\left(r\right)-\alpha_{0}\left(r\right)\right|^{2}dr\\
			& \ge\int_{\max\left(t-\xi,0\right)}^{\min\left(t+\xi,\tau\right)}\delta^{2}dr\\
			& \ge 2\xi\delta^{2}=2\eta^{2},
		\end{align*}
		contradicting the definition of $\mathcal{G}_{n,\eta}$.
		
		Condition \ref{enu:alpha_para_space} and the property
		of B-splines ensure the uniform Lipschitz continuity of $\alpha\in\mathcal{G}_{n,\eta}$.
		Let $M$ denote a Lipschitz constant for $\mathcal{G}_{n,\eta}$.
		Consequently, for any $t\in\left[0,\tau\right]$,  
		\begin{align*}
			\left|\alpha\left(t\right)-\alpha_{0}\left(t\right)\right| & \le\left|\alpha\left(t\right)-\alpha\left(r^{*}\right)\right|+\left|\alpha\left(r^{*}\right)-\alpha_{0}\left(r^{*}\right)\right|+\left|\alpha_{0}\left(r^{*}\right)-\alpha_{0}\left(t\right)\right|\\
			& \le M\xi+\delta+M\xi.
		\end{align*}
		Note that $M$, $\xi$ and $\delta$ are all independent of the choice of $t$, and $\alpha$ and  $M$ are irrelevant to $\xi$.
		Therefore, 
		$
		\left\Vert\alpha-\alpha_{0}\right\Vert_{\infty}\le2M\xi+\xi^{-1/2}\eta,
		$ for any small enough $\xi>0$ and $\alpha\in\mathcal{G}_{n,\eta}$.
		When $\xi=O\left(\eta^{2/3}\right)$, the order of upper bound of
		$\left\Vert\alpha-\alpha_{0}\right\Vert_{\infty}$ attains
		its minimum in terms of $\eta$. Then, we have the sharpest upper bound
		\[
		\sup_{t\in\left[0,\tau\right]}\left|\alpha\left(t\right)-\alpha_{0}\left(t\right)\right|\le O\left(\eta^{2/3}\right),
		\]
		which is valid for any $\alpha\in\mathcal{G}_{n,\eta}$.
		We then obtain the desired conclusion.
	\end{proof}
	\subsection{The proof of Lemma \ref{lem:CR_order_inequal}}
	
	\begin{proof}
		Recall $\theta_{0n}=\left(\beta_{0,},\alpha_{0n}\right)$ with $\left\Vert \alpha_{0n}-\alpha_{0}\right\Vert _{\infty}=O\left(n^{-\kappa\nu}\right)$ from the proof of Theorem~\ref{them:consistency}.
		Then,
		\begin{align}
			\mathbb{M}_{n,h_{n}}(\hat{\theta}_{n})-\mathbb{M}_{n,h_{n}}(\theta_{0})\ge & \mathbb{M}_{n,h_{n}}(\theta_{0n})-\mathbb{M}_{n,h_{n}}(\theta_{0})\nonumber \\
			= & \left(\mathbb{P}_{n}-\mathbf{P}\right)\left(m_{\theta_{0n},h_{n}}\left(O\right)-m_{\theta_{0},h_{n}}\left(O\right)\right) \label{eq:mtheta0ndiff_1} \\
			& +\left\{ \bM_{h_{n}}\left(\theta_{0n}\right)-\bM_{h_{n}}\left(\theta_{0}\right)-\left(\bM\left(\theta_{0n}\right)-\bM\left(\theta_{0}\right)\right)\right\} \label{eq:mtheta0ndiff_2}  \\
			& +\left\{ \bM\left(\theta_{0n}\right)-\bM\left(\theta_{0}\right)\right\}. \label{eq:mtheta0ndiff_3} 
		\end{align}
		We next shall find the lower founds of \eqref{eq:mtheta0ndiff_1}, \eqref{eq:mtheta0ndiff_2} and \eqref{eq:mtheta0ndiff_3}.
		
		For the first term \eqref{eq:mtheta0ndiff_1}, by Lemma~\ref{lem:hempridiffprop}, it follows from letting $\eta = O( n^{-k\nu})$ that
		\[
		E\left[\left(m_{\theta_{0n},h_{n}}-m_{\theta_{0},h_{n}}\right)^{2}\right]\lesssim\frac{1}{h_{n}}n^{-2\kappa v}.
		\]
		Then, by Chebyshev's inequality, for any $\epsilon>0$,
		\[
		\Pr\left(\frac{\left|\left(\mathbb{P}_{n}-\mathbf{P}\right)\left(m_{\theta_{0n},h_{n}}-m_{\theta_{0},h_{n}}\right)\right|}{n^{-\kappa v}\sqrt{q_{n}/\left(h_{n}n\right)}}\ge\epsilon\right) \le\frac{E\left[\left(m_{\theta_{0n},h_{n}}-m_{\theta_{0},h_{n}}\right)^{2}\right]}{\epsilon^{2}nn^{-2\kappa\nu}q_{n}/\left(h_{n}n\right)} \lesssim\frac{1}{\epsilon^{2}q_{n}}.
		\]
		Clearly, $1/q_{n}\to0$ as $n\to\infty$. Thus, for any $\epsilon>0$,
		\[
		\Pr\left(\frac{\left|\left(\mathbb{P}_{n}-\mathbf{P}\right)\left(m_{\theta_{0n},h_{n}}-m_{\theta_{0},h_{n}}\right)\right|}{n^{-\kappa v}\sqrt{q_{n}/\left(h_{n}n\right)}}\ge\epsilon\right)\to0,
		\]
		that is, 
		\[
		\left|\left(\mathbb{P}_{n}-\mathbf{P}\right)\left(m_{\theta_{0n},h_{n}}-m_{\theta_{0},h_{n}}\right)\right|=o_{p}\left(n^{-\kappa\nu}\sqrt{\frac{q_{n}}{nh_{n}}}\right).
		\]
		Then, 
		\begin{align*}
			r_{n}^{2}\left|\left(\mathbb{P}_{n}-\mathbf{P}\right)\left(m_{\theta_{0n},h_{n}}-m_{\theta_{0},h_{n}}\right)\right| = & o_p\left(n^{\min\left\{ \left(1-\nu-a\right)/2-\kappa\nu,-\left(1-\nu-a\right)/2+\kappa\nu\right\} }\right)\\
			=  & o_p\left(1\right).
		\end{align*}
		The last equation because $\min\left\{ \left(1-\nu-a\right)/2-\kappa\nu,-\left(1-\nu-a\right)/2+\kappa\nu\right\} $ must be nonpositive as the two terms in the $\min$ function have opposite signs or are zero.
		Therefore, 
		\[
		\left|\left(\mathbb{P}_{n}-\mathbf{P}\right)\left(m_{\theta_{0n},h_{n}}-m_{\theta_{0},h_{n}}\right)\right|\le o_{p}\left(r_{n}^{-2}\right),
		\]
		implying 
		\[
		\left(\mathbb{P}_{n}-\mathbf{P}\right)\left(m_{\theta_{0n},h}-m_{\theta_{0},h}\right)\ge-O_{p}\left(r_{n}^{-2}\right).
		\]
		
		For the second term \eqref{eq:mtheta0ndiff_2}, 
		by the similar argument to prove Lemma \ref{lem:wtiproperty} and \eqref{eq:kernelbiasorder},
		\[
		|\bM_{h_{n}}(\theta_{0n})-\bM_{h_{n}}(\theta_{0})-(\bM(\theta_{0n})-\bM(\theta_{0}))| \lesssim h_{n} \left\Vert \alpha_{0n}-\alpha_{0}\right\Vert _{\infty} \le h_{n} n^{-\kappa\nu}.
		\]
		We next show that $h_{n}n^{-\kappa\nu}\lesssim r_{n}^{-2}$, that is $n^{-\kappa\nu-a}\lesssim n^{-\min\left\{ 1-\nu-a,2\kappa\nu\right\} }$.
		It suffices to verify that either $2a+(\kappa+1)\nu-1$ or $a-\kappa\nu$ is larger
		than or equal to 0.
		If both $2a+(\kappa+1)\nu-1<0$ or $a-\kappa\nu<0$, we have $3a +\nu-1<0$ by adding the two inequalities.
		This further implies that $a < 1/3(1-\nu)$ which contradicts $0.4\left(1-\nu\right)\le a$ for $0.5>\nu>0$. 
		Consequently,
		\[
		|\bM_{h_{n}}(\theta_{0n})-\bM_{h_{n}}(\theta_{0})-(\bM(\theta_{0n})-\bM(\theta_{0}))| \lesssim h_{n}  \le  O\left(r_{n}^{-2}\right).
		\]
		and therefore, 
		\[
		\bM_{h_{n}}\left(\theta_{0n}\right)-\bM_{h_{n}}\left(\theta_{0}\right)-\left(\bM\left(\theta_{0n}\right)-\bM\left(\theta_{0}\right)\right)\ge-O\left(r_{n}^{-2}\right).
		\]

		We now turn to the third term \eqref{eq:mtheta0ndiff_3}. By an argument similar to Step 1 of the proof of Theorem \ref{them:consistency}, the third term \eqref{eq:mtheta0ndiff_3} becomes
		\begin{equation*}
			\bM\left(\theta_{0n}\right)-\bM\left(\theta_{0}\right)=
			-E\left[\int_{0}^{\tau}S_{T\wedge C|Z}\left(t\right)H\left(\theta_{0};t,Z\left(t\right)\right)\mu\left(t\right)q\left(\frac{H\left(\theta_{0n};t,Z\left(t\right)\right)}{H\left(\theta_{0};t,Z\left(t\right)\right)}\right)dt\right].
		\end{equation*}
		Since $-q\left(x\right)\ge-\left(x-1\right)^{2}$ for $0.5<x<1.5$, for large enough $n$,
		\begin{multline*}
			\bM\left(\theta_{0n}\right)-\bM\left(\theta_{0}\right) \ge\\
			-E\left[\int_{0}^{\tau}\frac{S_{T\wedge C|Z}\left(t\right)}{H\left(\theta_{0};t,Z\left(t\right)\right)}\mu\left(t\right)\left(H\left(\theta_{0n};t,Z\left(t\right)\right)-H\left(\theta_{0};t,Z\left(t\right)\right)\right)^{2}dt\right].
		\end{multline*}
		By the similar arguments in Step 1 of the proof of Theorem~\ref{them:rate}, 
		there exist $\xi^{*}\in\left(0,1\right)$, such that 
		\begin{align*}
			& \bM\left(\theta_{0n}\right)-\bM\left(\theta_{0}\right) \\
			\ge & -E\left[\int_{0}^{\tau}\frac{S_{T\wedge C|Z}\left(t\right)}{H\left(\theta_{0};t,Z\left(t\right)\right)}\mu\left(t\right)H^{\prime}\left(\left(\xi^{*}\alpha+\left(1-\xi^{*}\right)\alpha_{0},\beta_{0}\right);t,Z\left(t\right)\right)^{2}\left(\alpha_{0n}\left(t\right)-\alpha_{0}\left(t\right)\right)^{2}dt\right]\\
			\ge & -E\left[\int_{0}^{\tau}\frac{S_{T\wedge C|Z}\left(t\right)}{H\left(\theta_{0};t,Z\left(t\right)\right)}\mu\left(t\right)H^{\prime}\left(\left(\xi^{*}\alpha+\left(1-\xi^{*}\right)\alpha_{0},\beta_{0}\right);t,Z\left(t\right)\right)^{2}dt\right]\left\Vert \alpha_{0n}-\alpha_{0}\right\Vert _{\infty}^{2}\\
			\ge & -O\left(n^{-2\kappa\nu}\right)\\
			\ge & -O\left(r_{n}^{-2}\right).
		\end{align*}

		In sum, since \eqref{eq:mtheta0ndiff_1}-\eqref{eq:mtheta0ndiff_3} are all larger than $-O\left(r_{n}^{-2}\right)$, we complete the proof by
		\[
		\mathbb{M}_{n,h_{n}}\left(\hat{\theta}_{n}\right)-\mathbb{M}_{n,h_{n}}\left(\theta_{0}\right)\ge-O_{p}\left(r_{n}^{-2}\right).
		\]
	\end{proof}
	
	\subsection{The proof of Proposition \ref{prop:genasym}}
	\begin{proof}
		By Conditions (A4) and (A5) with $\mathbf{S}\left(\theta_{0}\right)=0$
		from Condition (A2),
		\begin{align*}
			-\sqrt{nh_{n}}\dot{\mathbf{S}}\left(\theta_{0}\right)\left(\hat{\theta}_{n}-\theta_{0}\right)\left[g_{1},g_{2}\right] & =-\sqrt{nh_{n}}\left(\mathbf{S}\left(\hat{\theta}_{n}\right)-\mathbf{S}\left(\theta_{0}\right)\right)\left[g_{1},g_{2}\right]+o_{p}\left(1\right)\\
			& =-\sqrt{nh_{n}}\mathbf{S}\left(\hat{\theta}_{n}\right)\left[g_{1},g_{2}\right]-\sqrt{nh_{n}}\mathbf{S}\left(\theta_{0}\right)\left[g_{1},g_{2}\right]+o_{p}\left(1\right)\\
			& =-\sqrt{nh_{n}}\mathbf{S}\left(\hat{\theta}_{n}\right)\left[g_{1},g_{2}\right]+o_{p}\left(1\right).
		\end{align*}
		Then by Condition (A1) and $\mathbb{S}_{n,h_{n}}\left(\hat{\theta}_{n}\right)=o_{p}\left(\left(nh_{n}\right)^{-1/2}\right)$
		in (A2), 
		\[
		-\sqrt{nh_{n}}\mathbf{S}\left(\hat{\theta}_{n}\right)\left[g_{1},g_{2}\right]=\sqrt{nh_{n}}\left(\mathbb{S}_{n,h_{n}}-\mathbf{S}\right)\left(\theta_{0}\right)\left[g_{1},g_{2}\right]+o_{p}\left(1\right).
		\]
		Thus, 
		\[
		-\sqrt{nh_{n}}\dot{\mathbf{S}}\left(\theta_{0}\right)\left(\hat{\theta}_{n}-\theta_{0}\right)\left[g_{1},g_{2}\right]=\sqrt{nh_{n}}\left(\mathbb{S}_{n,h_{n}}-\mathbf{S}\right)\left(\theta_{0}\right)\left[g_{1},g_{2}\right]+o_{p}\left(1\right)
		\]
		By Condition (A3) holds, Slutsky's theorem yields the asymptotic normality for each $(g_1,g_2)$.
	\end{proof}
	
	\subsection{The proof of Lemma \ref{lem:PsiSidiffprop}}

	Except the statements regarding the envelope function of $\Gamma_{n}$, the proofs are similar to those  of Lemma  \ref{lem:hempridiffprop}. The main difference is that the multiplier is $\sqrt{h}$ rather than $h$. Compared with Lemma  \ref{lem:hempridiffprop}, we need to divide most formula and conclusions by $\sqrt{h}$. Specially, the squared upper bounds are divided by $h$. The detailed proof is omitted.
	
	For the envelope function of $\Gamma_{n}$, by Conditions  \ref{enu:alpha_para_space}-\ref{enu:zbounded} and \ref{enu:H_cond} as well as the compactness of  $\mathcal{G}_1$ and  $\mathcal{G}_2$, for every $g_{1}$ and $g_{2}$, with probability one,
	\[
	|\sqrt{h_{n}}m_{\theta_{0},h_{n}}^{\prime}[g_{1},g_{2}]| \le F_{n},
	\]
	where
	\[
	F_{n}\coloneqq D_{6}\sqrt{h_{n}}\left\{ \int_{0}^{X}K_{h_{n}}\left(X-r\right)dN\left(r\right)+\int_{0}^{X}\int_{0}^{t}K_{h_{n}}\left(t-r\right)dN\left(r\right)dt\right\} ,
	\]
	for some positive constant $D_{6}$.
	We can then let $F_{n}$ be an envelope function for $\Gamma_{n}$.
	Similar to the proof of Lemma  \ref{lem:hempridiffprop}, we can show $\mathbf{P}F^{2}_{n} =O(1)$.
	
	Furthermore, by Conditions \ref{enu:kernel}, \ref{enu:ratefun} and \ref{enu:finiteobs}, with probability one,
	\[
	F_{n}\le\frac{D_{6}}{\sqrt{h_{n}}}J\bar{K}(1+\tau).
	\]
	Then, given $\sqrt{nh_{n}} \to \infty$, $\Pr(F_{n}\le \delta \sqrt{n})=1$ for any $\delta>0$ and large enough $n$.
	Therefore, for large enough $n$, $\mathbf{P}F_{n}^{2}I(F_{n}>\delta\sqrt{n})=\mathbf{P}F_{n}^{2}-\mathbf{P}F_{n}^{2}I(F_{n}<\delta\sqrt{n})=\mathbf{P}F_{n}^{2}-\mathbf{P}F_{n}^{2}=0$ for any $\delta > 0$. Thus, $\mathbf{P}F_{n}^{2}I(F_{n}>\delta\sqrt{n})\to0$ as $n\to\infty$.


	\subsection{The detailed derivation of $\sigma^2$}
	\label{supp:sigmaderive}
	
	We first introduce additional notation.
	Let 
	
	\[
	m_{\theta,h,1}^{\prime}\left[g_{1},g_{2}\right] \coloneqq \int_{0}^{X}
	K_{h}\left(X-r\right)\Delta\cdot H^{(1)}\left(\theta;X,Z(r)\right)\left(g_{2}\left(X\right)+g_{1}^{\top}Z\left(r\right)\right)dN\left(r\right),
	\]
	and
	\[
	m_{\theta,h,2}^{\prime}\left[g_{1},g_{2}\right] \coloneqq \int_{0}^{X}\int_{0}^{t}K_{h}\left(t-r\right)H^{\prime}\left(\alpha\left(t\right)+\beta^{\top}Z\left(r\right)\right)\left(g_{2}\left(t\right)+g_{1}^{\top}Z\left(r\right)\right)dN\left(r\right)dt.
	\]
	Clearly,
	$m_{\theta,h}^{\prime}[g_{1},g_{2}]=m_{\theta,h,1}^{\prime}[g_{1},g_{2}]-m_{\theta,h,2}^{\prime}[g_{1},g_{2}]$.
	%
	

To obtain the explicit form of $\sigma^{2}$, we first derive that

\begin{align*}
& hE\left[\left(m_{\theta_{0},h}^{\prime}\left[g_{1},g_{2}\right]\right)^{2}\right]\\
= & hE\left[\left(m_{\theta_{0},h,1}^{\prime}\left[g_{1},g_{2}\right]-m_{\theta_{0},h,2}^{\prime}\left[g_{1},g_{2}\right]\right)^{2}\right]\\
= & hE\left[\left(m_{\theta_{0},h,1}^{\prime}\left[g_{1},g_{2}\right]\right)^{2}\right]\\
& -2hE\left[m_{\theta_{0},h,1}^{\prime}\left[g_{1},g_{2}\right]m_{\theta_{0},h,2}^{\prime}\left[g_{1},g_{2}\right]\right]+hE\left[\left(m_{\theta_{0},h,2}^{\prime}\left[g_{1},g_{2}\right]\right)^{2}\right]\\
= & \bV_{1}-2\bV_{2}+\bV_{3}.
\end{align*}
We then calculate the respective limits of $\mathbf{V}_{1}$, $\mathbf{V}_{2}$ and
$\mathbf{V}_{3}$ as $h \to 0$.

First, $\mathbf{V}_{1}=\mathbf{V}_{11}+\mathbf{V}_{12}$
where
\begin{multline*}
\mathbf{V}_{11} \coloneqq hE\left[\int_{0}^{\infty}K_{h}^{2}\left(X-r\right)I\left(r\le X\right)\Delta\cdot\left\{ \frac{H^{\prime}\left(\alpha\left(X\right)+\beta^{\top}Z\left(r\right)\right)}{H\left(\alpha\left(X\right)+\beta^{\top}Z\left(r\right)\right)}\right\} ^{2}\right.\\
\Bigg.\times\left(g_{2}\left(X\right)+g_{1}^{\top}Z\left(r\right)\right)^{2}\mu\left(r\right)dr\Bigg],
\end{multline*}
and
\begin{align*}
\mathbf{V}_{12} \coloneqq  & hE\left\{\int_{0}^{\infty}\int_{0}^{\infty}K_{h}\left(X-r_{1}\right)I\left(r_{1}\le X\right)\Delta\cdot\frac{H^{\prime}\left(\alpha\left(X\right)+\beta^{\top}Z\left(r_{1}\right)\right)}{H\left(\alpha\left(X\right)+\beta^{\top}Z\left(r_{1}\right)\right)}\right.\\
& \times\left(g_{2}\left(X\right)+g_{1}^{\top}Z\left(r_{1}\right)\right)\mu\left(r_{1}\right)K_{h}\left(X-r_{2}\right)I\left(r_{2}\le X\right)\\
& \times\left.\frac{H^{\prime}\left(\alpha\left(X\right)+\beta^{\top}Z\left(r_{2}\right)\right)}{H\left(\alpha\left(X\right)+\beta^{\top}Z\left(r_{2}\right)\right)}\left(g_{2}\left(X\right)+g_{1}^{\top}Z\left(r_{2}\right)\right)\mu\left(r_{2}\right)dr_{1}dr_{2}\right\}.
\end{align*}
By Conditions \ref{enu:kernel}, \ref{enu:alpha_para_space},\ref{enu:beta_para_space},
\ref{enu:zbounded}, \ref{enu:H_cond} and \ref{enu:ratefun}, if $h\to0$,
\begin{align*}
\mathbf{V}_{11}= & 4E\Bigg[\int_{0}^{\infty}\int_{0}^{\tau}K^{2}\left(\left(t-r\right)/h\right)/hI\left(r\le t\right)S_{T\wedge C|Z}\left(t\right)H\left(\alpha_{0}\left(t\right)+\beta_{0}^{\top}Z\left(t\right)\right)\Bigg.\\
& \times\left.\left\{ \frac{H^{\prime}\left(\alpha\left(t\right)+\beta^{\top}Z\left(r\right)\right)}{H\left(\alpha\left(t\right)+\beta^{\top}Z\left(r\right)\right)}\right\} ^{2}\left(g_{2}\left(t\right)+g_{1}^{\top}Z\left(r\right)\right)^{2}\mu\left(r\right)dtdr\right]\\
= & 4E\Bigg[\int_{0}^{\infty}\int_{0}^{\tau}K^{2}\left(u\right)I\left(u\le t/h\right)S_{T\wedge C|Z}\left(t\right)H\left(\alpha_{0}\left(t\right)+\beta_{0}^{\top}Z\left(t\right)\right)\Bigg.\\
& \times\left\{ \frac{H^{\prime}\left(\alpha\left(t\right)+\beta^{\top}Z\left(t-uh\right)\right)}{H\left(\alpha\left(t\right)+\beta^{\top}Z\left(t-uh\right)\right)}\right\} ^{2}\left(g_{2}\left(t\right)+g_{1}^{\top}Z\left(t-uh\right)\right)^{2}\mu\left(t-uh\right)dtdu\\
\to & 4\int_{0}^{\tau}K^{2}\left(u\right)duE\Bigg[\int_{0}^{\infty}S_{T\wedge C|Z}\left(t\right)H\left(\alpha_{0}\left(t\right)+\beta_{0}^{\top}Z\left(t\right)\right)\Bigg.\\
& \left.\times\left\{ \frac{H^{\prime}\left(\alpha\left(t\right)+\beta^{\top}Z\left(t\right)\right)}{H\left(\alpha\left(t\right)+\beta^{\top}Z\left(t\right)\right)}\right\} ^{2}\left(g_{2}\left(t\right)+g_{1}^{\top}Z\left(t\right)\right)^{2}\mu\left(t\right)dt\right],
\end{align*}
and
\begin{align*}
\mathbf{V}_{12}= & 4hE\left\{\int_{0}^{\infty}\int_{0}^{\infty}\int_{0}^{\tau}K\left(\left(t-r_{1}\right)/h\right)/hI\left(r_{1}\le t\right)S_{T\wedge C|Z}\left(t\right)\right.\\
& \times H\left(\alpha_{0}\left(t\right)+\beta_{0}^{\top}Z\left(t\right)\right)\frac{H^{\prime}\left(\alpha\left(t\right)+\beta^{\top}Z\left(r_{1}\right)\right)}{H\left(\alpha\left(t\right)+\beta^{\top}Z\left(r_{1}\right)\right)}\left(g_{2}\left(t\right)+g_{1}^{\top}Z\left(r_{1}\right)\right)\mu\left(r_{1}\right)\\
& \times K\left(\left(t-r_{2}\right)/h\right)/hI\left(r_{2}\le t\right)S_{T\wedge C|Z}\left(t\right)H\left(\alpha_{0}\left(t\right)+\beta_{0}^{\top}Z\left(t\right)\right)\\
& \times\left.\frac{H^{\prime}\left(\alpha\left(t\right)+\beta^{\top}Z\left(r_{2}\right)\right)}{H\left(\alpha\left(t\right)+\beta^{\top}Z\left(r_{2}\right)\right)}\left(g_{2}\left(t\right)+g_{1}^{\top}Z\left(r_{2}\right)\right)\mu\left(r_{2}\right)dtdr_{1}dr_{2}\right\}\\
= & 4hE\left\{\int_{0}^{\infty}\int_{0}^{\infty}\int_{0}^{\tau}K\left(u_{1}\right)I\left(u_{1}\le t/h\right)S_{T\wedge C|Z}\left(t\right)\right.\\
& \times H\left(\alpha_{0}\left(t\right)+\beta_{0}^{\top}Z\left(t\right)\right)\frac{H^{\prime}\left(\alpha\left(t\right)+\beta^{\top}Z\left(t-u_{1}h\right)\right)}{H\left(\alpha\left(t\right)+\beta^{\top}Z\left(t-u_{1}h\right)\right)}\left(g_{2}\left(t\right)+g_{1}^{\top}Z\left(t-u_{1}h\right)\right)\\
& \times \mu\left(t-u_{1}h\right) K\left(u_{2}\right)I\left(u_{2}\le t/h\right)S_{T\wedge C|Z}\left(t\right)H\left(\alpha_{0}\left(t\right)+\beta_{0}^{\top}Z\left(t\right)\right)\\
& \times\left.\frac{H^{\prime}\left(\alpha\left(t\right)+\beta^{\top}Z\left(t-u_{2}h\right)\right)}{H\left(\alpha\left(t\right)+\beta^{\top}Z\left(t-u_{2}h\right)\right)}\left(g_{2}\left(t\right)+g_{1}^{\top}Z\left(t-u_{2}h\right)\right)\mu\left(t-u_{2}h\right)dtdu_{1}du_{2}\right\}\\
\to & 0.
\end{align*}
We
then have
\begin{align*}
\mathbf{V}_{1}=&hE\left[\left(m_{\theta,h,1}^{\prime}\left[g_{1},g_{2}\right]\right)^{2}\right]\\
\to & 4\int_{0}^{\tau}K^{2}\left(u\right)duE\left[\int_{0}^{\infty}S_{T\wedge C|Z}\left(t\right)H\left(\alpha_{0}\left(t\right)+\beta_{0}^{\top}Z\left(t\right)\right)\right.\\
& \times\left.\left\{ \frac{H^{\prime}\left(\alpha_{0}\left(t\right)+\beta_{0}^{\top}Z\left(t\right)\right)}{H\left(\alpha_{0}\left(t\right)+\beta_{0}^{\top}Z\left(t\right)\right)}\right\} ^{2}\left(g_{2}\left(t\right)+g_{1}^{\top}Z\left(t\right)\right)^{2}\mu\left(t\right)dt\right].
\end{align*}

For the second term $\mathbf{V}_{2}$, by the similar idea of Lemma \ref{lem:integralvar},
$\mathbf{V}_{2}=\mathbf{V}_{21}+\mathbf{V}_{22}$ where
\begin{multline*}
\mathbf{V}_{21} \coloneqq hE\left[\int_{0}^{\infty}K_{h}\left(X-r\right)I\left(r\le X\right)\Delta\cdot\frac{H^{\prime}\left(\alpha\left(X\right)+\beta^{\top}Z\left(r\right)\right)}{H\left(\alpha\left(X\right)+\beta^{\top}Z\left(r\right)\right)}\left(g_{2}\left(X\right)+g_{1}^{\top}Z\left(r\right)\right)\right.\\
\times\left\{ \int_{0}^{\tau}I\left(t\le X\right)K_{h}\left(t-r\right)I\left(t\ge r\right)H^{\prime}\left(\alpha\left(t\right)+\beta^{\top}Z\left(r\right)\right)\left(g_{2}\left(t\right)+g_{1}^{\top}Z\left(r\right)\right)dt\right\}\\
\times \bigg.\mu\left(r\right)dr\bigg],
\end{multline*}
and 
\begin{multline*}
\mathbf{V}_{22} \coloneqq hE\left\{\int_{0}^{\infty}K_{h}\left(X-r_{1}\right)I\left(r_{1}\le X\right)\Delta\cdot\frac{H^{\prime}\left(\alpha\left(X\right)+\beta^{\top}Z\left(r_{1}\right)\right)}{H\left(\alpha\left(X\right)+\beta^{\top}Z\left(r_{1}\right)\right)}\right.\\
\times \left(g_{2}\left(X\right)+g_{1}^{\top}Z\left(r_{1}\right)\right)\mu\left(r_{1}\right)dr_{1}\\
\times\int_{0}^{\tau}\int_{0}^{\infty}I\left(t\le X\right)K_{h}\left(t-r_{2}\right)I\left(t\ge r_{2}\right)H^{\prime}\left(\alpha\left(t\right)+\beta^{\top}Z\left(r_{2}\right)\right)\\
\bigg.\times \left(g_{2}\left(t\right)+g_{1}^{\top}Z\left(r_{2}\right)\right)dt\mu\left(r_{2}\right)dr_{2}\bigg\}.
\end{multline*}
We then have that
\begin{align*}
\mathbf{V}_{21}= & 4hE\left\{\int_{0}^{\infty}\int_{0}^{\infty}\int_{0}^{\infty}I\left(t\le s\right)K_{h}\left(t-r\right)I\left(r\le t\right)K_{h}\left(s-r\right)I\left(r\le s\right)S_{c}\left(s|Z\right)\right.\\
& \times S_{T|Z}\left(s\right)H^{\prime}\left(\alpha_{0}\left(t\right)+\beta_{0}^{\top}Z\left(r\right)\right)H^{\prime}\left(\alpha_{0}\left(s\right)+\beta_{0}^{\top}Z\left(r\right)\right)\left(g_{2}\left(s\right)+g_{1}^{\top}Z\left(r\right)\right)\\
& \times\bigg.\left(g_{2}\left(t\right)+g_{1}^{\top}Z\left(r\right)\right)dtds\mu\left(r\right)dr\bigg\},
\end{align*}
and $\mathbf{V}_{22}\to0$ for $h\to0$ as $\mathbf{V}_{12}$ is calculated.

For the third term,$\mathbf{V}_{3}=\mathbf{V}_{31}+\mathbf{V}_{32}$
where
\begin{multline*}
\mathbf{V}_{31} \coloneqq hE\left\{\int_{0}^{\tau}\int_{0}^{\infty}\int_{0}^{\infty}I\left(t\le X\right)K_{h}\left(t-r\right)I\left(t\ge r\right)H^{\prime}\left(\alpha\left(t\right)+\beta^{\top}Z\left(r\right)\right)\right.\\
\times\left(g_{2}\left(t\right)+g_{1}^{\top}Z\left(r\right)\right)I\left(s\le X\right)K_{h}\left(s-r\right)I\left(s\ge r\right)H^{\prime}\left(\alpha\left(s\right)+\beta^{\top}Z\left(r\right)\right)\\
\bigg.\times\left(g_{2}\left(s\right)+g_{1}^{\top}Z\left(r\right)\right)dsdt\mu\left(r\right)dr\bigg\},
\end{multline*}
and 
\begin{align*}
\mathbf{V}_{32} \coloneqq & hE\left[\left\{ \int_{0}^{\tau}\int_{0}^{\infty}I\left(t\le X\right)K_{h}\left(t-r\right)I\left(t\ge r\right)H^{\prime}\left(\alpha\left(t\right)+\beta^{\top}Z\left(r\right)\right)\right.\right.\\
& \left.\left.\times\left(g_{2}\left(t\right)+g_{1}^{\top}Z\left(r\right)\right)dt\mu\left(r\right)dr\right\} ^{2}\right].
\end{align*}
After some calculation, $\mathbf{V}_{32}\to0$. Note that $1=I\left(t\le s\right)+t\left(t>s\right)$. 
We then have
\begin{align*}
\mathbf{V}_{31}= & hE\left[\int_{0}^{\tau}\int_{0}^{\infty}\int_{0}^{\infty}\left\{ S_{T\wedge C|Z}\left(t\right)I\left(t\le s\right)+S_{T\wedge C|Z}\left(s\right)I\left(t>s\right)\right\} \right.\\
& \times K_{h}\left(t-r\right)I\left(t>r\right) H^{\prime}\left(\alpha\left(t\right)+\beta^{\top}Z\left(r\right)\right)\left(g_{2}\left(t\right)+g_{1}^{\top}Z\left(r\right)\right)K_{h}\left(s-r\right)I\left(s\ge r\right)\\
& \left.\times H^{\prime}\left(\alpha\left(s\right)+\beta^{\top}Z\left(r\right)\right)\left(g_{2}\left(s\right)+g_{1}^{\top}Z\left(r\right)\right)dsdt\mu\left(r\right)dr\right].
\end{align*}
Since $t$ and $s$ are interchangeable in $\mathbf{V}_{21}$, some algebra yields that $\mathbf{V}_{31}=2\mathbf{V}_{21}$.
It follows that $\lim_{h\to0}\mathbf{V}_{3}-2\mathbf{V}_{2}=0$ and hence $\lim_{h\to0}\mathbf{V}_{11}$ dominates $\sigma^{2}$.
As a result, 
\begin{align*}
\sigma^{2}[g_1,g_2]= & \lim_{h\to0}hE\left\{\left(m_{\theta_{0},h,1}^{\prime}\left[g_{1},g_{2}\right]-m_{\theta_{0},h,2}^{\prime}\left[g_{1},g_{2}\right]\right)^{2}\right\}\\
= & 4\int_{0}^{\infty}K^{2}\left(u\right)duE\left\{\int_{0}^{\tau}S_{T\wedge C|Z}\left(t\right)\right.\\
& \times\left.\frac{H^{\prime}\left(\alpha_{0}\left(t\right)+\beta_{0}^{\top}Z\left(t\right)\right)^{2}}{H\left(\alpha_{0}\left(t\right)+\beta_{0}^{\top}Z\left(t\right)\right)}\left(g_{2}\left(t\right)+g_{1}^{\top}Z\left(t\right)\right)^{2}\mu\left(t\right)\right\}dt.
\end{align*}

\newpage

\section{Technical lemmas}
\label{supp:teclem}

We first present some technical lemmas for establishing the asymptotic
properties of $\hat{\theta}_{n}$.
Lemma \ref{lem:maximial-inequality}
provides a maximal inequality of the empirical process, which plays
a vital role in proving the consistency of $\hat{\theta}_{n}$.
Particularly, Lemma \ref{lem:maximial-inequality} is a specific implication of Lemma 19.36 in \citet{Vaart1998} for our problem and the second statement of \ref{lem:maximial-inequality} directly implies \eqref{eq:epbound_raw}, \eqref{eq:empiricalmaximal} and other similar conclusions.
Lemma \ref{lem:GZLip} shows that a Lipschitz transformation of $Z\left(t\right)$
preserves its continuity as in Condition \ref{enu:zbounded}.
Lemma \ref{lem:GZLip} is mainly used for proving the Lipschitz continuity of $\tilde{g}\left(\theta;t,r\right)$ in $r$ and $\tilde{w}\left(\theta,t,r\right)$ in $r$, where $\tilde{g}\left(\theta;t,r\right)$ and $\tilde{w}\left(\theta,t,r\right)$ are defined in \eqref{eq:gtileddef} and \eqref{eq:wtildedef}, respectively.
Lemma \ref{lem:integralvar} gives a general way to calculate the $\mathcal{L}_{2}\left(\mathbf{P}\right)$
norm of an integral with respect to a counting process.
Lemma \ref{lem:splinebracket} provides a foundation for deriving the bracketing numbers in Lemma \ref{lem:Wbracketing}.
\begin{lem}
\label{lem:maximial-inequality}Let $\mathcal{G}$ be a class of pointwise
measurable and bounded functions $g:\mathcal{X}\mapsto\mathbf{\mathbb{R}}$.
Suppose that the following conditions
hold for some positive constants $U$, $D$, $\xi$ and $v\ge1$:
\begin{enumerate}[label=(L\arabic*)]
	\item $\sup_{g\in\mathcal{G}}\left\Vert g\right\Vert _{\infty}\le U$;
	\item $\sup_{g\in\mathcal{G}}\mathbf{P}g^{2}\le\xi^{2}$;
	\item $N_{[\,]}\left(\epsilon,\mathcal{G},\mathcal{L}_{2}\left(\mathbf{P}\right)\right)\le D\left(\eta/\epsilon\right)^{v}$,
	$0\le\epsilon\le\eta$ for some $D>1$ and any $\eta\ge\xi$.
\end{enumerate}
Then, for $D_{1}=D^{1/v}\vee e$,
\[
\sqrt{n}E\left\Vert \mathbb{G}_{n}\right\Vert _{\mathcal{G}} \lesssim \xi\sqrt{nv\log\left(D_{1}\frac{\eta}{\xi}\right)}+vU\log\left(D_{1}\frac{\eta}{\xi}\right).
\]
Furthermore, if we can choose $\eta=\xi$, 
\[	
\sqrt{n}E\left\Vert \mathbb{G}_{n}\right\Vert _{\mathcal{G}} \lesssim \xi\sqrt{nv\log\left(D_{1}\right)}+vU\log\left(D_{1}\right).
\]
\end{lem}
\begin{proof}
Under Conditions (L1)-(L3), by Lemma 19.36 in \citet{Vaart1998}, 
\begin{equation}
	\sqrt{n}E\left\Vert \mathbb{G}_{n}\right\Vert _{\mathcal{G}}\lesssim\sqrt{n}J_{[\,]}\left(\xi,\mathcal{G},\mathcal{L}_{2}\left(\mathbf{P}\right)\right)\left\{1+\frac{J_{[\,]}\left(\xi,\mathcal{G},\mathcal{L}_{2}\left(\mathbf{P}\right)\right)}{\xi^{2}\sqrt{n}}U\right\}.\label{eq:Vaart1998-1936}
\end{equation}

Here,
\begin{align*}
	J_{[\,]}\left(\xi,\mathcal{G},\mathcal{L}_{2}\left(\mathbf{P}\right)\right) & \coloneqq \int_{0}^{\xi}\sqrt{\log\left(N_{[\,]}\left(\epsilon,\mathcal{G},\mathcal{L}_{2}\left(\mathbf{P}\right)\right)\right)}d\epsilon\\
	& \le\int_{0}^{\xi}\sqrt{\log\left(D\left(\eta/\epsilon\right)^{v}\right)}d\epsilon\\
	& =\sqrt{v}\int_{0}^{\xi}\sqrt{\log\left(D^{1/v}\eta/\epsilon\right)}d\epsilon\\
	& \le\sqrt{v}\int_{0}^{\xi}\sqrt{\log\left(D_{1}\eta/\epsilon\right)}d\epsilon.
\end{align*}
Let $u=\sqrt{\log\left(D_{1}\eta/\epsilon\right)}$. 
Then, using integration by parts, 
\begin{align*}
	& \int_{0}^{\xi}\sqrt{\log\left(D_{1}\eta/\epsilon\right)}d\epsilon\\
	= & \int_{\sqrt{\log\left(D_{1}\eta/\xi\right)}}^{\infty}2u^{2}\eta D_{1}\exp\left(-u^{2}\right)du\\
	= & \eta D_{1}\left\{ -\int_{\sqrt{\log\left(D_{1}\eta/\xi\right)}}^{\infty}ud\exp\left(-u^{2}\right)\right\} \\
	= & \eta D_{1}\left\{ \sqrt{\log\left(D_{1}\eta/\xi\right)}\exp\left(-\log\left(D_{1}\eta/\xi\right)\right)+\int_{\sqrt{\log\left(C^{1/v}\eta/\xi\right)}}^{\infty}\exp\left(-u^{2}\right)du\right\} \\
	= & \xi\sqrt{\log\left(D_{1}\eta/\xi\right)}+\frac{1}{2}D_{1}\eta\sqrt{\pi}\mathrm{Erfc}\left(\sqrt{\log\left(D_{1}\eta/\xi\right)}\right),
\end{align*}
where $\mathrm{Erfc}$ is the complementary error function defined as
\[
\mathrm{Erfc}\left(x\right)=\frac{2}{\sqrt{\pi}}\int_{x}^{\infty}e^{-t^{2}}dt.
\]
It then follows that
\begin{align*}
	J_{[\,]}\left(\xi,\mathcal{G},\mathcal{L}_{2}\left(\mathbf{P}\right)\right) & =\sqrt{v}\left(\frac{1}{2}D_{1}\eta\sqrt{\pi}\mathrm{Erfc}\left(\sqrt{\log\left(D_{1}\eta/\xi\right)}\right)+\xi\sqrt{\log\left(D_{1}\eta/\xi\right)}\right),
\end{align*}
In addition, for $x>0$, 
\[
\mathrm{Erfc}\left(x\right)=\frac{2}{\sqrt{\pi}}\int_{x}^{\infty}e^{-t^{2}}dt\le\frac{2}{\sqrt{\pi}}\int_{x}^{\infty}\frac{t}{x}e^{-t^{2}}dt=\frac{1}{\sqrt{\pi}}\frac{e^{-x^{2}}}{x}
\]
and then
\[
\frac{1}{2}D_{1}\eta\sqrt{\pi}\mathrm{Erfc}\left(\sqrt{\log\left(\frac{D_{1}\eta}{\xi}\right)}\right) \le \frac{\xi}{2\sqrt{\log\left(\frac{D_{1}\eta}{\xi}\right)}}.
\]
Clearly, $D_{1} \eta/\xi \ge e > e^{1/2}$.
It then follows that 
\[
\frac{\xi}{2\sqrt{\log\left(\frac{D_{1}\eta}{\xi}\right)}}\le\xi\sqrt{\log\left(\frac{D_{1}\eta}{\xi}\right)}.
\]
and
\[
J_{[\,]}\left(\xi,\mathcal{G},\mathcal{L}_{2}\left(\mathbf{P}\right)\right)\le2\sqrt{v}\xi\sqrt{\log\left(D_{1}\frac{\eta}{\xi}\right)}.
\]
Then, \eqref{eq:Vaart1998-1936} yields that
\[
\sqrt{n}E\left\Vert \mathbb{G}_{n}\right\Vert _{\mathcal{G}}\lesssim \xi\sqrt{nv\log\left(D_{1}\frac{\eta}{\xi}\right)}+vU\log\left(D_{1}\frac{\eta}{\xi}\right).
\]
Letting $\eta = \xi$, the second statement of the lemma clearly holds.
\end{proof}
\begin{lem}
\label{lem:GZLip}Let $f$ be a differentiable function such that $f^{\prime}\left(x\right)$
is continuous and bounded.
Under Condition \ref{enu:zbounded}, $|\beta^{\top}E[Z(t)-Z(r)]|\le L\sup_{\beta\in\mathcal{B}}\left\Vert \beta\right\Vert_{2}|t-r|$ and there exists $\xi^{*} \in (0,1)$ such that
\begin{multline*}
	\left|E\left[f\left(\beta^{\top}Z\left(t\right)\right)\right]-E\left[f\left(\beta^{\top}Z\left(r\right)\right)\right]\right|\\
	\le L\sup_{\beta\in\mathcal{B}}\left\Vert \beta\right\Vert _{2}\sqrt{E\left[\left|f^{\prime}\left(\beta^{\top}\left\{\xi^{*}Z\left(t\right)+\left(1-\xi^{*}\right)Z\left(r\right)\right\}\right)\right|^{2}\right]}|t-r|.
\end{multline*}
\end{lem}
\begin{proof}
It is seen that 
\begin{align*}
	& \left|\beta^{\top}E\left[Z\left(t\right)-Z\left(r\right)\right]\right|\\
	= & \sqrt{\left|\beta^{\top}E\left[Z\left(t\right)-Z\left(r\right)\right]\right|^{2}}\\
	= & \sqrt{\mathrm{tr}\left(\beta^{\top}E\left[Z\left(t\right)-Z\left(r\right)\right]E\left[Z\left(t\right)-Z\left(r\right)\right]^{\top}\beta\right)}\\
	= & \sqrt{\mathrm{tr}\left(\beta\beta^{\top}E\left[Z\left(t\right)-Z\left(r\right)\right]E\left[Z\left(t\right)-Z\left(r\right)\right]^{\top}\right)}.
\end{align*}
By the Cauchy-Schwarz inequality, 
\begin{align*}
	& \mathrm{tr}\left(\beta\beta^{\top}E\left[Z\left(t\right)-Z\left(r\right)\right]E\left[Z\left(t\right)-Z\left(r\right)\right]^{\top}\right)\\
	\le & \sqrt{\mathrm{tr}\left(\left(\beta\beta^{\top}\right)^{2}\right)}\sqrt{\mathrm{tr}\left(\left(E\left[Z\left(t\right)-Z\left(r\right)\right]E\left[Z\left(t\right)-Z\left(r\right)\right]^{\top}\right)^{2}\right)}\\
	= & \beta^{\top}\beta\left(E\left[Z\left(t\right)-Z\left(r\right)\right]^{\top}E\left[Z\left(t\right)-Z\left(r\right)\right]\right).
\end{align*}
By \ref{enu:zbounded},
\begin{align*}
	\left|\beta^{\top}E\left[Z\left(t\right)-Z\left(r\right)\right]\right| & \le\left\Vert \beta\right\Vert _{2}\sqrt{E\left\Vert Z\left(t\right)-Z\left(r\right)\right\Vert _{2}^{2}}\\
	& \le\left\Vert \beta\right\Vert _{2}L\left|t-r\right|.
\end{align*}

By the mean value theorem, there exists $\xi^{*}\in\left[0,1\right]$,
such that
\begin{align*}
	& \left|E\left[f\left(\beta^{\top}Z\left(t\right)\right)\right]-E\left[f\left(\beta^{\top}Z\left(r\right)\right)\right]\right|\\
	= & \left|E\left[f^{\prime}\left(\beta^{\top}\left\{\xi^{*}Z\left(t\right)+\left(1-\xi^{*}\right)Z\left(r\right)\right\}\right)\left\{ \beta^{\top}\left(Z\left(t\right)-Z\left(r\right)\right)\right\} \right]\right|\\
	\le & E\left[\left|f^{\prime}\left(\beta^{\top}\left\{\xi^{*}Z\left(t\right)+\left(1-\xi^{*}\right)Z\left(r\right)\right\}\right)\right|\left|\beta^{\top}\left(Z\left(t\right)-Z\left(r\right)\right)\right|\right]\\
	= & \sqrt{E\left[\left|f^{\prime}\left(\beta^{\top}\left\{\xi^{*}Z\left(t\right)+\left(1-\xi^{*}\right)Z\left(r\right)\right\}\right)\right|^{2}\right]}\sqrt{E\left[\left|\beta^{\top}\left(Z\left(t\right)-Z\left(r\right)\right)\right|^{2}\right]}\\
	\le & L\sup_{\beta\in\mathcal{B}}\left\Vert \beta\right\Vert _{2}\sqrt{E\left[\left|f^{\prime}\left(\beta^{\top}\left\{\xi^{*}Z\left(t\right)+\left(1-\xi^{*}\right)Z\left(r\right)\right\}\right)\right|^{2}\right]}|t-r|.
\end{align*}
\end{proof}
\begin{lem}
\label{lem:integralvar}Let $\{\mathcal{F}_{t}: t \ge 0 \}$ be a filtration, $f$ be a bounded stochastic
process and $N$ be a bounded counting process for $t\in\mathbb{R}^{+}$.
Suppose that (i) $f$ is $\mathcal{F}_{t}$-predictable;
(ii) the $\mathcal{F}_{t}$-martingale $M=N-\Lambda$
satisfies that $EM^{2}<\infty$ for $t\in\mathbb{R}^{+}$ and
$\Lambda$ is the unique deterministic commentator for $N$; and
(iii) $f$ and $N$ are independent. Then, $E\left[\left(\int_{0}^{\infty}fdN\right)^{2}\right]=E\left[\int_{0}^{\infty}f^{2}d\Lambda\right]+E\left[\left(\int_{0}^{\infty}fd\Lambda\right)^{2}\right]$.
\end{lem}
\begin{proof}
By the definition of $M$, $E\left[\int_{0}^{\infty}fdM\right]=0$
and
\begin{align*}
	E\left[\left(\int_{0}^{\infty}fdN\right)^{2}\right]= & E\left[\left(\int_{0}^{\infty}fdM+\int_{0}^{\infty}fd\Lambda\right)^{2}\right]\\
	= & E\left[\left(\int_{0}^{\infty}fdM\right)^{2}\right]+2E\left[\int_{0}^{\infty}fdM\int_{0}^{\infty}fd\Lambda\right]\\
	& +E\left[\left(\int_{0}^{\infty}fd\Lambda\right)^{2}\right]\\
	= & E\left[\left(\int_{0}^{\infty}fdM\right)^{2}\right]+E\left[\left(\int_{0}^{\infty}fd\Lambda\right)^{2}\right].
\end{align*}
The last equality holds because
\begin{align*}
	E\left[\int_{0}^{\infty}fdM\int_{0}^{\infty}fd\Lambda\right] & =E\left[\int_{0}^{\infty}fE\left[dM|f\right]\int_{0}^{\infty}fd\Lambda\right]\\
	& =E\left[\int_{0}^{\infty}fE\left[dM\right]\int_{0}^{\infty}fd\Lambda\right]\\
	& =0,
\end{align*}
since $f$ and $N$ are independent. By Corollary 2.3.2, Theorem 2.4.3
and Theorem 2.5.1 in \citet{Fleming1991},
\begin{align*}
	E\left[\left(\int fdM\right)^{2}\right] & =E\left\langle \int fdM,\int fdM\right\rangle\\
	& =E\int f^{2}d\left\langle M,M\right\rangle\\
	& =E\left[\int f^{2}d\Lambda\right].
\end{align*}
Hence, the lemma follows. 
\end{proof}
\begin{rem}
The filtration $\mathcal{F}_{t}$ in Condition \ref{enu:ratefun}
satisfies the requirements in Lemma \ref{lem:integralvar} for $\mathbb{M}_{n,h,1}\left(\theta\right)$
and $\mathbb{M}_{n,h,2}\left(\theta\right)$. The independence
condition could be relaxed by that $f$ and $N$ are independent conditionally
on $\mathcal{F}_{t-}$.
\end{rem}

\begin{lem}
\label{lem:splinebracket}Under Condition~\ref{enu:sieveoder}, for any  $0<\epsilon<1$, there exists a constant $c>0$, 
\[
N_{[\,]}\left(\epsilon,\mathcal{A}_{n,l},\Vert\cdot\Vert_{2
}\right) \lesssim \left(1/\epsilon\right)^{cq_{n}},
\]
where $q_{n}=m_n+l$ as shown in \eqref{eq:seivespace} and Condition~\ref{enu:sieveoder}.
\end{lem}
\begin{proof}
By the similar bracketing number calculation in
\citet{Shen1994} and Lemma A1 in \citet{Lu2007}, the lemma follows.
\end{proof}

\newpage

\bibliographystyle{apalike}

\bibliography{asyntransbib}